\documentclass[12pt,preprint]{aastex}
\usepackage{graphics}

\begin{document}
\slugcomment{Submitted to ApJ Supplement}

\title{Ultra-luminous X-ray Sources in nearby galaxies from ROSAT HRI
observations I. data analysis}

\author{Ji-Feng Liu and Joel N. Bregman}
\affil{Astronomy Department, University of Michigan, MI 48109}

\begin{abstract}

X-ray observations have revealed in other galaxies a class
of extra-nuclear X-ray point sources with X-ray luminosities of
$10^{39}$--$10^{41}$ erg/sec, exceeding the Eddington luminosity for stellar
mass X-ray binaries.
These ultra-luminous X-ray sources (ULXs) may be powered by intermediate mass
black hole of a few thousand $M_\odot$ or stellar mass black holes with special
radiation processes. 
In this paper, we present a survey of ULXs in 313 nearby galaxies with
$D_{25}$$>$$1^\prime$ within 40 Mpc with 467 ROSAT HRI archival observations. 
The HRI observations are reduced with uniform procedures, refined by 
simulations that help define the point source detection algorithm
employed in this survey.
A sample of 562 extragalactic X-ray point sources with $L_X = 10^{38}$--$10^{43}$
erg/sec is extracted from 173 survey galaxies, including 106 ULX candidates
within the $D_{25}$ isophotes of 63 galaxies and 110 ULX candidates between
1--2$\times D_{25}$ of 64 galaxies, from which a clean sample of 109 ULXs is
constructed to minimize the contamination from foreground or background
objects.
The strong connection between ULXs and star formation is confirmed based on the
striking preference of ULXs to occur in late-type galaxies, especially in
star forming regions such as spiral arms. 
ULXs are variable on time scales over days to years, and exhibit a
variety of long term variability patterns.
The identifications of ULXs in the clean sample show some ULXs identified as
supernovae (remnants), HII regions/nebulae, or young massive stars in star
forming regions, and a few other ULXs identified as old  globular clusters.
In a subsequent paper, the statistic properties of the survey  will be studied
to calculate the occurrence frequencies and luminosity functions for ULXs in
different types of galaxies  to shed light on the nature of these enigmatic
sources.

\end{abstract}

\keywords{catalogs -- galaxies: general -- X-rays: binaries -- X-rays: galaxies}

\section{INTRODUCTION}


Ultra-luminous X-ray sources (ULXs) are extra-nuclear sources with luminosities
in the range of $ 10^{39}-10^{41}$ erg/sec in other galaxies, and have been
observed by ROSAT, ASCA, recently by XMM-Newton and Chandra Observatory in
large numbers.
As compared to the cases of the X-ray binaries in our Galaxy, which are powered
by accretion onto neutron stars or stellar mass black holes and have
luminosities of $ 10^{33}-10^{38}$ erg/sec, the luminosities of ULXs require
accreting compact objects of masses  $10^3$ -- $10^4$ $M_\odot$ if they emit at
$10^{-2}$ of the Eddington luminosity, typical of Galactic X-ray binaries.
While the required masses could be much larger if they emit at much less than
$10^{-2}$ of the Eddington luminosity, as in the cases of some low luminosity
active galactic nuclei (AGN), the masses cannot be much greater than $10^5$
$M_\odot$ for these extra-nuclear sources to survive the dynamic friction over
a few Gigayears (Colbert 1999).
Such intermediate mass compact objects can only be black holes, and if they
exist, are the missing links between stellar mass black holes and super-massive
black holes in the nuclei of galaxies.

While the explanation with intermediate mass black holes is simple, intriguing
and astrophysically interesting, such black holes are not predicted by ordinary
stellar evolution theories. It is suggested that black holes of a few hundred
$M_\odot$ can form from the death of Pop III stars, and more massive ones might
form from multiple stellar interactions in some dense clusters, hencing
manifest as ultra-luminous X-ray sources (Portegies Zwart et al.  2002).
Alternatively, these X-ray sources could be stellar mass black holes or neutron
stars whose apparent super-Eddington luminosities are due to some special
processes. One suggestion is that radiation pressure-dominated accretion disks
with photon-bubble instabilities are  able to emit truly super-Eddington
luminosities (Begelman 2002). Another suggestion is that beaming effects can
produce the observed luminosities of ULXs (King et al. 2001).


The leading goals in ULX studies are to determine the masses of the primary, to
understand how and where they form, and to find out how they emit at such high
luminosities.
In the last few years many efforts have been made to address these questions,
and important clues have been revealed.
However, these studies mainly focus on some well-known objects and galaxies
(e.g., M81 X-9, NGC5204 X-1, Antenna galaxy, Cartwheel galaxy) except for a few
works (e.g., Colbert \& Ptak, 2002, hereafter CP2002; Swartz et al. 2004; Ptak
\& Colbert 2004), and now it is time to define a complete sample of ULXs to
study the properties of the phenomenon and test critical ideas.
%
One resource to rely on for this purpose is the data archive of the ROSAT High
Resolution Imager (HRI), which includes 5403 observations in the ten years from
June 1990 to February 1999.
The HRI is a multichannel plate detector of $38^\prime \times 38^\prime$ square
field of view, large enough to contain all individual nearby galaxies other
than LMC, SMC, M31, and M33 in our Local Group.
Its spatial resolution is suitable for extra-galactic point source studies,
with on-axis FWHM $<5^{\prime\prime}$ and a pixel scale of
$0.^{\prime\prime}5$/pixel, adequate to resolve point sources in most cases.
Also, the archival observations have reasonable sky coverage for survey
purposes.  For example, the 5393 pointed HRI observations used in the First
ROSAT HRI Source Catalog (1RXH; ROSAT scientific team, 2000) covers about 2\%
of the sky.

The large database of ROSAT HRI observations has not been thoroughly exploited
for complete samples of ULXs in nearby galaxies. 
Roberts \& Warwick (2000; hereafter RW2000) have used the HRI archive to study
the X-ray properties of nearby galaxies, and detected in 83 galaxies 187
discrete X-ray sources of all luminosity ranges, among which 27 non-nuclear sources
have $L_X$$>$$10^{39}$ erg/sec and can be taken as ULXs.
They used the nearby galaxy sample by Ho, Filippenko \& Sargent (1995), which
was constructed to  search for dwarf Seyfert nuclei in nearby galaxies and
contains 486 bright northern galaxies. Many nearby galaxies with HRI
observations are not included in this sample, and the HRI archive is far from
being fully utilized for surveying ULX.
Recently Colbert \& Ptak (2002) made an effort to analyze the
HRI observations to search for ULXs in a sample of 9999 galaxies in the Third
Reference Catalog of galaxies (RC3; de Vaucouleurs et al. 1991) with $cz$$<$$5000$
km/sec.  
They found  87 ULXs in 54 galaxies, with 37 in early-type galaxies.
However, many ULXs in the CP2002 catalog are projected far from the host
galaxies, and may be false ULXs from foreground stars or background
AGN/QSOs. For example, Irwin et al. (2004) pointed out that the radial
distribution of ULXs in early-type galaxies in the CP2002 catalog is consistent
with a random distribution, thus these ULXs are probably not physically
associated with these early-type galaxies.


Here we present our study of ULXs in nearby galaxies with the wealth of HRI
archive.
To fully utilize the HRI archive, we choose all observations associated with
any RC3 galaxies within 40 Mpc with isophotal diameters $>$1 arcminute. The RC3
galaxy sample, the selected HRI observations and the survey galaxies are
described in section 2. 
In our analysis a wavelet algorithm is used for point source
detection, and in section 3 we discuss its performance  on HRI
images through simulations.
In section 4, we describe the analysis procedures applied on the data,
including the point source detection, the variability tests, astrometric
corrections, associations with galaxies and computation of luminosities. 
These efforts lead to a catalog of 562 extragalactic X-ray sources above
$3\sigma$ detection within $2\times D_{25}$ of 173 out of the 313 survey
galaxies, including 106 ULX candidates within the $D_{25}$ isophotes of 63
galaxies and 110 ULX candidates between 1--2 $\times D_{25}$ of 64 galaxies.
The identifications of these sources are given in section 5.
As a final product, we construct a clean sample of ULXs to include ULXs within,
and a few outside the $D_{25}$ isophotes, with identified foreground stars and
background QSOs excluded.  The comments and finding charts are given for these
ULXs in section 6.
In section 7 we summarize the procedures and preliminary results of the survey.
In a subsequent paper, we will study the occurrence rates and luminosity
functions for ULXs in different types of galaxies.

\section{ROSAT HRI survey of nearby galaxies}

To search for the ultra-luminous X-ray sources in the nearby galaxies from the
ROSAT HRI archival observations, we cross correlated a nearby galaxy sample
with the list of HRI archival observations to select observations with nearby
galaxies within their field of views. Here we describe the nearby galaxy
sample, the selected HRI observations, and the survey galaxy sample.

The nearby galaxy sample is extracted from RC3, one of the most complete
catalog for nearby galaxies having apparent diameters larger than 1 arcminute at
the $D_{25}$ isophotal level, total B-band magnitudes $B_T$ brighter than 15.5,
and redshifts not in excess of 15,000 km/sec. 
Distances to galaxies are collected from the literature, including the HST Key
Project (designated as KP; Freedman et al. 2002), the surface brightness
fluctuation method (SBF; Tonry et al. 2001), the nearby galaxy flow model by
Tully (1992; T92), and the Nearby Galaxy Catalog (T88; Tully 1988). Distances
are also computed using the Hubble relation $v = H_0D$ with $H_0=75$ km/s/Mpc.
The NED service provides the recessional velocity $v$ for most galaxies, and
distances computed from them are designated as NED. Another source of $v$ is
RC3, and the distances computed are designated as V3K, since we use the
velocities corrected to the rest frame defined by the 3K cosmic microwave
background.  When multiple distance measurements are available we use the best
distance measurement, which is KP, followed by SBF, T92, T88, NED, and V3K.

The positions of the galaxies are taken from RC3. NED also gives galaxy
positions that are consistent with RC3 to a few arcseconds for most galaxies.
In the rare occasions when the positions from the two sources differ by
$>$$1^\prime$, we verify the positions by overlaying the isophotal ellipses on
the optical images from the Digital Sky Survey (DSS). The  $D_{25}$ isophotes,
the elliptical contour best corresponding to the 25 mag/arcsec$^2$ blue
isophote, are also taken from RC3.  The $D_{25}$ isophote is a quantitative
description of the domain of a galaxy, though the galaxies can extend beyond
$D_{25}$ in some cases but usually within $2 \times D_{25}$ as seen from DSS
images.

For our ULX study, we choose galaxies with isophotal diameters $>$1 arcminute,
for which RC3 is reasonably complete for bright nearby galaxies. Given HRI
position uncertainties typically below $10^{\prime\prime}$, this choice of
galaxy size leaves space for extra-nuclear sources  to be distinguished from
nuclear ones.
%
%
For a short HRI observation of 10 kiloseconds, the count threshold for
3$\sigma$ detection for on-axis point sources is $\sim12$ photons based on the
simulations in section 3, and for the X-ray spectrum used in section 4 this
corresponds to a flux level of $\sim8\times10^{-14}$ erg/sec/cm$^2$, which in
turn corresponds to a luminosity of $\sim10^{39}$ erg/sec for sources in
galaxies at 10 Mpc. 
Sources with extremely high luminosities, e.g., $L_X$$>$$10^{40}$ erg/sec, are
usually very revealing about the nature of ULXs, and can still be detected in
galaxies at larger distances.  
Here we choose to survey galaxies within 40 Mpc, with the hope of detecting
these extreme ultra-luminous X-ray sources from the large number of galaxies
between 10 Mpc and 40 Mpc.
Galaxies in our Local Group usually have large projected areas and have severe
contamination problems from foreground stars, we thus exclude them from our
study here.
Our above selection criteria lead to a sample of 4434 RC3 galaxies.


A complete list of ROSAT HRI observations was extracted from the MPE ROSAT
site\footnote{http://wave.xray.mpe.mpg.de/rosat/}. 
The list of HRI observations with exposures $>$1000 seconds was cross
correlated with the above RC3 galaxy list with a correlation radius of
$15^\prime$. 
Galaxies only partly observed by the HRI are excluded from our survey for their
incompleteness and large off-axis angles.  Also excluded are galaxies behind
the Galactic plane or the Magellanic Clouds for their dense foreground stellar
fields to reduce the confusion problems for identifying X-ray sources.  In
total, our survey includes 467 observations for 313 galaxies.
These observations were downloaded from the MPE ftp
site\footnote{ftp://ftp.xray.mpe.mpg.de/rosat/archive/} and processed with the
uniform procedures described in section 4. 
%


In the HRI galaxy sample, 313 galaxies were surveyed by 467 HRI observations,
about 7.1\% of the RC3 galaxy sample.
This fraction is comparable to that in the work of Ptak \& Colbert (2004), in
which 766 ($\sim$8\%) of the 9452 RC3 galaxies with $cz$$<$$5000$ km/sec are
surveyed by HRI observations.
In Table 1 the survey galaxies are listed with galactic positions, sizes, the
Galactic HI column densities, the distances, the galaxy types, the B-band, the
IRAS 60$\mu$m and FIR luminosities.
This HRI galaxy sample is compared to the RC3 galaxy sample to reveal possible
preferences in the survey galaxies.
The blue luminosity distribution of the survey galaxies resembles that of the
RC3 galaxy sample to a large extent, as shown in Figure 1, except for a slight
over-sampling of very bright galaxies.
This over-sampling is related to the fact that the galaxies surveyed by the HRI
tend to be larger and closer ones, as demonstrated in Figure 2 and Figure 3.
The morphological type distributions of the galaxies are compared between these
two samples, and an over-abundance of ellipticals, and over-abundances of
lenticulars and S0/a--Sc early spirals to a lesser extent, are found in the HRI
galaxy sample with respect to the RC3 galaxy sample as shown in Figure 4. In
contrast, there is an under-abundance of dwarf spirals and irregulars in the
survey sample, reflecting
a bias in selecting the HRI targets that is against the dwarf spirals and
irregulars.
%
To summarize, the HRI survey galaxy sample is representative for the galaxies
within 40 Mpc, although there are preferences for galaxies that are larger,
closer, optically or X-ray brighter.

\section{simulations of the detection method}


A point source detection method suitable for the HRI images is needed to
study the X-ray point sources in HRI observations.
Two packages for two detection algorithms are widely used in the X-ray
astronomy community.
One is celldetect, which uses the sliding cell method to check for local maxima
and detect unresolved sources.  This was the method adopted by 1RXH.
Another popular package is wavdetect, which employs a wavelet method to detect
point sources.  
For our purpose of detecting X-ray point sources in the HRI images, it is more
effective to use wavdetect rather than celldetect, since celldetect does not
treat diffuse emission well, and tends to split diffuse emission into multiple
point sources.  
For example, celldetect reports 67 point sources for the 10 kilo-second HRI
observation RH600463A02 of an elliptical galaxy NGC 4406, while there is really
only one nuclear source plus diffuse X-ray emission, as correctly reported by
wavdetect.

The wavdetect package was developed for use with the Chandra Observatory data,
and can also be used with other X-ray instruments, including the ROSAT HRI.
However, when using wavdetect with HRI images, we need to address the questions
about the false detection rate, the count (rate) threshold for $3\sigma$
detection, and the correction factor between the detected counts and the true
counts.
Here we study these questions by running wavdetect on simulated HRI images.

For this study, the HRI images are simulated with the task QPSIM in the PROS
package. 
The PSF shape is chosen to be ROSHRI, the in-flight PSF for the HRI detector. 
The background rate is set to a constant of 5 count/second, which is typical
for HRI observations and equivalent to $2.5\times 10^{-7}$ count/second/pixel. 
Artificial sources are evenly spaced on the images, with off-axis angles ($\theta$)
ranging from 0 to $20^\prime$. 
An exposure time of $T$ seconds and a source count are assigned to all
artificial sources on one image. For our simulations, we use a series of source
counts from 3 to $10^4$, and $T$ from $10^3$ to $10^6$ seconds.
The central 4098 pixels ($\sim34^\prime$ for a pixel scale of $0\farcs5$/pixel)
of the simulated HRI images are binned by a factor of four to get a 1024x1024
image to run wavdetect on. 
During the wavdetect runs, the parameter of significance threshold ({\tt
sigthresh}) is set to {\tt 1e-6} for the 1024x1024 images, which would cause
about one out of the total {\tt 1e6} pixels be erroneously identified with a
source.

\subsection{The Scale Set for HRI Images}

The result of a wavdetect run depends on the scale set used in search of point
sources.
The scales chosen should include the range of source sizes across the image, since
a source cannot be found by scales much larger or smaller than the source size.
For ROSAT HRI images, the source size changes with the off-axis angle. 
One estimate of the source size is $R_{50}$, the radius encircling 50\% of the
total source count, which can be expressed in a closed form
as\footnote{http://hea-www.harvard.edu/rosat/rsdc\_www/HRI\_CAL\_REPORT/node12.html}
$R_{50} = 1.175[\sigma_{HRI}^2+\sigma_{aspect}^2+(\sigma_{mirror}+a
\theta^b)^2]^{0.5} arcsec$, with $\sigma_{HRI} = 0\farcs74$,
$\sigma_{aspect} = 1\farcs0$, $\sigma_{mirror} = 1\farcs3$,
$a = 0.0205$, and $b = 2.349$.  This radius changes from an on-axis $2\farcs1$
to $23\farcs0$ at an off-axis angle of $18^\prime$.  Note that the source size
is supposed to enclose (almost) all source counts, and should be larger than
$R_{50}$ by a factor of a few.
Here five sets of scales (in binned pixels), i.e., (1 2 4 8), (2 4 8 16), (3 6
12), (4 8 16), and (5 10 20), are tried to determine the best scale set for HRI
images.

One indicator for choosing the best scale set is the false detection rate. 
A source detected $>$$25^{\prime\prime}$ away from any artificial sources is
defined as a false source.
Our simulations show that there are fewer false detections for larger scales.
In Figure 5 we show the false detection percentages for five scale sets. 
While none of the five scale sets give false detections above $6\sigma$, the
first three scale sets give many more false detections below $6\sigma$ than the
last two.
The total numbers of detected sources from the same HRI images also depend on
the scale sets, with smaller scale sets detecting fewer sources. 
This trend, shown in Figure 6, arises because smaller scales are sensative to
smaller source regions and tend to miss more source counts, leading to a lower
detection significance or even a non-detection. 
These results make the last two scale sets favorable over the first three of
smaller scales.
Here we choose (4 8 16) as the best scale set to use in following simulations
and data analysis.
While the scale set (5 10 20) is also favorable, we choose (4 8 16) since
scales of $2^N$ are computationally convenient for the wavelet algorithm.
For the chosen scale set, the false detection rate is about 2\% for sources
with 3--4$\sigma$, 0.4\% for sources with 4--5$\sigma$, and there is no false
detections above 5$\sigma$ (Figure 5). The false detection rate is $>$10\% for
detections below 3$\sigma$, which are discarded in our analysis of the HRI
observations.

\subsection{Count Thresholds}

An important quantity to determine for an observation is the count threshold
for $3\sigma$ detection.
Completely different from the parameter of significance threshold in wavdetect,
this quantity could mean the count level at which a source will be detected in
a 99.7\% chance, or the count level at which a source will be detected with a
significance of $3\sigma$, i.e., with a source count equal to 3 times the
background error.
%
Though related, these two count levels are not always the same, and in our
analysis we use the latter since it is easier to calculate both theoretically
and through simulations.
By comparing the thresholds from simulations with the theoretical thresholds as
a function of the source size and the background rate, one would be able to
study how the source size changes across the images, with which the count
thresholds across the image of an observation can be computed given its
background rate.

The count threshold  can be computed theoretically given the background level
and the source region.
As adopted by the wavdetect package, the Gehrels error $\sigma_G$ of the
background count enclosed in a source region with an area of $S$ pixels is
$\sigma_G = 1+\sqrt{0.75+dTS}$ for a uniform background rate of $d$
count/second/pixel and an exposure of $T$ seconds.
The source count threshold  for the $3\sigma$ detection is then
\begin{equation} \label{threshold} 3\sigma = 3\times(1+\sqrt{0.75+dTS})
\end{equation}
At large off-axis angles the source region deviates from a circle and is
usually described as an ellipse due to the elongation of the PSF. However, for
the purpose of computing the background count, only the total area of the
source region matters for a uniform background rate.
The background rates in most of the observations in our HRI survey are
reasonably uniform across the HRI images, because the background vignetting is
mild for the central $34^\prime$$\times$$34^\prime$ square field of view of the
HRI detector, and observations with foreground diffuse emission from the
Galactic plane, LMC and SMC are excluded from our survey.
Thus, for the purpose of calculating the area, the source region can be
approximated as a circle with radius in unit of $R_{50}$ as $R = n R_{50}$, and
the source area $S = \pi R^2 = \pi (nR_{50})^2$.  Here $n$ should be $>1$, and
its exact value may vary for different off-axis angles.

The count threshold can also be obtained by simulation runs on images of
artificial sources spanning a large range of count levels.  
In our simulations, when the artificial source counts change from the smallest
3 counts to more and more counts, the sources change from being missed, to
being detected with 1$\sigma$, to 2$\sigma$, to 3$\sigma$, to a few hundred
$\sigma$.  In Figure 7 we show how this count threshold is determined with
a series of count values for certain off-axis angles and background levels. 
These sources are sorted in ascending order of counts and connected in the
count--$\sigma$ plane, the point where the 3$\sigma$ line crosses this
connecting line marks the count value for 3$\sigma$ detection.
For artificial sources with the same counts that have different detection
$\sigma$, two methods are available to sort these sources, one in ascending
order of detection $\sigma$ and another in descending order of detection
$\sigma$.  Since the two methods may lead to different results, both methods
are tried and their average is used as the method-independent count value of
3$\sigma$ detection.
The situation can be rather complex, in which there are more than one crossing,
i.e., detections above 3$\sigma$ and detections below 3$\sigma$ or
non-detections alternate in a range of count values due to the complexity of
the implementation of the detection algorithm. In this situation, the simulated
count threshold is taken as the average of count values from all crossings.

The count thresholds are obtained from the simulations for a grid of
off-axis angles and background levels to determine the dependencies on these
two parameters.  
For a fixed off-axis angle $\theta$, the simulated count thresholds change with
the background level the same way as thresholds computed for a fixed source
region, shown in Figure 8, and consistent with the expectation that the source
region is fixed for a fixed off-axis angle.
With the source region approximated as a circle and its area computed as $S =
\pi (nR_{50})^2$, the value of $n$ can be obtained by fitting the simulated
count thresholds to the count thresholds computed using equation (1) for that
off-axis angle with a least square method.  For example, for the off-axis angle
of $10^\prime.23$ in Figure 8, the best fit $n$ is 2.6, and the equivalent
source radius is $2.6R_{50}(\theta=10^\prime.23)$ = $20\farcs8$.
The fitted $n$ shows a dependency on off-axis angle $\theta$ that can be fitted
to a 4-th polynomial, $n(\theta) =
7.46+0.1184\theta-0.1949\theta^2+0.01813\theta^3-0.00046\theta^4$. 
With this analytical form, the equivalent radius, thus the area, of the source
region can be computed for any off-axis angle, as shown in Figure 9.
The equivalent source radius does not change much within $10^\prime$,
outside which the source radius increases rapidly with the off-axis angle.
Note that the change can be attributed to mostly the change of point source
sizes and partly the change of the effective area as described by the
vignetting function.
Equipped with the source area, the count thresholds can be computed with
equation (1) given the background rate and the exposure time.
In Figure 10, the count rate thresholds for $3\sigma$ detection, or the
observation sensitivity, for a few typical exposures are calculated and
plotted.
Thus computed thresholds are consistent with the simulated ones to within 20\%.

\subsection{Correction Factors}

The detected source counts are usually not the true source counts due to a
number of reasons, e.g., the mismatching between the source region sizes and
the wavelet scales used by wavdetect, vignetting and other nonuniformities of
the detector.  To correct the detected counts to the true counts, one must
apply a correction factor ($C$), which we define as $C = {detected\ counts
\over true\ counts}$.  Our simulations here, with known true source count,
detection significance and detected source count for each source, provide us a
chance to study the correction factor that can be applied to real data.


The simulations reveal strong dependencies of the correction factor on the
off-axis angle and the detection significance.
The dependency of the correction factor upon the detection significance for a
fixed off-axis angle is shown in Figure 11.
For sources with high detection significance, the factor is close to unity and
is approximately constant.
It begins to decrease below a certain significance and drops to a few tenth at
$<3\sigma$.
The details of this dependency vary with the off-axis angle $\theta$. For
sources on the edge of the image, the factor begins to drop at $\sim100\sigma$,
and drops to $\sim0.2$ at $<3\sigma$. For sources at smaller off-axis angles,
the factor begins to drop at smaller significances, and drops to higher values.
In contrast, there is no significant drop in the factor for on-axis sources.

For a fixed off-axis angle $\theta$ and a fixed detection significance
$\sigma$, sources show slightly different correction factors in the
simulations.
An average correction factor $C(\theta,\sigma)$ can be computed from these
simulated sources, with its error representing the dispersion of the simulated
factors.
Based on the simulations, average correction factors are computed for a grid of
off-axis angles and detection significances.  
For a source with any combination of $\theta$ and $\sigma$, the correction
factor  $C(\theta,\sigma)$ can be interpolated from this grid, and applied to
the detected count to obtain the true count.
The interpolated correction factors, plotted as a function of off-axis angles
for $\sigma=3$,10, and 100 in Figure 12, are approximately constant within
$10^\prime$, and decrease outward with the off-axis angle.
This decrease means more source photons are missed by the used wavelet scales
at larger off-axis angles, which can be attributed to the increase of point
source sizes and the vignetting function with the off-axis angle.

The interpolated correction factors are tested by applying them to compute
``corrected'' counts from detected counts and comparing to the true counts of
the simulated sources. The results are shown in Figure 12.
For sources with a high detection significance, the detected counts can be
corrected to within a few percents of the true counts. For example, the
detected counts for 67\% ($1\sigma$) of the simulated sources can be corrected
to within 1\% of their true counts for detection $\sigma \ge 150$, i.e., the
($1\sigma$) fractional error of the corrected counts is 1\%. This error is 1.2\%
for sources with $\sigma=100$, 2.7\% for $\sigma=50$, and 7\% for $\sigma=20$.
The error is considerably larger for sources with lower $\sigma$, e.g., 12\%
for $\sigma=10$, 20\% for $\sigma=5$, and up to 30\% for $\sigma=3$.  
For sources with $\sigma<3$, the corrected counts are significantly lower than
their true counts. This systematic inaccuracy is caused by the small number of
low $\sigma$ sources in the simulation, and will not affect the applicability
of the interpolated correction factors on our HRI data, since we only work on
sources above $3\sigma$.

\section{Data Analysis of HRI observations}


The selected HRI observations are processed to extract point sources with
uniform procedures in light of the simulations in section 3.
Each HRI observation is binned by a factor of 4 into a 1024x1024 image,
on which wavdetect is run with the scale set (4 8 16), i.e.,
(8$^{\prime\prime}$ 16$^{\prime\prime}$ 32$^{\prime\prime}$).
The source list from wavdetect includes the position, detection significance
$\sigma$ and the detected net count for each source.

The off-axis angle $\theta$ is computed for a source from its position on the
image, and $C(\theta,\sigma)$ is computed based on the simulations to correct
the detected count to the true count. 
The corrected count rate, computed as the corrected count divided by the
exposure time, is converted to flux with the conversion factor computed with
PIMMS\footnote{http://heasarc.gsfc.nasa.gov/Tools/w3pimms.html} by assuming a
power-law spectrum between 0.3--8 KeV with a photon index of 1.7, with the
Galactic HI column density derived from the HI map by Dickey \& Lockman (1990).
Such a spectrum, also used by CP2002, was adopted to enable comparison with
recent Chandra observations.
For each source, a Kolmogorov-Smirnov test is also run to check for the
variability during the observation.

For each observation, efforts have been made to improve the source positions by
cross correlating X-ray sources with sources in optical, radio or infrared.
Sources are also cross correlated to galaxy isophotal ellipses to determine
whether they are associated with galaxies, and if they are, their luminosities
are calculated.
%

In the following subsections we describe the data analysis procedures, which
result in a catalog of point sources in other galaxies (Table 2), and a list of
individual observations for ULXs (Table 3).

\subsection{Variability During An Observation}

The temporal variations of an X-ray source can be visualized by its light
curve.
In our study, the background-subtracted light curves are constructed for bright
sources.
The event list for an X-ray source is extracted from the $3\sigma$ elliptical
source region as taken from wavdetect results, which contains 95\% of the
source count for an assumed 2-D Gaussian distribution.
In Figure 13 we show the light curve for a ULX in the Circinus galaxy, in
comparison to the scaled background.
For this observation, multiple observation intervals (OBIs) of total $\sim13$
hours scatter over a $\sim22$ day period, and only a few days with significant
exposure time are plotted. The light curve shows clear variations over these
days.
However, some temporal features are missing from the light curves of HRI
sources due to the discontinuous observation scheme. The ULX in Figure 13 is an
X-ray eclipsing source with a period of 7.5 hours as revealed by Chandra
observations (Bauer et al. 2001), yet the periodicity is not obvious in the HRI
light curve due to the wide separations between OBIs.

The variability can be described quantitatively with the Kolmogorov-Smirnov test.
The Kolmogorov-Smirnov test is a widely used nonparametric method to test
whether two distributions differ significantly, and here it is used to test the
null hypothesis that the source is constant during an observation.
A source can be viewed as constant if the null hypothesis probability (listed
in Table 3) is of the order of 1. 
A source can be viewed as variable if  the probability is much smaller, for
example, $\le$0.1. To be conservative, we define a source as variable if the
null hypothesis probability is  $<$0.01, i.e., the source is variable with a significance 
of $>$99\%.  The timescale for this variability is the duration of
the observation, which ranges from a few hours to a few months.
In Figure 14, we plot for the above mentioned ULX the cumulative probability
curves of the Kolmogorov-Smirnov tests. The null hypothesis probability is 11\%
for the ULX to be constant over the 22 day observation for this source, and the
probability for it to be variable is 89\%.
We caution here that even if the source is constant by a high probability
during OBIs, the source could be variable during the wide separations between
OBIs.

\subsection{Astrometric Corrections}

The positional accuracy of an X-ray source is important in our study for
determining whether the source is associated with a galaxy and whether it is a
nuclear source, and for identifying the source in other wavelengths.
For sources from HRI observations, the positions usually have an uncertainty of
$\sim4^{\prime\prime}$  attributable to the smearing and elongation of HRI PSF
due to residual errors in the aspect solutions associated with the ROSAT wobble
or the reacquisition of the guide star (Morse, 1994).  The detected positions
may offset from the true positions by  up to $\sim10^{\prime\prime}$ in rare
occasions of very bad aspect solutions.

The astrometric solution for an observation can be improved if some X-ray
sources can be identified with objects whose positions have been accurately
catalogued.
In our study, the X-ray sources are cross correlated with the two micron all
sky survey (2MASS; Cutri et al. 2003) and the Faint Images of the Radio Sky at
Twenty centimeters (FIRST) survey catalog (Becker et al. 2003) to identify
optical and radio counterparts.
The correlation procedure works iteratively by re-correlating after adjusting
the source positions based on the identified sources in the previous
correlation, until the number of identified sources and the adjusted positions
stabilize. 
The identifications are visually checked after the iterative procedure
stabilizes.
This process was carried out for 416 observations, and the distribution of the offsets
$\alpha$ between the HRI and catalogued positions for all identified sources is
plotted in Figure 15.
Assuming a 2-D normal distribution of the HRI positions relative to the
catalogued positions, $P(\alpha) = {1 \over 2\pi\sigma^2}exp(-{\alpha^2 \over
2\sigma^2})$, a least square fit to the plotted distribution results in $\sigma
= 3\farcs62\pm0\farcs02$. This is consistent with what was found by Morse
(1994).

With the identified sources registered to their accurate positions, the
positions of unidentified sources are corrected using their positions relative
to the identified ones on the X-ray image. 
Such corrections are carried out when there are X-ray sources identified with
objects within $5^{\prime\prime}$ of their uncorrected positions to reduce the
misidentifications.
This process removes the possible large offsets induced by aspect solution
errors. The position error after correction is mainly attributeable to
statistical errors of centroiding and plate scale variations.
In Table 2 the number of identified sources used to correct the position is
listed for each source. A number $\gg$0 indicates the source position error is
less than $4^{\prime\prime}$. In the rare occasions when this number is zero,
the position error could be up to $10^{\prime\prime}$. 
Note that the identification based merely on positional coincidence may be
insecure, and the astrometric correction based on the identifications may be in
error. Extra cautions should be exercised when there are only one such
identification for an observation, especially when this identification is for
the ULX in the obaservation.


After the above astrometric corrections, multiple observations for the same
galaxies are aligned to identify the same source in different observations. 
The position of a source, as listed in Table 2, is computed from averaging the
positions in these observations. 
If a source is not detected but within the field of view of an observation, the
$3\sigma$ upper limit at its supposed position is computed based on the
simulations and listed in Table 3.

\subsection{Sources Associated with Galaxies}

One question for studying the ULXs in other galaxies is to determine whether an
X-ray source is associated with a galaxy, and further whether it is an
extra-nuclear source.
Once a source is known to be associated with a galaxy, its luminosity can be
calculated from its flux using the distance of the galaxy it is associated
with.

The association of a source with a galaxy is determined by means of the blue
$D_{25}$ isophote in our study.
The separation $\alpha$ between the galaxy center and the source is computed
and compared to the elliptical radius $R_{25}$ of the $D_{25}$ isophotal
ellipse along the great arc connecting the galaxy center and the source. 
This elliptical radius $R_{25}$ has a minimum value of the length of the
semi-minor axis when the source is along the minor axis, and has a maximum
value of the length of the semi-major axis when the source is along the major
axis. 
A source with $\alpha<R_{25}$, i.e., within the $D_{25}$ isophote, is
considered as associated with the galaxy.
Visual inspections of the DSS images show that the $D_{25}$ isophotes are a
good delimiter of the optical domain of the galaxies, though galactic features
extend apparently beyond $D_{25}$, but within 2 $\times D_{25}$ of some
galaxies.
For example, in NGC1313 there are dusty star forming regions between 1 -- 2
$\times D_{25}$, and in NGC1316 there are dust loops and strips between 1 -- 2
$\times D_{25}$.
To not miss any ULXs, a source with $R_{25} < \alpha < 2\times R_{25}$ is
tentatively associated with the galaxy. 
In follow-up studies for the statistical properties, sources within the
$D_{25}$ isophote and sources between 1 -- 2 $\times D_{25}$ are treated
differently.

Nuclear X-ray sources powered by accretion onto central supermassive black
holes of galaxies should be excluded from the class of ULXs.
In this work the source positions after astrometric corrections are better than
$5^{\prime\prime}$, and the positions of optical nuclei are better than a few
arcseconds for most galaxies.  
An offset limit of $10^{\prime\prime}$ is thus chosen to tentatively identify
nuclear and non-nuclear sources.
The identification is verified by examining the X-ray positions on DSS images,
to account for possible large errors in the positions of optical nuclei.
A nuclear source is labeled as `N' in Table 2.
%
%
We note that RW2000 used $25^{\prime\prime}$ to identify nuclear sources based
on the positions without astrometric corrections. This will miss ULXs near the
nuclear region. For example, the ULX in NGC 5204 is $\sim17^{\prime\prime}$
from the nucleus, and was classified as nuclear source in RW2000.

\section{A Catalog of Extragalactic X-ray sources}

Data analysis of the 467 HRI observations in our survey as described in section
4 leads to 562 X-ray sources with detection significance above $3\sigma$ within
$2\times D_{25}$ of 173 out of the 313 survey galaxies.
These include 371 sources within the $D_{25}$ isophotes in 155 galaxies, and
191 sources between 1--2 $\times D_{25}$ in 88 galaxies.

These extragalactic sources are listed in Table 2 by the host galaxy and the
source number within the galaxy in the order of the nuclear offsets.
The positions are corrected with the help of $N_c$ (col 6) X-ray sources
identified with objects that have accurate cataloged positions in the same
field, and are generally reliable to better than $5^{\prime\prime}$ for $N_C > 0$.
The position of a source with respect to its host galaxy is described by the
nuclear offsets in arcseconds and in unit of the elliptical radius $R_{25}$.
For each source, also listed are the maximum detection significance, the
maximum luminosity, and number of observations during which the source was
variable by a probability $>99$\%.
Inspection of the source environments shows that some sources are located in
dusty regions such as dust rings/loops/patches/strips/lanes or spiral arms that
usually harbor star-forming activities and young massive stars. Sources in such
dusty environments are flagged as `D' in the table. Particularly, those on
spiral arms are flagged as `S', and those on dust rings are flagged as `R'.

Efforts have been made to identify the sources in other wavelengths by cross
correlating with the databases of SIMBAD, NED and VizieR.
While based primarily on positional coincidence, some identifications are
strengthened by the rarity and strong X-ray emission power of the objects. 
The X-ray sources are identified to a variety of objects.
Some sources are identified with objects not physically associated with the
host galaxy,  such as background QSO/AGNs, or bright foreground stars (denoted
as `*', prefixed with their spectral type if available).
Some sources are identified with objects within the host galaxy, such as
supernovae (remnants), globular clusters (GC), young massive stars, or nuclei
of the host galaxies. These identifications are given in Table 2 enclosed by
parenthesis.  
Many sources are identified with objects for which the current limited
knowledge can not dictate whether they are foreground stars, objects in the
galaxies, or background QSO/AGNs.
These include faint point sources from optical catalogs such as the USNO-B1
catalog (Monet et al. 2003), which are denoted as `P' in Table 2, or point
sources from radio catalogs such as the FIRST survey catalog, which are denoted
as `Pr'. The sources are denoted as `mP' when there are multiple faint objects
close to the X-ray position. 
Usually magnitudes for these point sources are quoted from the
catalogs\footnote{The B1/B2 magnitudes are taken from USNO-B1; the J/H/K/B/V
magnitudes are from 2MASS; the BT/VT magnitudes are from the Tycho-2 catalog
(Hog et al. 2000).} to illustrate their optical/infrared brightness.
Visual inspections of the optical images reveal some sources positionally
coincident with faint fuzzy non-pointlike features, and they are denoted as
`Z/mZ'.
These P/Pr/Z sources make promising targets for future identification programs
with better positional accuracies and spatial resolutions with instruments such
as Chandra and HST to better understand their nature.

Here we define ULX candidates as those with $L_X>10^{39}$ erg/sec in at least
one observation, and with X-ray positions offset from optical galactic nuclei
by $>10^{\prime\prime}$ .
Note that for a few X-ray nuclear sources the nuclear offset listed in Table 2
are larger than $10^{\prime\prime}$ due to the inaccuracy of optical nuclear
positions in RC3.
ULXs are grouped by their nuclear offsets. ULXs within the $D_{25}$ isophote
are defined as `1ULX', and those between 1--2 $\times R_{25}$ defined as
`2ULX'.
The most extreme ULXs which show $L_X>10^{40}$ erg/sec are defined as `EULX'.
If a source is above $10^{39}$ erg/sec in the observations with the deepest
sensitivity for the host galaxies, it is defined as `ULXd'.
In our survey there are 216 ULXs found in 95 galaxies, with 106 `1ULX' (of
which 14 are `EULX') in the $D_{25}$ isophotes of 63 galaxies, and 110 `2ULX'
(of which 19 are `EULX') between  1--2 $\times R_{25}$ of 64 galaxies.

The individual observations for the 216 ULX candidates from Table 2 are listed
in Table 3.
For an individual observation of a ULX candidate, we list the observation date,
the exposure time, the duration of the observation, the detection $\sigma$, the
correction factor $C(\theta,\sigma)$, the corrected count and its percentage
error, the flux, the luminosity, and the probability of being constant during
the observation from Kolmogorov-Smirnov tests. If a source is not detected in
an observation, we list the upper limits computed from the count thresholds
for $3\sigma$ detection based on the simulations.
The information in this table can be used for long term variability studies,
possibly along with recent observations from instruments on the Chandra and
XMM-Newton observatories.

\section{A Clean Sample of ULXs}

The ULX sample defined in the above section is based on the proximity of the
sources to the host galaxies, and is subject to contamination from background
or foreground objects, as demonstrated by the identifications of some ULXs to
background QSO/AGNs or foreground stars.
To form a basis for future studies on ULXs' environments and optical
identifications, here we construct a clean sample of ULXs to minimize 
contamination.
For this clean sample, we exclude identified QSOs and foreground stars. Since
the ULX candidates between 1--2 $\times D_{25}$ of host galaxies are outside 
the optical domain of most galaxies and less likely to associate with the
galaxies, they are also excluded from the clean sample, except for a few which
show apparent connections to the host galaxies by means of dust loops/strips or
spiral arms.
In total, this clean sample includes 109 ULXs within 61 galaxies, for which the
names are given as `UName' in Table 2, and the finding charts are shown in
Figures 16--75.
In the following we comments on the host galaxies, the environments, possible
identifications and variabilities for these ULXs. The IXO number for a ULX is
given if it is listed in CP2002.

\subsection{NGC253}

This Sc spiral galaxy (at a distance of 3.0 Mpc) is a prototype starburst
galaxy. 
ULX1 is on the outer edge of the galaxy. During seven years, the source was
detected once at 2.4$\pm$0.1 $\times 10^{39}$ erg/sec, and it was below
0.3$\times 10^{39}$ erg/sec in the other five observations.

\subsection{NGC891}

NGC891 is an  edge-on Sb galaxy at a distance of 8.36 Mpc. 
ULX1 is close to but not the nucleus and on the edge of the dust lane. ULX2 is on the
dust lane and is identified as SN1986J.  ULX3 (IXO 3) is on the tip of the dust lane,
and showed a steady luminosity increase from $1.4\times10^{39}$ to
$2.4\times10^{39}$ erg/sec in three years.

\subsection{NGC1042}

ULX1 (IXO 4 in CP2002) is on the tip of the spiral arm of this face-on
Scd spiral galaxy, for which the distance is 8.4 Mpc.

\subsection{NGC1068}

NGC1068 at a distance of 14.4 Mpc is a Sb spiral galaxy with an outer ring.
The luminosities of ULX1 changed from below $1.5\times 10^{39}$ erg/sec to
$\sim4\times10^{39}$ erg/sec between four observations over five years.  ULX2
is positioned on the outer ring, and showed variability during an 18 day
observation.

\subsection{NGC1073}

NGC1073 (at a distance of 15.2 Mpc) is a barred Sc spiral galaxy.  ULX1 (IXO 5)
is close to a star forming knot on the spiral arm, and its luminosities
increased by more than 50\% between two observations separated by half a year.

\subsection{NGC1291}

ULX1 (IXO 6) is located on the outer ring-like spiral arm of
this barred S0/a spiral galaxy, for which the distance is 8.6 Mpc.

\subsection{NGC1313}

Three ULXs are associated with the barred Sd galaxy NGC1313 (at a distance of 3.7
Mpc), which shows scattered star-forming regions outside its $D_{25}$ isophote.  
ULX1 (IXO 7) is close to but definitely not the nucleus, and it exhibits extremely
high luminosities with $L_X>10^{40}$ erg/s.  During six years of observations,
its luminosities vary by more than 50\% and showed a dramatic decrease from
$10^{40}$ to $\sim4\times10^{39}$ erg/sec in less than a month.  Pakull \&
Mirioni (2002; hereafter PM2002) found a $H_\alpha$ nebula around this ULX.
ULX2 is identified as SN1978K in a star forming region outside the $D_{25}$
isophote.
ULX3 (IXO 8) is in another star forming region outside the $D_{25}$ isophote.
During the observations, it has shown variations of $>50$\%. It has been
identified with a R=21.6 mag stellar-like object (Zampieri et al. 2003) within a
bubble nebula (PM2002).
Both ULX1 and ULX3 are variable on time scales of a few tens of days.

\subsection{NGC1316}

This peculiar S0 (lenticular) radio galaxy at a distance of 21.48 Mpc is a member of
the Fornax cluster, and shows pronounced dusty patches/loops/shells outside its
$D_{25}$ isophote, reminiscent of recent merger events and possibly star
forming activities.
ULX1 (IXO 11) and ULX2 (IXO 10) are within the isophote. ULX2 is close to a
faint point source with B2\footnote{The blue photographic magnitude from the J
plate used in the second Palomar sky survey.} = 20.24 mag and R2\footnote{The
red photographic magnitude from the F plate in the second Palomar sky survey.}
= 19.90 mag. Assuming V$\sim$20, the X-ray-to-optical ratio $log(f_x/F_v)
\equiv logf_x(0.3-3.5 KeV)+V/2.5+5.37 \sim -0.2$, and indicates it might be a
background AGN (Stocke et al. 1991). However, due to the large uncertainties in
the optical magnitudes (up to 0.5 mag) and the X-ray spectral shape, this could
be a red old cluster with $M_{B2} \sim -10$ mag in NGC1316, or a M dwarf in our
Galaxy.
Five ULXs are on the dusty features outside the $D_{25}$ isophote. ULX3 is IXO
12 in CP2002.  ULX4 is close to a faint point source with B2 = 20.25 mag and R2
= 19.52 mag.  ULX6 (IXO 13) is close to a faint point source with B2 = 21.56
mag and R2 = 19.93 mag.  Multiple faint optical objects are found around ULX7.

\subsection{NGC1365}

NGC1365 is a two armed Sb galaxy with a giant bar. 
ULX1 is on the edge of the bar. Its luminosity dropped from
$7\times10^{39}$ to below $3\times10^{39}$ erg/sec in one year.
Two ULXs are outside the $D_{25}$ isophote but lie on the extension of
a spiral arm. ULX2 (IXO 16) is close to a faint point source
with B2 = 18.73 and R2 = 19.32. This could be a very young massive star cluster
($M_{B2}\sim -12$ mag) in NGC1365, or a
background QSO.  The luminosity of ULX3 (IXO 15) dropped
from $6\times10^{39}$ to below $3\times10^{39}$ erg/sec in one year.

\subsection{NGC1380}

NGC1380 is a lenticular galaxy at a distance of 7.2 Mpc. ULX1 is on the edge of
the galaxy.

\subsection{NGC1399}

This peculiar E1 elliptical is the central and brightest galaxy in the Fornax
cluster, and is known to have globular clusters four times more abundant than
a typical elliptical and 15 times more abundant than a typical spiral.
ULX1 (IXO 18) and ULX2 (IXO 17) are identified with globular clusters from
Chandra and WFPC2 observations (Angelini et al. 2001).
ULX1 showed a change in luminosities from $\sim8\times10^{39}$ to below
$2\times10^{39}$ erg/sec in less than a year.  
ULX3 is located around some faint fuzzy features.

\subsection{NGC1427A}

This magellanic irregular galaxy is an asymmetric barred galaxy with many small
knots at a distance of 16.9 Mpc in the Fornax I cluster. ULX1 (IXO 21) is
located on the dusty bar.

\subsection{PGC13826}

PGC13826 (IC342) is a face-on Scd spiral at a distance of 3.9 Mpc with  a very
bright nucleus and weak surface brightness spiral arms.
Three ULXs are all on spiral arms. ULX1 is close to but not the second nuclear
source. ULX2 was highly variable during the two day HRI observation. ULX3 (IXO
22) is coincident with a tooth-shaped nebula (PM2002), and is quite variable
and showed spectral state transitions reminiscent of black hole candidates
during ASCA observations (Kobuta et al. 2001).

\subsection{NGC1553}

ULX1 and ULX2 are within the $D_{25}$ isophote in this lenticular galaxy at a distance
of 18.54 Mpc.

\subsection{NGC1559}

ULX1 and ULX2 are both nearby knots on the dusty spiral arms of this Scd spiral
galaxy at a distance of 13.6 Mpc.

\subsection{NGC1566}

This face-on barred Sbc spiral galaxy (at a distance of 13.4 Mpc) shows clear
spiral patterns.
ULX1 (IXO 24) and ULX2 are on the thin spiral arms, and ULX3 is on the
edge of a spiral arm.
During five years of observations, all three ULXs showed variations in luminosity
by more than 50\%.
ULX3 showed variability during a five day observation.

\subsection{NGC1672}

ULX1 and ULX2 (IXO 26 and 27) are on the spiral arms of this barred
Sb spiral galaxy at a distance of 14.5 Mpc.

\subsection{NGC1792}

NGC1792 is a Sbc spiral galaxy at a distance of 10.1 Mpc.  ULX1 and ULX2 (IXO
28) are both close to knots on the spiral arms. Between two
observations in two years, ULX1 showed a change in luminosity from below
$0.7\times10^{39}$ to $3\times10^{39}$ erg/s.

\subsection{NGC2276}

This Sc spiral galaxy at a distance of 36.8 Mpc shows unusual asymmetric spiral
patterns perhaps because of a tidal encounter with its E3 companion NGC2300.
There are many bright knots on the spiral arms, some of them are known to be
recent supernovae.  ULX1 is an extreme ULX  on the spiral arm and close to a
bright knot. Its luminosity increased from 24$\pm$4 $\times 10^{39}$ to
34$\pm$8 $\times 10^{39}$ erg/sec in half a year.

\subsection{NGC2403}

NGC2403 is a Sc galaxy with open spiral arms at a distance of 3.133 Mpc.
ULX1 is on the amorphous spiral arms.

\subsection{PGC23324}

PGC23324 (Holmberg II) is a magellanic irregular galaxy at a distance of 4.5
Mpc with numerous HII regions and blue stellar complexes.
ULX1 (IXO 31) is an extreme ULX with $L_X = 28.4\times10^{39}$ erg/s. It is
located in an HII complex that shows strong He II $\lambda$4686 that requires
ionizing X-ray emission of $\sim1\times10^{40}$ erg/sec, thus arguing against
beaming of the X-ray emission (PM2002). This region also exhibits flat spectrum
radio emission indicative of a supernova remnant (Tongue \& Westpfahl 1995).

\subsection{NGC2775}

This nearly face-on ring Sab spiral galaxy (at a distance of 17.0 Mpc) shows
multiple arms of fine structure. 
ULX1 is on the outer edge of the galaxy and close to a faint object with B2 = 21.58
mag and R2 = 20.15 mag.

\subsection{NGC2782}

This ringed Sa spiral galaxy (at a distance of 37.3 Mpc) is a starburst galaxy
with tidal plumes/tails. ULX1 is an extreme ULX with $L_X \approx 2 \times
10^{40}$ erg/sec, and is located on the dusty plume near the disk.

\subsection{NGC2903}

This barred Sbc spiral is a starburst galaxy at a distance of 7.4 Mpc. 
ULX1 is on the edge of a spiral arm.

\subsection{NGC3031}

NGC3031 (M81) is an Sc spiral galaxy at a distance of 3.42 Mpc.  ULX1 and ULX2
are both on a thin spiral arm. ULX1 is identified with SN1993J.  ULX2 is
identified as a black hole binary system with a O8V secondary based on Chandra
and HST/WFPC2 observations (Liu et al. 2003).

\subsection{PGC28757}

PGC28757 (Holmberg IX) is a dwarf irregular companion to M81 with a low surface
brightness and very blue color, suggesting this is a young galaxy just
condensing out of the HI tidal material. The ULX (IXO 34) is outside
the $D_{25}$  isophote, and surrounded by a barrel shaped $H_\alpha$ emission nebula
(PM2002);  this source is M81-X9 (Fabbiano et al. 1989). It is an
extreme ULX with luminosities varying between 5--13 $\times10^{39}$ erg/sec
during six years of observations. Its luminosity once dropped from 13
$\times10^{39}$ to 5 $\times10^{39}$ erg/sec between two observations in half a
year. La Perola et al. (2001) showed that the X-ray emission is highly variable
over 20 years with spectral changes reminiscent of black hole candidates.

\subsection{NGC3310}

This nearly face-on Sbc spiral galaxy at a distance of 18.7 Mpc shows spiral
arms with sporadic star forming knots. ULX1 and ULX2 (IXO 38) are both around
knots on spiral arms.

\subsection{NGC3623}

NGC3623 (M65) is an Sa spiral galaxy at a distance of 12.3 Mpc with a
pronounced dust arm in its silhouette against the disk.  ULX1 is positioned
on the dust arm.

\subsection{NGC3627}

NGC3627 (M66) is an Sb spiral galaxy at a distance of 8.75 Mpc with great
quantity of dust throughout the disk.
ULX1 is on the outer edge of a spiral arm and
close to a very faint feature on the DSS image.

\subsection{NGC3628}

NGC3628 is an edge-on Sb spiral galaxy at the distance of 10.6 Mpc with a
starburst nucleus, and distorted dust lanes due to its interaction with other
Leo Triplet galaxies NGC3627 and NGC3623.
ULX1 is located on the tip of the dust lane where distortions occur due to
tidal interactions.
Note that the aspect solution of the HRI observation for this galaxy was offset
by up to $15^{\prime\prime}$.

\subsection{NGC4088}

This asymmetric Sbc spiral galaxy at a distance of 17.0 Mpc shows distorted
massive knotty arms with many HII regions.
ULX1 (IXO 42) is located at a bright knot on the spiral arm, and is coincident
with a radio point source from the FIRST survey catalog.

\subsection{NGC4136}

NGC4136 is a face-on Sc spiral galaxy at a distance of 9.7 Mpc.
ULX1 is located on a spiral arm and surrounded by several knots of emission.

\subsection{NGC4151}

This almost face-on ringed Sab galaxy (at a distance of 20.3 Mpc) hosts a
prototype Seyfert 1.5 AGN, and its low surface brightness spiral arms extend
slightly beyond the isophote.
ULX1 (IXO 44) is within the $D_{25}$ isophote and close to a faint point
source with B1\footnote{The blue photographic magnitude from the O plate in the
first Palomar sky survey.} = 19.70 mag and R1\footnote{The red photographic
magnitude from the E plate used in the first Palomar sky survey.} = 19.40 mag.
Three ULXs are on the spiral arms outside the $D_{25}$  isophote. Multiple
faint features are around ULX4 on the DSS image. ULX4 showed a luminosity drop
from $\sim 10\times10^{39}$ to below $1\times10^{39}$ erg/sec between two
observations separated by two years.

\subsection{NGC4190}

This peculiar magellanic irregular galaxy at a distance of 2.8 Mpc show a
moderately high surface brightness indicative of a high recent star-formation
rate.
ULX1 is located $\sim10^{\prime\prime}$ east of the optical nucleus.

\subsection{NGC4254}

NGC4254 (M99) is a face-on Sc grand design spiral galaxy at a distance of 16.8
Mpc in the Virgo cluster.
ULX1 is located on a thin spiral arm, while ULX2 (IXO 46) is  an extreme ULX
with $L_X \approx 15 \times 10^{39}$ erg/sec which is located  $\sim
10^{\prime\prime}$ north of a foreground star.

\subsection{NGC4258}

NGC4258 (M106) is a large barred Sbc spiral galaxy at the distance of 7.727 Mpc
which hosts a weak Seyfert 2 nucleus.  
ULX1 is on the edge of a spiral arm, and ULX2 is on a thin spiral arm. The
luminosities of both ULX1 and ULX2 changed from below $0.3\times10^{39}$ to
$\sim1.6\times10^{39}$ erg/sec in 500 days. ULX1 showed variability during an 20
day observation.

\subsection{NGC4303}

NGC4303 (M61) is a face-on Sc  spiral galaxy with a Seyfert 2 nucleus at a
distance of 15.2 Mpc in the Virgo cluster.
ULX1 is located on a thin knotty spiral arm.

\subsection{NGC4321}

NGC4321 (M100) is a grand-design Sbc spiral galaxy at a distance of 14.13 Mpc.
ULX1 is identified with SN1979C.
ULX2 is on the tenuous tip of a spiral arm.

\subsection{NGC4395}

This is a magellanic spiral galaxy at a distance of 3.6 Mpc hosting a Seyfert 1
nucleus.
ULX1 (IXO 53) is located on a knotty spiral arm.

\subsection{NGC4485 and NGC4490}

The magellanic irregular NGC4485 (at a distance of 9.3 Mpc) and the Sd spiral
NGC4490 (at a distance of 7.8 Mpc) are interacting, and their morphologies are
distorted.
There are three ULXs in NGC4490, all located on irregular dusty regions.
Within the isophote of NGC4485 is NGC4485-ULX1 (IXO 62), located on a dust
lane.  Its luminosity decreased steadily from $4\times10^{39}$ to
$1.5\times10^{39}$ erg/sec between three observations separated by 600 days.

\subsection{NGC4501}

NGC4501 (M88) is a Sb spiral galaxy at a distance of 16.8 Mpc.
ULX1 is an extreme ULX  with $L_X \approx 15 \times 10^{39}$ erg/s.  This ULX
is located on a spiral arm, and showed a luminosity drop from $15 \times
10^{39}$ to $8 \times 10^{39}$ erg/sec between two observations separated by half a year.

\subsection{NGC4559}

This Scd spiral galaxy, at a distance of 5.8 Mpc, hosts two ULXs, NGC4559-ULX1
and ULX2.
ULX1 (IXO 66) is $10^{\prime\prime}$ east of the optical nucleus, while ULX2
(IXO 65) is coincident with some blue fuzzy objects on the extension of a
spiral arm.

\subsection{NGC4565}

NGC4565 is an edge-on Sb spiral galaxy at a distance of 17.46 Mpc.  ULX1 (IXO 67) is an
extreme ULX with $L_X \approx 25 \times 10^{39}$ erg/s. It has been identified
to a globular cluster in the halo of NGC4565 with Chandra and HST/WFPC2
observations (Wu et al. 2002).

\subsection{NGC4594}

NGC4594 is a nearly edge-on Sa spiral galaxy with pronounced dust lanes at a
distance of 9.77 Mpc. 
ULX1 and ULX2 are both on the dust lane.

\subsection{NGC4631}

NGC4631 is an edge-on Sd galaxy at a distance of 6.9 Mpc.  ULX1 (IXO 68) is is
located on the dust lanes.

\subsection{NGC4656 and NGC4657}

The magellanic spiral NGC4656 at a distance of 7.2 Mpc is interacting with its
companion NGC4657 to its north-east. 
ULX1 is on the tenuous tip of the tidal tail, and is
coincident with a very faint object in the DSS image.

\subsection{NGC4697}

This E6 elliptical galaxy at a distance of 11.75 Mpc has an appreciable
globular cluster population.
ULX1 is on the outer edge of the galaxy and  coincident with a faint object
with B1 = 18.81 mag, R1 = 19.15 mag, B2 = 19.39 mag, and R2 = 19.04 mag.  It
showed a luminosity drop from $3.4\times10^{39}$ to below $0.4\times10^{39}$
erg/sec between two observations over one year.

\subsection{NGC4861}

This magellanic spiral galaxy at a distance of 17.8 Mpc has no regular structure
but appears to be a strand of HII regions, with a huge HII complex or OB
association as the bright spot at the south end.
ULX1 (IXO 73) is on the strand of HII regions, and ULX2 (IXO 72) is coincident
with the OB association.
ULX1 is an extreme ULX, and its luminosity increased from $\sim10\times10^{39}$
to $30\times10^{39}$ erg/sec between two observations in half a year.

\subsection{NGC5055}

This Sbc spiral galaxy at a distance of 8.5 Mpc show flocculent spiral arms in
the DSS image.
ULX1 is close to but not the nucleus. ULX2 and ULX3  are on the spiral arms.
ULX4 (IXO 74) is on the outer edge of the galaxy. ULX4 showed a luminosity drop
from $9\times10^{39}$ to $5\times10^{39}$ erg/sec between two observations separated by 
three years.

\subsection{NGC5128}

NGC5128 (Centarus A) is a peculiar lenticular galaxy at a distance of 4.21 Mpc
with a warped lane of gas and dust. It is also a bright radio galaxy with an
X-ray jet.
ULX1 (IXO 76) is located on the edge of the bulge.  This ULX showed luminosities
$\sim8\times10^{39}$ erg/sec in five observations in a ten day time window, but
it was below $0.2\times10^{39}$ erg/sec in the other three observations over eight
years.

\subsection{NGC5194}

NGC5194 (M51) is a grand-design Sbc galaxy at a distance of 7.7 Mpc which is
interacting with its companion NGC5195.
Six ULXs are associated with the galaxy, with ULX6 on the outer edge of a
spiral arm, and all the other five right on the thin knotty spiral arms.
ULX1 (IXO 79), ULX2 and ULX3 (IXO 80) are near bright knots on the DSS image
that are probably young star clusters.
All ULXs have exhibited variations by more than 50\% during six years of
observations.
ULX3 showed a luminosity drop from $4\times10^{39}$ to below $0.7\times10^{39}$
erg/sec in the observations. This source appears to have two-hour periodic
variations in a recent Chandra observation (Liu et al. 2003).
ULX4 is IXO 78 in CP2002.  ULX5 (IXO 81) showed variability during a six day
observation.

\subsection{NGC5204}

NGC5204 is a magellanic irregular galaxy at a distance of 4.3 Mpc.
ULX1 (IXO 77) is positioned  $\sim17^{\prime\prime}$ east of the
optical nucleus.
It showed high variability within a one day observation.
This ULX has been identified with a B0Ib supergiant presumably as a secondary in
a black hole X-ray binary from Chandra and HST observations (Liu et al. 2004).

\subsection{NGC5236}

NGC5236 (M83) is a starburst barred Sc spiral galaxy at a distance of 4.7 Mpc
with very luminous spiral arms. 
ULX1 (IXO 82) is located on the edge of a spiral arm, and its luminosity showed an
increase from $1.5\times10^{39}$ to $2.5\times10^{39}$ erg/sec between two
observations separated by  600 days.

\subsection{NGC5457}

NGC5457 (M101) is a face-on prototype Sc spiral galaxy at a distance of 6.855
Mpc.
ULX1 is coincident with an object with B = 17.80 mag and R = 15.00 mag, and
showed variability within a two day observation.
ULX2 and ULX3 (IXO 83) are both on spiral arms, both showed variability within a 26 day
observation, and the luminosities of both varied by more than 50\% during the
four observations in five years.

\subsection{NGC5774}

NGC5774 is a face-on Sd spiral with bright blue knots in the arms at a distance
of 26.8 Mpc.
ULX1 (IXO 84) is close to knotty features on a chain of knots across the disk.

\subsection{NGC6946}

NGC6946 is a Scd spiral at a distance of 5.5 Mpc with recent star formation
throughout the spiral arm structure and a mild starburst at the center.
ULX1 (IXO 85) and ULX2 are on knotty spiral arms.  ULX1 showed a steady luminosity
decrease from $1.8\times10^{39}$ to below $0.7\times10^{39}$ erg/sec in 800 days.
ULX3 is on the edge of a spiral arm and coincident with a cocoon shaped
supernova remnant (Dunne et al. 2001).  Its extreme luminosity $L_X = 11\times10^{39}$ erg/sec either
comes from the colliding SNR or from accretion onto a newborn black hole
(Roberts \& Colbert, 2003).

\subsection{NGC7314}

This Sbc spiral, at a distance of 12.6 Mpc, has a Seyfert 1.9 nucleus and HII
regions all over its spiral arms. ULX1 (IXO 86) is on the outer edge of the galaxy.

\subsection{NGC7590}

This Sbc spiral at a distance of 17.3 Mpc has a Seyfert 2 nucleus and very
bright spiral arms in the inner part. ULX1 (IXO 87) is close to knotty features on the
spiral arm.  Its luminosity showed a drop from $6.5\times10^{39}$ to
$3.4\times10^{39}$ erg/sec in a year.

\subsection{NGC7714}

This Sc spiral at a distance of 36.9 Mpc is interacting with its dwarf
irregular companion NGC7715. ULX1 is an extreme ULX with $L_X \sim
40\times10^{39}$ erg/sec, and is located on the tenuous plume on the outer
edge of a spiral arm in NGC7714.

\subsection{NGC7742}

This Sb spiral at a distance of 22.2 Mpc has an extremely bright nucleus and many
poorly resolved knotty arms.
ULX1 is on the outer edge of the galaxy.

\section{Summary}

In this paper, we report our archival ROSAT HRI survey of extragalactic X-ray
point sources and ultraluminous X-ray sources in  a sample of 313 nearby
galaxies with $D_{25} > 1^\prime$ within 40 Mpc.

The survey was carried out with well defined data reduction procedures.
To detect point sources from HRI observations we utilize the wavdetect package,
an implementation of a wavelet algorithm, for which simulations were run to
better understand its characteristics such as the count thresholds for
$3\sigma$ detection and correction factors.
For sources detected above $3\sigma$ from the HRI observations in our survey,
we compute their count rate and fluxes, and test whether they are variable
during the observations with Kolmogorov-Smirnov tests.
The X-ray positions were corrected by registering X-ray sources to the accurate
positions of their counterparts in other wavelengths, and the positional
accuracy is better than $5^{\prime\prime}$ for most X-ray sources.
X-ray Sources within $2\times D_{25}$ of a galaxy are considered associated
with the galaxy and their luminosities are computed by placing them at the
distance for the galaxy.
The uniform data reduction procedures lead to 562 extragalactic X-ray sources
in 173 nearby galaxies spanning a luminosity range of $10^{38}$ -- $10^{43}$
erg/sec, with 371 sources within the $D_{25}$ isophotes in 155 galaxies, and
191 sources between 1--2 $\times D_{25}$ in 88 galaxies.

In our survey we define those extra-nuclear sources with $L_X > 10^{39}$ erg/sec
as ULX candidates. This leads to a sample of 216 ULX candidates, which includes
a sample of 106 ULXs within the $D_{25}$ isophotes in 63 galaxies (1ULX), and a
sample of 116 ULXs between 1--2 $\times D_{25}$ in 64 galaxies (2ULX).
Thus defined ULX candidates are possibly not in the host galaxies, and ten (16)
in the 1ULX (2ULX) sample are already identified as background QSOs or
foreground stars.
To minimize such contaminations, we constructed a clean sample of 109 ULXs in
61 galaxies by excluding those identified stars and QSOs, and excluding most of
the 2ULX sample since they are more likely contaminating sources due to their
large separations from the galaxies.
This clean sample forms a good basis for studies on ULX's environments and
identifications.

A few conclusions can be drawn from examinations of the clean sample. 
ULXs are preferentially found  in late-type galaxies.  for example, 49 out of
181 spiral galaxies in our survey host 89 ULXs, while 4 out of 93 early-type
galaxies host 7 ULXs if we exclude the 8 ULXs in two peculiar lenticular
galaxies NGC1316 and NGC5128.
There is also a strong tendency for ULXs to occur in dusty regions of star
formation, with 84 of the 109 ULXs in the clean sample in such regions.
Specifically, 51 ULXs occur on thin dusty spiral arms. Also note that the seven
ULXs in NGC1316 are associated with dust loops/shells/patches reminiscent of
star forming activities.
The above statistics reveals a strong connection between the ULX phenomenon and star
formation, as reported by previous authors.

Some ULXs have demonstrated great variability over a range of time scales
from days to up to 10 years.
Fifteen out of 109 ULXs in the clean sample have shown variability during at
least one observation lasting from days to months. These ULXs are all located in
star forming regions in late-type galaxies.
For ULXs with more than one HRI observations, a variety of temporal behaviors
have been exhibited.
These include sudden bursts with small duty cycles (e.g., NGC5128-ULX1), steady
increases (e.g., NGC891-ULX3), steady decreases (e.g., NGC6946-ULX1), or
occasional sharp drops from almost constant high luminosity (e.g.,
NGC1313-ULX1).

ULXs show a very diverse nature seen from the identifications for the clean
sample. 
Five ULXs are identified as recent supernovae (remnants). 
Many ULXs on knotty spiral arms in the DSS images are clearly identified with HII
regions/nebulae in recent observations with better resolutions (PM2002).
There are two confirmed cases in which the ULXs are identified with very young
massive stars: NGC3031-ULX2 with an O8V star (Liu et al. 2003); and
NGC5204-ULX1 with a B0Ib star (Liu et al. 2004). 
These identifications are consistent with the strong link between ULXs and star
formation.
However, there are also ULXs identified with old globular clusters lacking
recent star formation.  For example, NGC4565-ULX1 is identified with a halo
globular cluster, and two ULXs are identified with globular clusters in the
elliptical galaxy NGC1399.
These ULXs in the globular clusters may have been formed differently from those
in star forming regions.


While many qualitative properties have been revealed from studying the clean
sample of ULXs, the quantative properties such as the occurrence rates of ULXs
derived from the clean sample are not strictly accurate because the clean
sample is not a complete sample in the statistical sense, and the
contamination, though small, is not known.
In a subsequent paper, statistical properties, such as the occurrence
frequencies and luminosity functions, of ULXs in different types of galaxies
will be studied quantatively with the contaimnation calrefully calculated and
subtracted, in comparison to statistical studies of ULXs in 766 nearby galaxies
with $cz$$<$$5000$ km/sec in HRI observations by Ptak \& Colbert (2004) and of
ULXs in 82 nearby galaxies in Chandra observations by Swartz et al.  (2004).

\acknowledgements

We are grateful to the NED, VizieR services. We would like to thank Dan Harris,
Samantha Stevenson, Renato Dupke, James Irwin, Ed Lloyd-Daves and Eric Miller
for helpful discussions. We thank the referee for his constructive suggestions.
We gratefully acknowledge support for this work from NASA under grants
HST-GO-09073.


\clearpage

\begin{deluxetable}{lrrlllrlrrlll}
\tabletypesize{\tiny}
\tablecaption{Properties of surveyed galaxies}
\tablehead{
   \colhead{galaxy} & \colhead{GLON\tablenotemark{a}} & \colhead{GLAT\tablenotemark{b}} & \colhead{D25\tablenotemark{c}} & \colhead{d25\tablenotemark{d}} & \colhead{nH\tablenotemark{e}} & \colhead{Distance\tablenotemark{f}} & \colhead{Morphol.\tablenotemark{g}} & \colhead{Hubble\tablenotemark{h}} & \colhead{$L_B$\tablenotemark{i}} & \colhead{$f_\nu(60\mu m)$\tablenotemark{j}} & \colhead{$M_{FIR}$\tablenotemark{k}} & \colhead{Starburst\tablenotemark{l}} \\
   \colhead{      } & \colhead{(deg)} & \colhead{(deg)} & \colhead{($\prime$)} & \colhead{($\prime$)} & \colhead{(cm$^{-2}$)} & \colhead{(Mpc)} & \colhead{Type} & \colhead{T} & \colhead{($L_\odot$)} & \colhead{(Jy)} & \colhead{(mag)} & \colhead{} \\
}

\startdata
NGC247 & 113.95 & -83.56 & 21.4 & 6.9 & 1.55E20 & 2.8[T92] & SAB(s)d & 7 & 1.6E09 & 7.8e-01 & -16.7 & \nodata \\
NGC253 &  97.36 & -87.96 & 27.5 & 6.8 & 1.42E20 & 3.0[T92] & SAB(s)c & 5 & 6.8E09 & 7.6e+02 & -21.8 & Sbrst \\
NGC300 & 299.20 & -79.42 & 21.9 & 15.5 & 3.23E20 & 2.023[KP] & SA(s)d & 7 & 2E09 & $<$6.8e+00 & -17.1 & \nodata \\
NGC404 & 127.03 & -27.01 & 3.5 & 3.5 & 5.31E20 & 3.27[SBF] & SA(s)0- & -3 & 5.9E08 & 2.3e+00 & -15.4 & \nodata \\
NGC520 & 138.70 & -58.06 & 4.5 & 1.8 & 3.27E20 & 27.8[T88] & peculiar & 99 & 1.9E10 & 3.2e+01 & -22.8 & \nodata \\
NGC613 & 229.07 & -80.30 & 5.5 & 4.2 & 1.63E20 & 26.7[T92] & SB(rs)bc & 4 & 5.6E10 & 2.0e+01 & -22.4 & \nodata \\
NGC720 & 173.02 & -70.36 & 4.7 & 2.4 & 1.54E20 & 27.67[SBF] & E5 & -5 & 4.1E10 & $<$2.2e-01 & $<$-17.3 & \nodata \\
NGC891 & 140.38 & -17.42 & 13.5 & 2.5 & 7.64E20 & 8.36[SBF] & SA(s)b & 3 & 6.3E09 & 3.4e+01 & -21.3 & HII \\
\enddata
\end{deluxetable}


\begin{deluxetable}{lllrrrrcllllll}
\tabletypesize{\tiny}
\tablecaption{X-ray sources in surveyed galaxies}
\tablehead{
   \colhead{Galaxy} & \colhead{Name} & \colhead{R.A.} & \colhead{Decl.} & \colhead{$N_c$\tablenotemark{a}}  & \colhead{r\tablenotemark{b}} & \colhead{r/R25\tablenotemark{c}} & \colhead{$\sigma_{Max}$\tablenotemark{d}} & \colhead{$L/10^{39}$\tablenotemark{e}} &\colhead{$N_{v}$\tablenotemark{f}} & \colhead{Env\tablenotemark{g}} & \colhead{ID\tablenotemark{h}} & \colhead{ULX\tablenotemark{i}} & \colhead{UName\tablenotemark{j}}  \\
   \colhead{      } & \colhead{    } &  \colhead{(hh:mm:ss)} & \colhead{(dd:mm:ss)} & \colhead{} & \colhead{($\prime\prime$)} & \colhead{} & \colhead{} & \colhead{(erg/s)} & \colhead{} & \colhead{} & \colhead{} & \colhead{} & \colhead{    } \\
}

\startdata
NGC247 & X1 & 00:47:10.30 & -20:47:10.7 & 0 & 95.4 & 0.16 & 5.9 & 0.133 &   &   &   &   &   \\
NGC247 & X2 & 00:47:04.00 & -20:47:45.7 & 1 & 143.8 & 0.42 & 32.4 & 0.853 &   & D  &   &   &   \\
NGC247 & X3 & 00:46:51.78 & -20:43:27.2 & 1 & 271.1 & 1.10 & 17.9 & 0.427 &   &   &  QSO PHL6625 &   &   \\
NGC247 & X4 & 00:47:27.57 & -20:47:24.6 & 1 & 285.2 & 1.23 & 12.3 & 0.265 &   &   &  P B2/18.83 &   &   \\
NGC247 & X5 & 00:47:03.69 & -20:40:15.5 & 0 & 330.1 & 0.54 & 3.7 & 0.0628 &   &   &   &   &   \\
NGC247 & X6 & 00:46:47.98 & -20:38:36.9 & 1 & 511.8 & 1.37 & 5.3 & 0.095 &   &   &  mZ &   &   \\
NGC247 & X7 & 00:47:03.31 & -20:36:57.1 & 1 & 526.4 & 0.83 & 9.3 & 0.243 &   & D  &   &   &   \\
NGC253 & X1 & 00:47:33.55 & -25:17:24.4 & 1 & 8.9 & 0.04 & 15.4 & 1.74 &   &   & (NGC253) &   N &   \\
\enddata

\tablenotetext{a}{The number of X-ray sources idetified with objects in
astrometric catalogs used to correct the X-ray positions. }

\tablenotetext{b}{The nuclear offset in arcseconds.}

\tablenotetext{c}{The nuclear offset in unit of elliptical radius $R_{25}$}

\tablenotetext{d}{The maximum detection significance  among all observations.}

\tablenotetext{e}{The maximum luminosity among all observations. The X-ray
luminosities are computed using a power-law spectrum between 0.3--8 KeV with a
photon index of 1.7.   }

\tablenotetext{f}{The number of observations in which the probability for the
Kolmogorov-Smirnov test of the source being constant is less than 0.01,
i.e., the source is highly variable during the observation.}

\tablenotetext{g}{The environments of the X-ray source. A flag 'D' indicates
the source is in a dusty region usually associative of star forming activities.
A flag 'S' means the source is associated with spiral arms. }

\tablenotetext{h}{The identification of the X-ray source. Flag '*' means the
source is a bright foreground star, 'P'/('Pr') means the source is positionally
coincident with a point source in optical/IR/(Radio), 'mP' means the source is
close to multiple such point sources and unique identification requires better
positions, 'Z' means the source is close to faint fuzzy features on the optical
image. The identification is enclosed with brackets when it is within the host
galaxy. }

\tablenotetext{i}{This indicates whether the source is a nuclear source (N) or
a ULX candicate (ULX). Prefix '1' stands for ULXs within the $D_{25}$ isophote
of the galaxy, '2' for ULXs  between 1 -- 2 $\times D_{25}$, and 'E' for ULXs
with $L_X > 10^{40}$ erg/s. Suffix 'd' stands for ULXs that are above $10^{39}$
erg/s in the deepest observation for the galaxy. }

\tablenotetext{i}{The ULX name in the clean ULX sample. }

\end{deluxetable}


\begin{deluxetable}{llrrrrrrrrrrcc}
\tabletypesize{\tiny}
\tablecaption{Individual observations for ULX candidates}
\tablehead{
   \colhead{Name} & \colhead{Obs.Src\tablenotemark{a}} & \colhead{Date\tablenotemark{b}} & \colhead{T\tablenotemark{c}} & \colhead{$\Delta D$\tablenotemark{d}} & \colhead{$\sigma$\tablenotemark{e}} & \colhead{$\theta$\tablenotemark{f}} & \colhead{$C(\theta,\sigma)$\tablenotemark{g}} & \colhead{CorCnt\tablenotemark{h}} & \colhead{pErr\tablenotemark{i}} & \colhead{Flux\tablenotemark{j}} & \colhead{$L_X$\tablenotemark{k}} & \colhead{$P_{KS}$\tablenotemark{l}} \\
   \colhead{} & \colhead{} & \colhead{(MJD)} & \colhead{ksec} & \colhead{} & \colhead{} & \colhead{($^\prime$)} & \colhead{} & \colhead{(count)} & \colhead{(\%)} & \colhead{(cgs)} & \colhead{($10^{39}$erg/s)} & \colhead{} \\
}

\startdata
NGC253-ULX1 & rh600088a00.u & 48599 & 3.1 & 1.2d & $<$3 &   &   & $<$12.5 &   & $<$2.64e-13 & $<$0.285 &   \\
  & rh600088a01.u+ & 48779 & 25.4 & 1.2d & $<$3 &   &   & $<$30.2 &   & $<$7.88e-14 & $<$0.0851 &   \\
  & rh600714n00.u & 49720 & 10.9 & 4.1d & $<$3 &   &   & $<$19.7 &   & $<$1.19e-13 & $<$0.129 &   \\
  & rh600714a01.u & 49882 & 19.7 & 23.9d & $<$3 &   &   & $<$23.0 &   & $<$7.72e-14 & $<$0.0834 &   \\
  & rh601111n00.1 & 50802 & 17.3 & 6.9d & 38.6 & 15.49 & 0.85 & 566.1 & 3.92 & 2.19e-12 & 2.37 & 0.42 \\
  & rh601113n00.u & 50996 & 2.1 & 0.6h & $<$3 &   &   & $<$16.5 &   & $<$5.36e-13 & $<$0.579 &   \\
\hline
NGC720-X2 & rh600472n00.15+ & 49360 & 56.8 & 4.4d & 9.3 & 2.61 & 0.95 & 116.3 & 13.56 & 1.38e-13 & 12.7 & 0.67 \\
\hline
NGC720-X3 & rh600472n00.17+ & 49360 & 56.8 & 4.4d & 4.5 & 3.36 & 0.96 & 53.8 & 26.40 & 6.38e-14 & 5.86 & 0.52 \\
\enddata

\tablenotetext{a}{Individual observation. Suffixed is the sequence number in
the source list from wavdetect. 'U' is suffixed when the source is below
$3\sigma$ detection in the observation. '+' is suffixed if the observation has
the deepest sensitivity for the host galaxy. }

\tablenotetext{b}{Observation date in MJD.}

\tablenotetext{a}{Exposure time in kilo-seconds.}

\tablenotetext{d}{Duration of the observation.}

\tablenotetext{e}{Detection significance. }
\tablenotetext{f}{Off-axis angle in arcminutes. }

\tablenotetext{g}{The correction factor $C(\theta,\sigma)$ computed based on
the simulations.}

\tablenotetext{h}{The source count corrected by applying $C(\theta,\sigma)$. }

\tablenotetext{i}{The fractional error for the corrected count. }

\tablenotetext{j}{The flux in erg/s/cm$^2$ computed with the corrected source
count, by assuming a power-law spectrum between 0.3--8 KeV with a photon index of 1.7. }

\tablenotetext{k}{The luminosity computed from the flux by assuming a distance
to the host galaxy.}

\tablenotetext{m}{The probability for the Kolmogorov-Smirnov test null
hypothesis that the source is constant during the observation.  }

\end{deluxetable}



\begin{figure}
\plotone{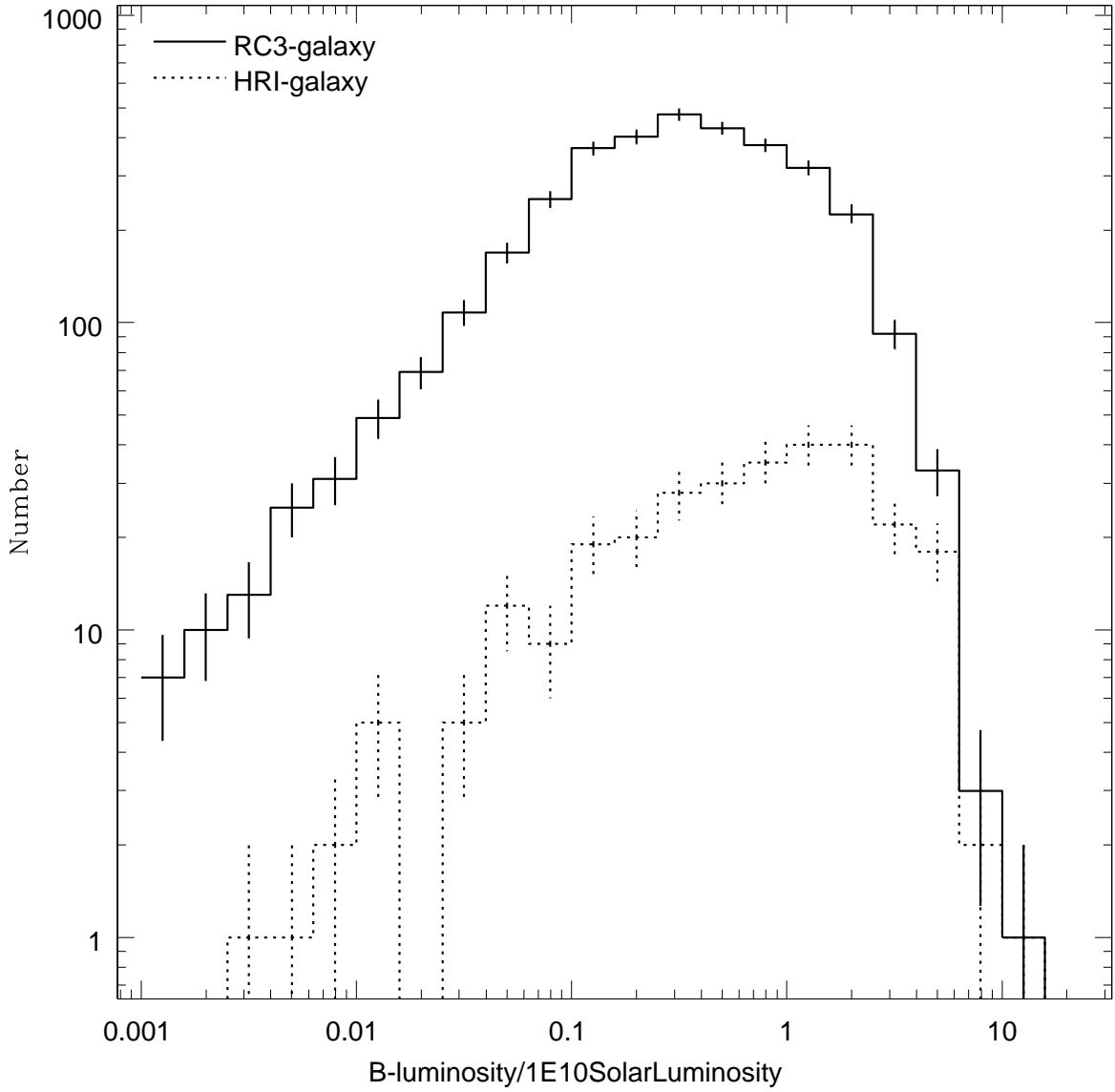}

\caption{The distribution of blue luminosities for the HRI survey galaxy sample
in comparison with the RC3 galaxy sample. More bright galaxies, presumably
larger in size and closer in distance, were surveyed by the HRI than faint
galaxies. }
\end{figure}

\begin{figure}
\plotone{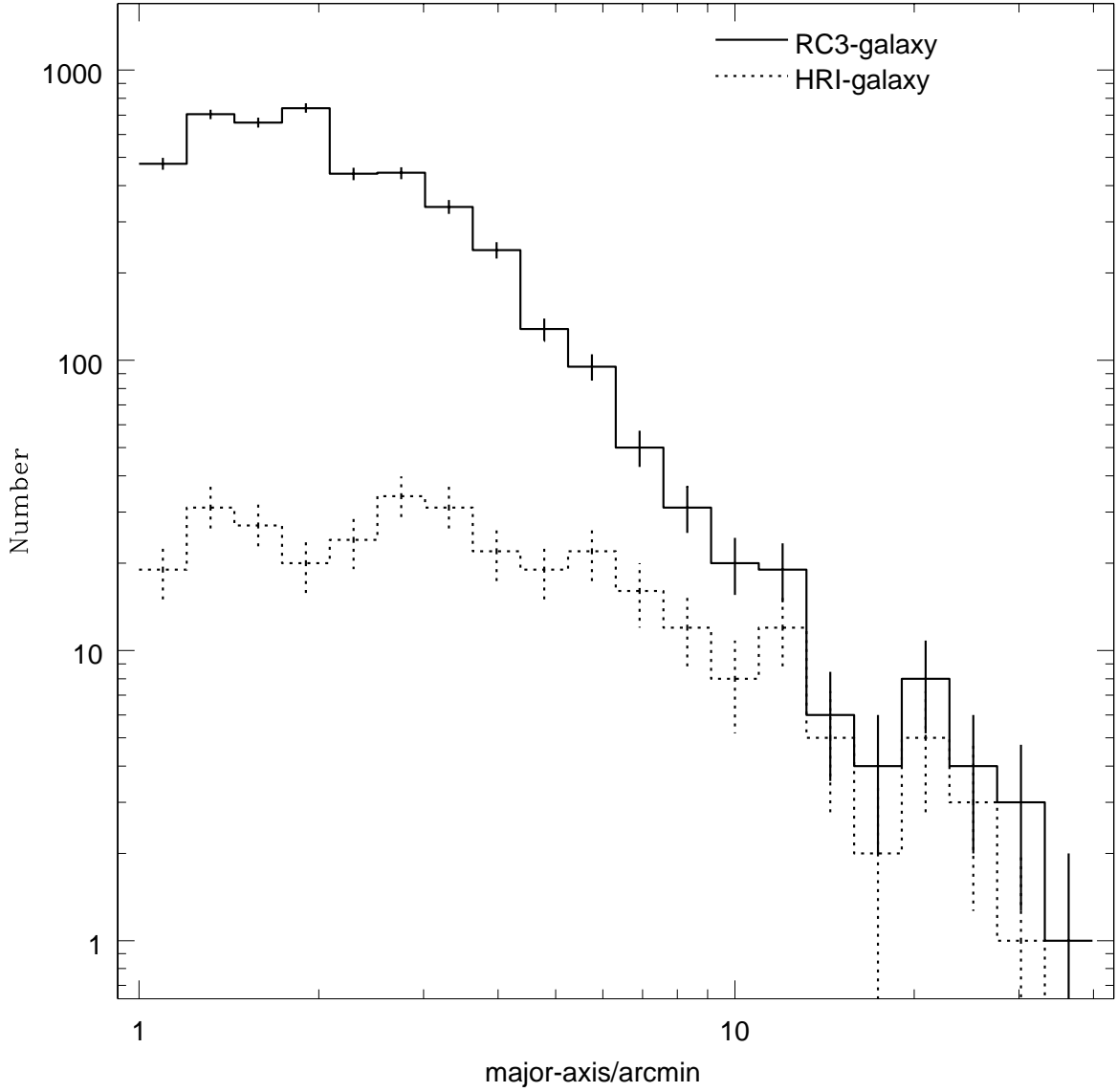}
\caption{Galaxy size distribution of the HRI survey galaxy sample in comparison
with the RC3 galaxy sample. Slightly more large galaxies were surveyed by the
HRI than small galaxies.}

\end{figure}

\begin{figure}
\plotone{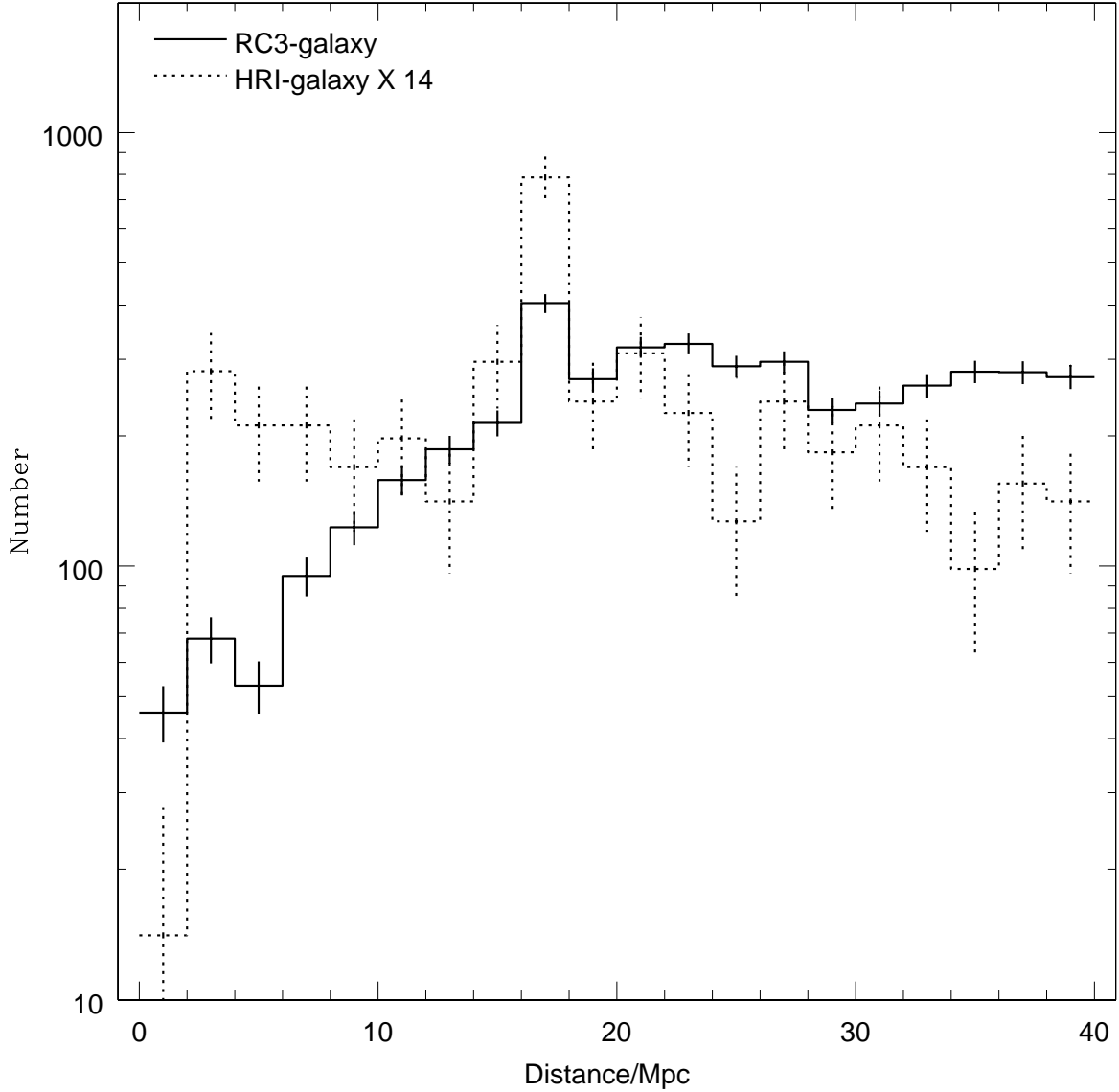}

\caption{Galaxy distance distribution of the HRI galaxy sample in comparison
with the RC3 galaxy sample. More nearby galaxies ($<15$Mpc) were surveyed than
those distant galaxies ($>15$ Mpc). }
\end{figure}

\begin{figure}

\plotone{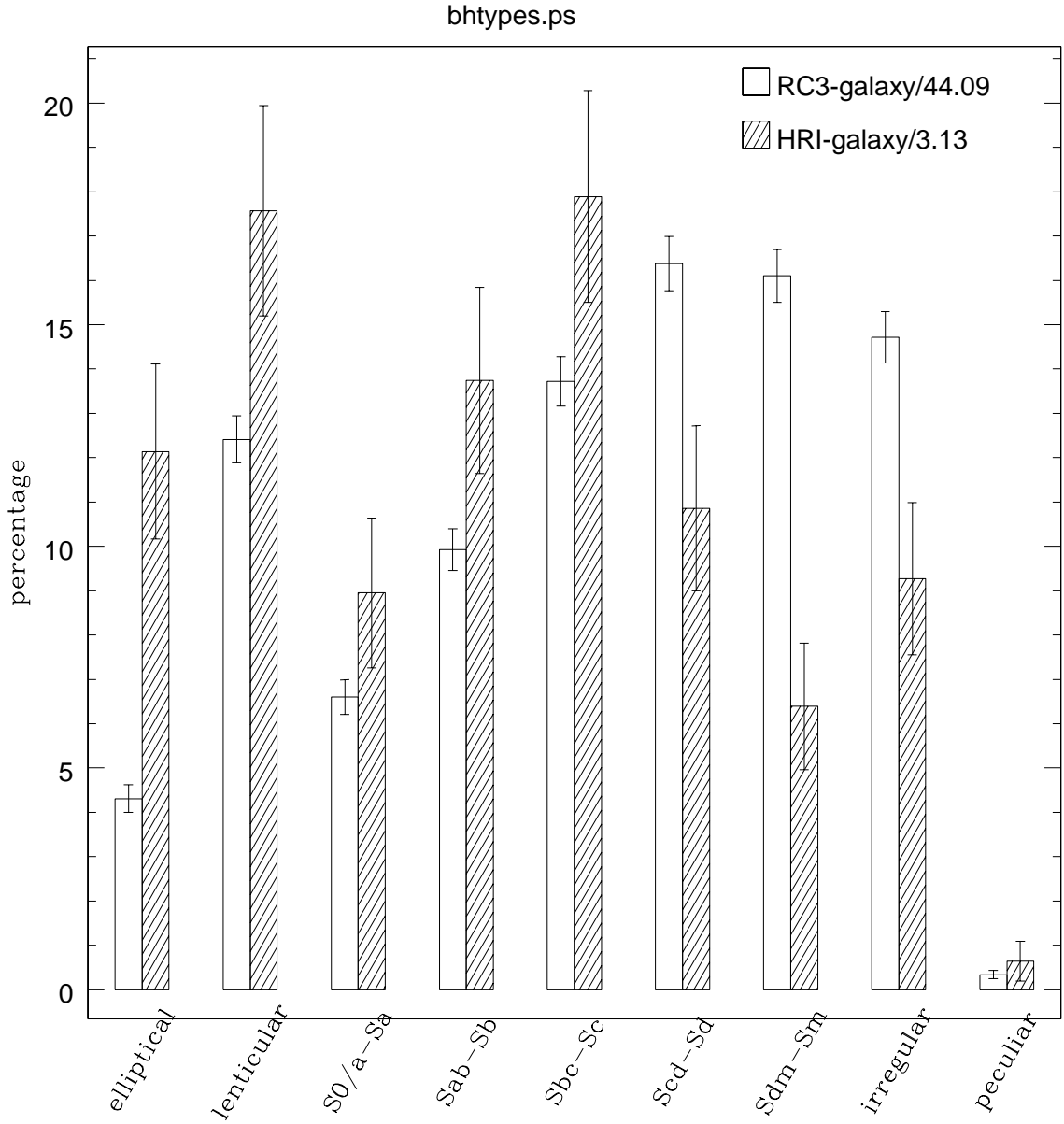}
\caption{The distribution of morphological types of the HRI galaxy sample in
comparison with the RC3 galaxy sample. More early-types galaxies and Sa--Sc
spiral galaxies were surveyed than other types. }

\end{figure}

\begin{figure}
\plotone{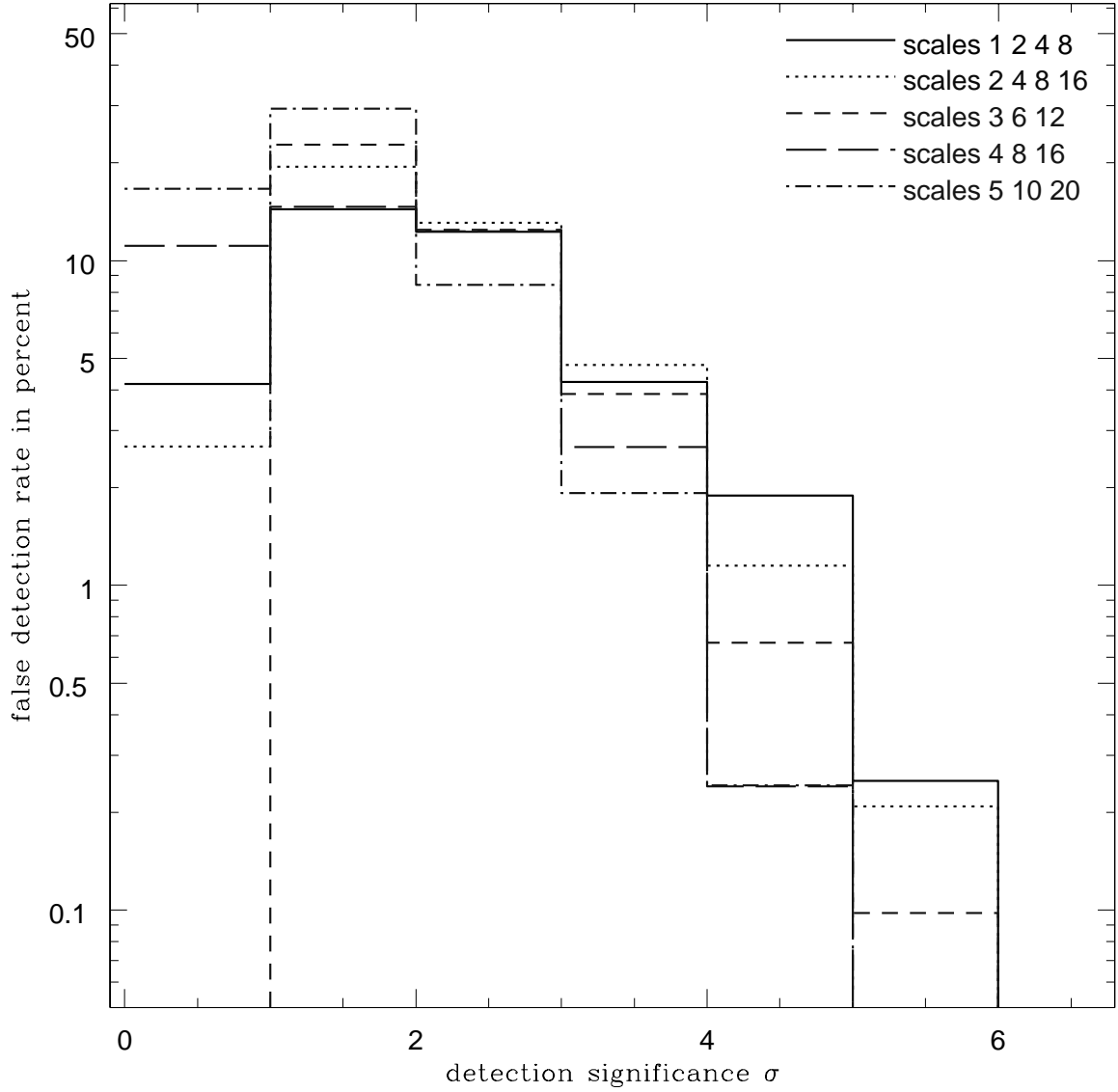}

\caption{False detections in simulations on HRI images with different scale
sets. The first three scale sets have larger false detection rates than the
latter two. The detections below 3$\sigma$ have a false detection rate $>$10\%
and are discarded in our data analysis. }

\end{figure}

\begin{figure}
\plotone{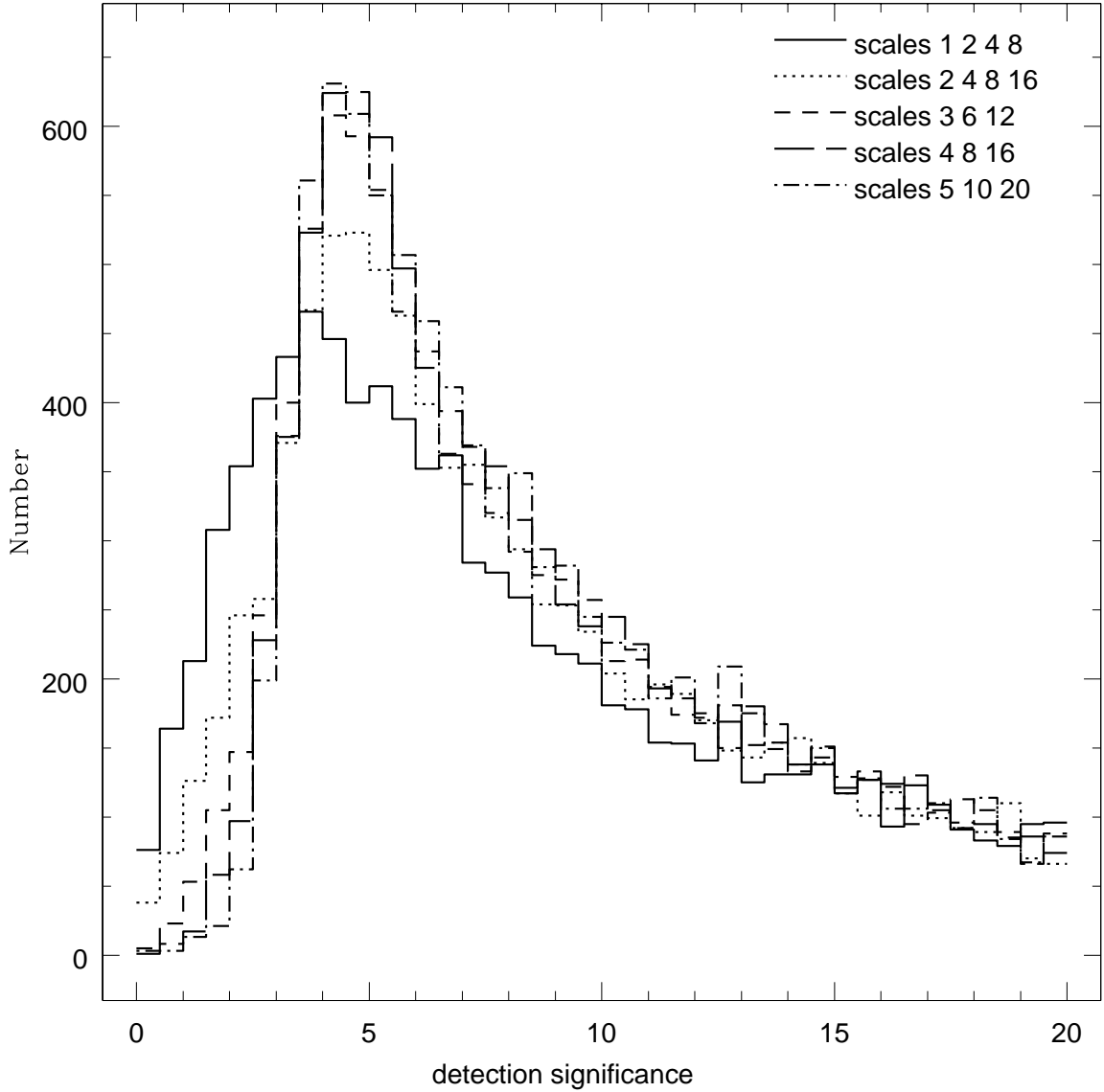}

\caption{Detections in simulations on HRI images with different scale sets.
There are more detections for the last two scale sets than the first three, and
there are more detections with large $\sigma$ for the last two scale sets than
for the first three.  }

\end{figure}

\begin{figure}
\plotone{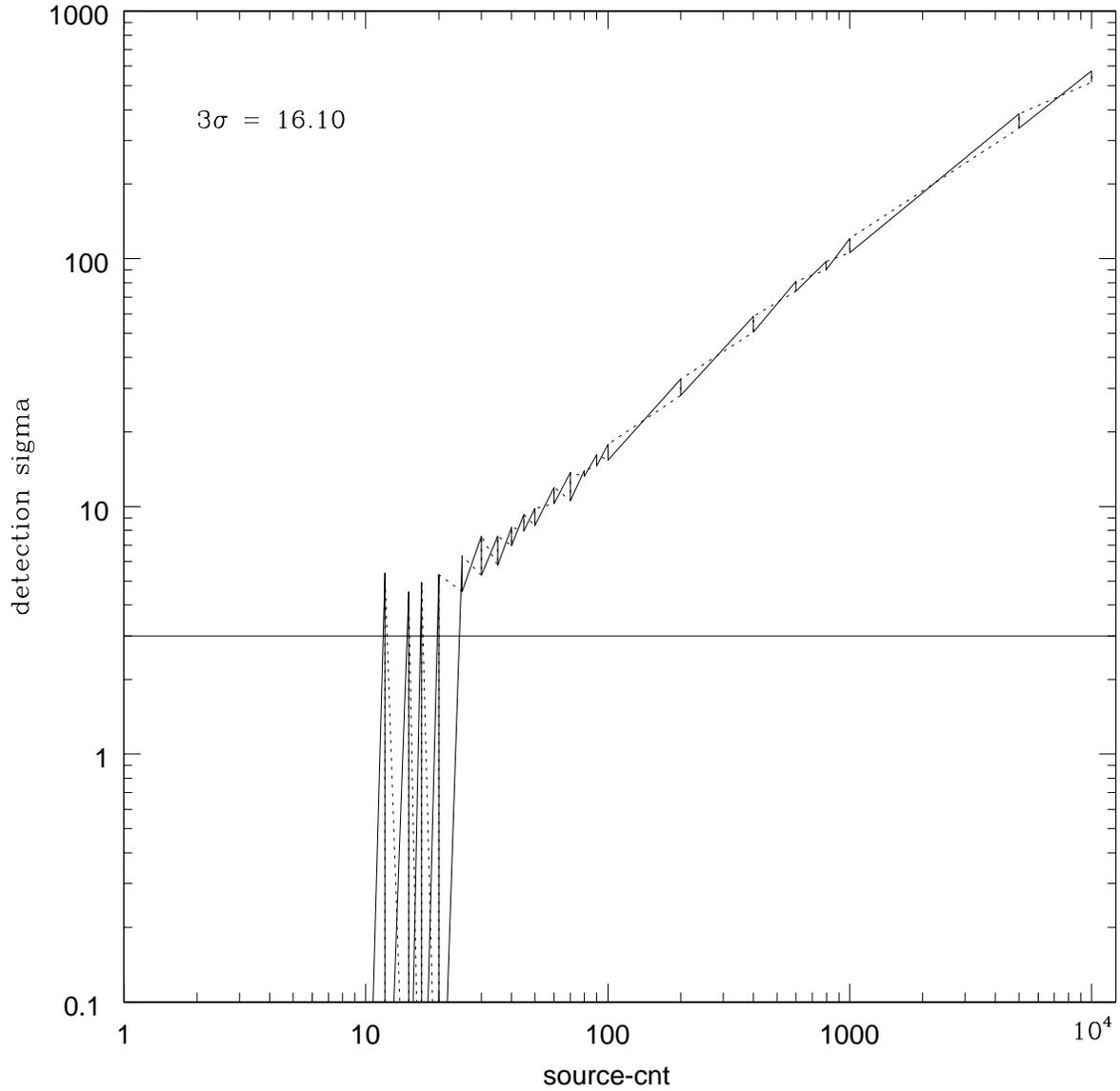}

\caption{Determination of the count threshold for 3$\sigma$ detection from
simulations. The example is for an off-axis angle of $10^\prime$ and an
exposure time of 10 kiloseconds. The solid and dashed connecting lines show two
ways to sort sources of same counts, one in ascending order of detection
$\sigma$ and another in descending order. The average of counts from the two
gives a method-independent count threshold. }

\end{figure}
\begin{figure}

\plotone{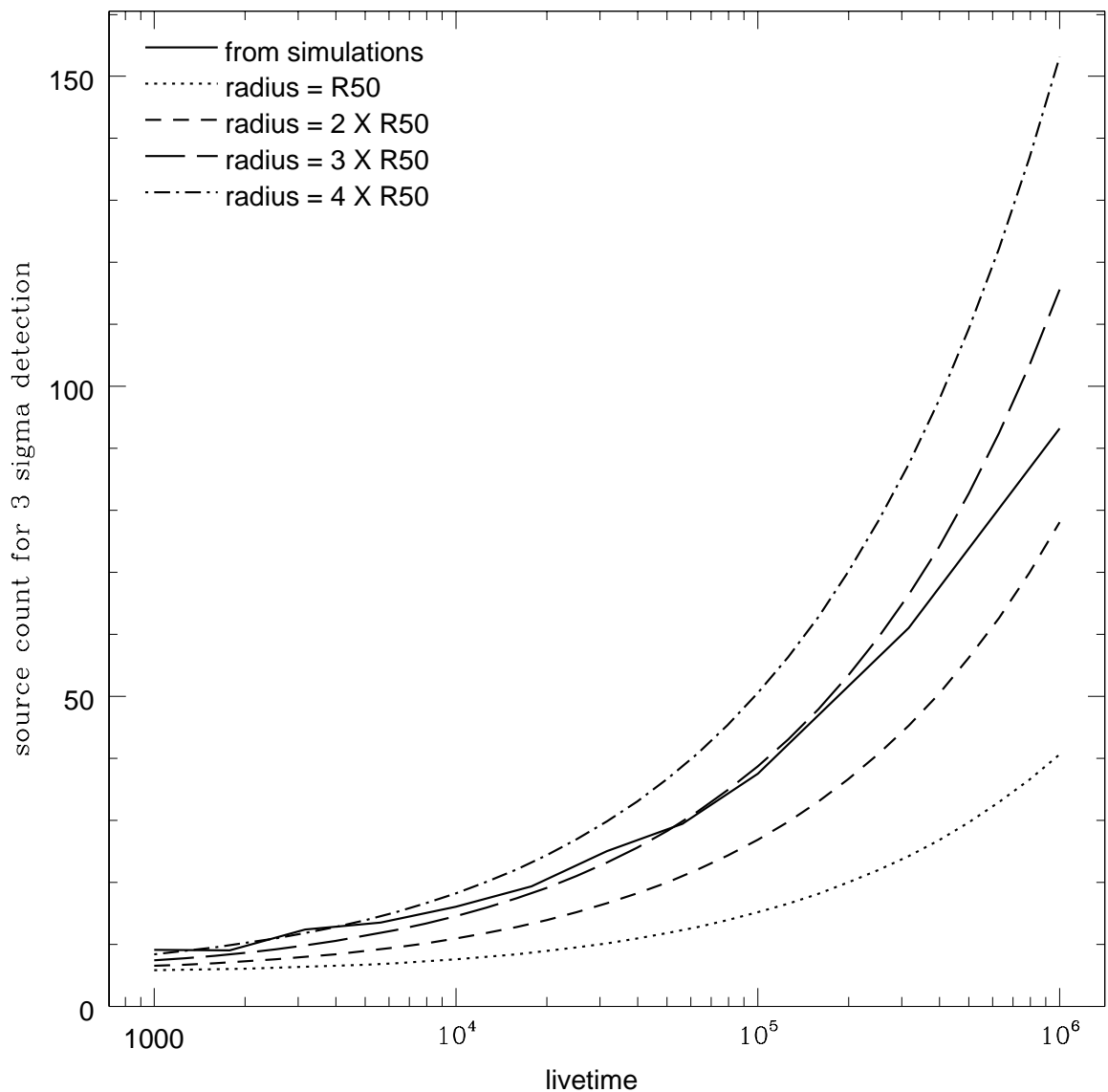}

\caption{The source count thresholds for $3\sigma$ detection with changes in the background
levels. The solid line is for thresholds derived from the
simulations, and the other lines for thresholds computed from source
regions of various sizes.  The simulated thresholds follow closely the computed
thresholds for a source region with a radius of $3\times R_{50}$ at an off-axis
angle of $10^\prime$ in this example. }

\end{figure}

\begin{figure}
\plottwo{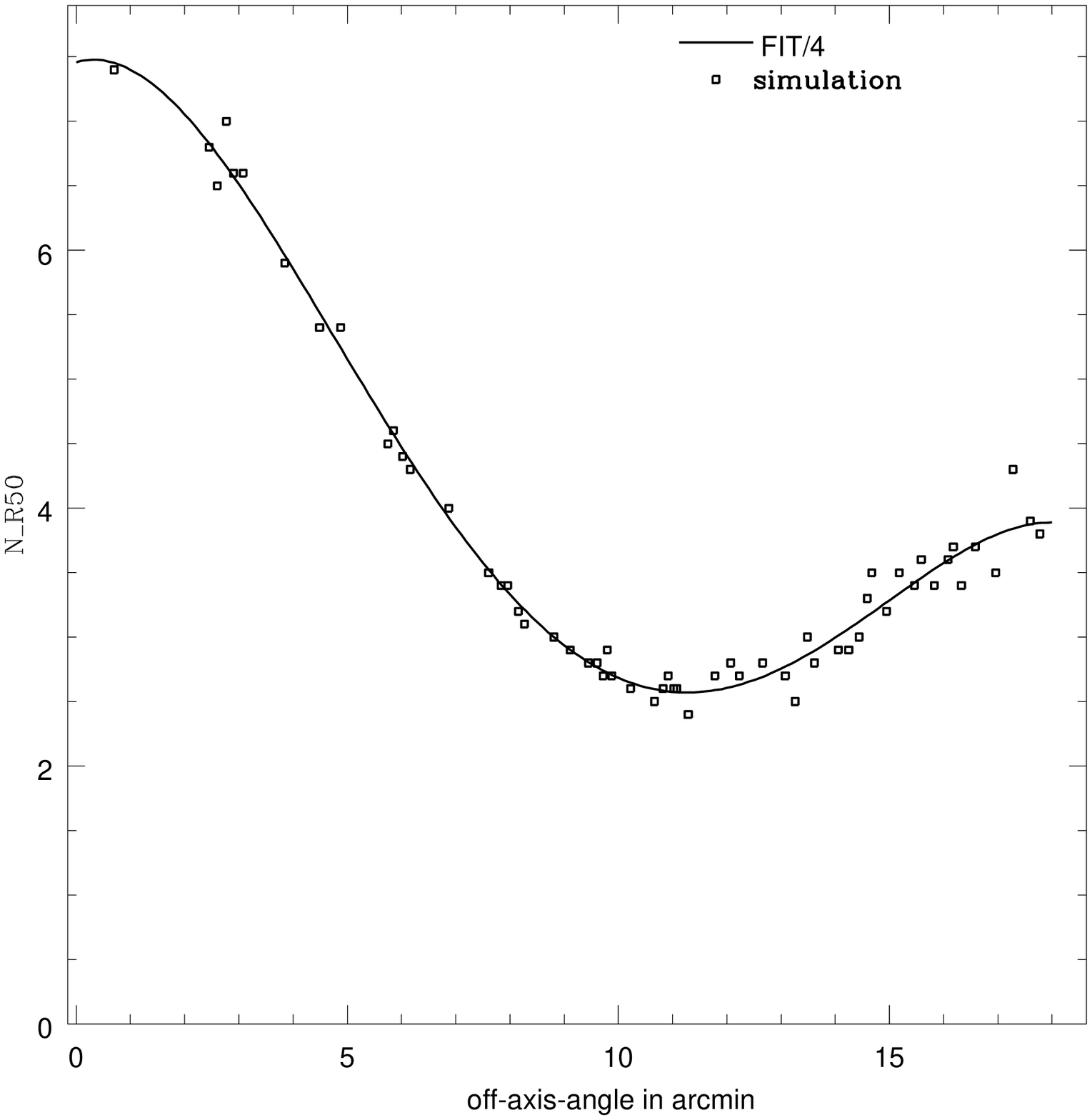}{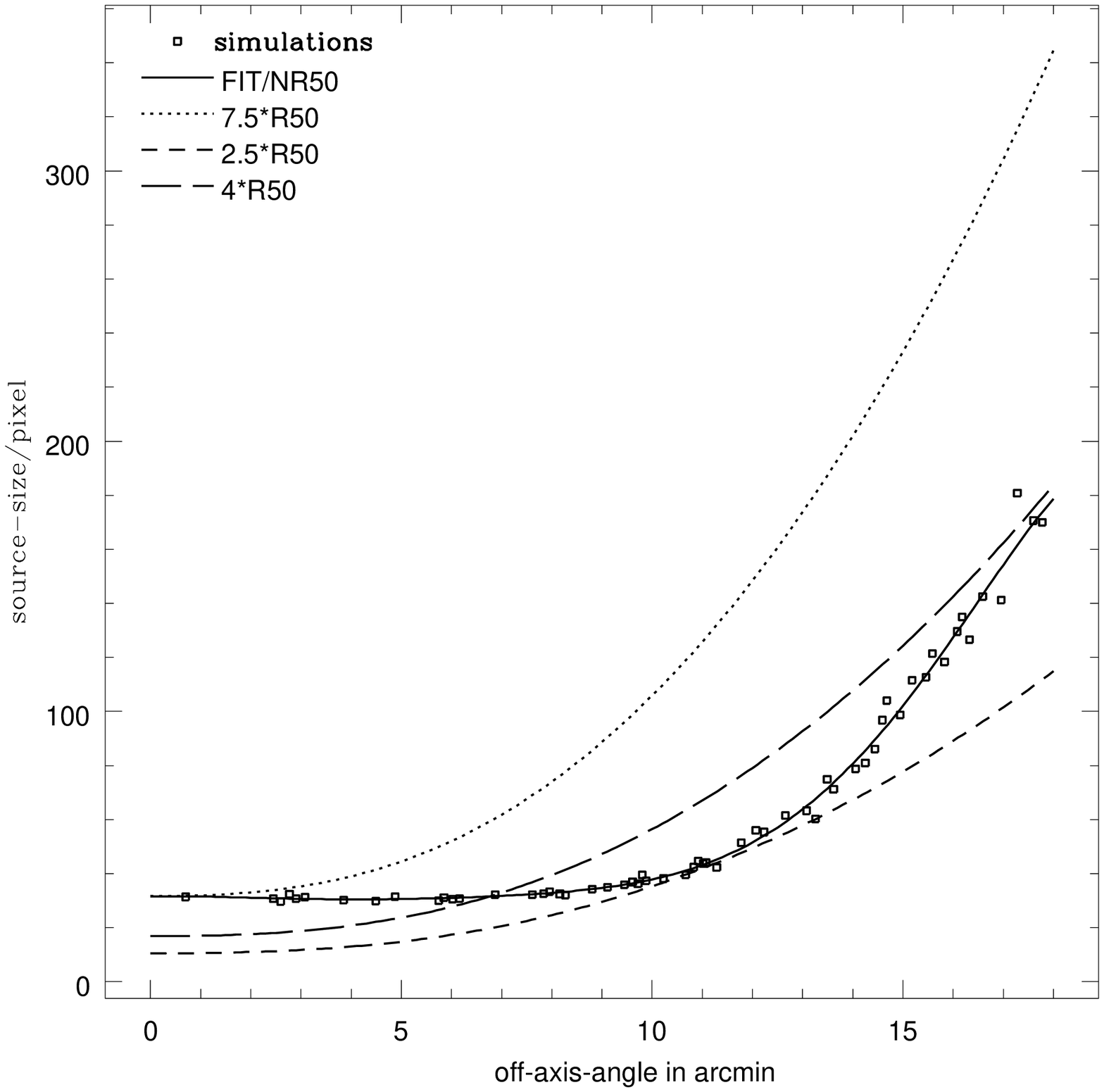}

\caption{The equivalent source region for computing count thresholds as a
function of off-axis angles. The left panel shows the best fit source size in
unit of $R_{50}$. The right panel shows the source size in pixels. The points
are source sizes derived from the simulations, and the solid line is a 4-th
polynomial fit to the simulations. }

\end{figure}

\begin{figure}
\plotone{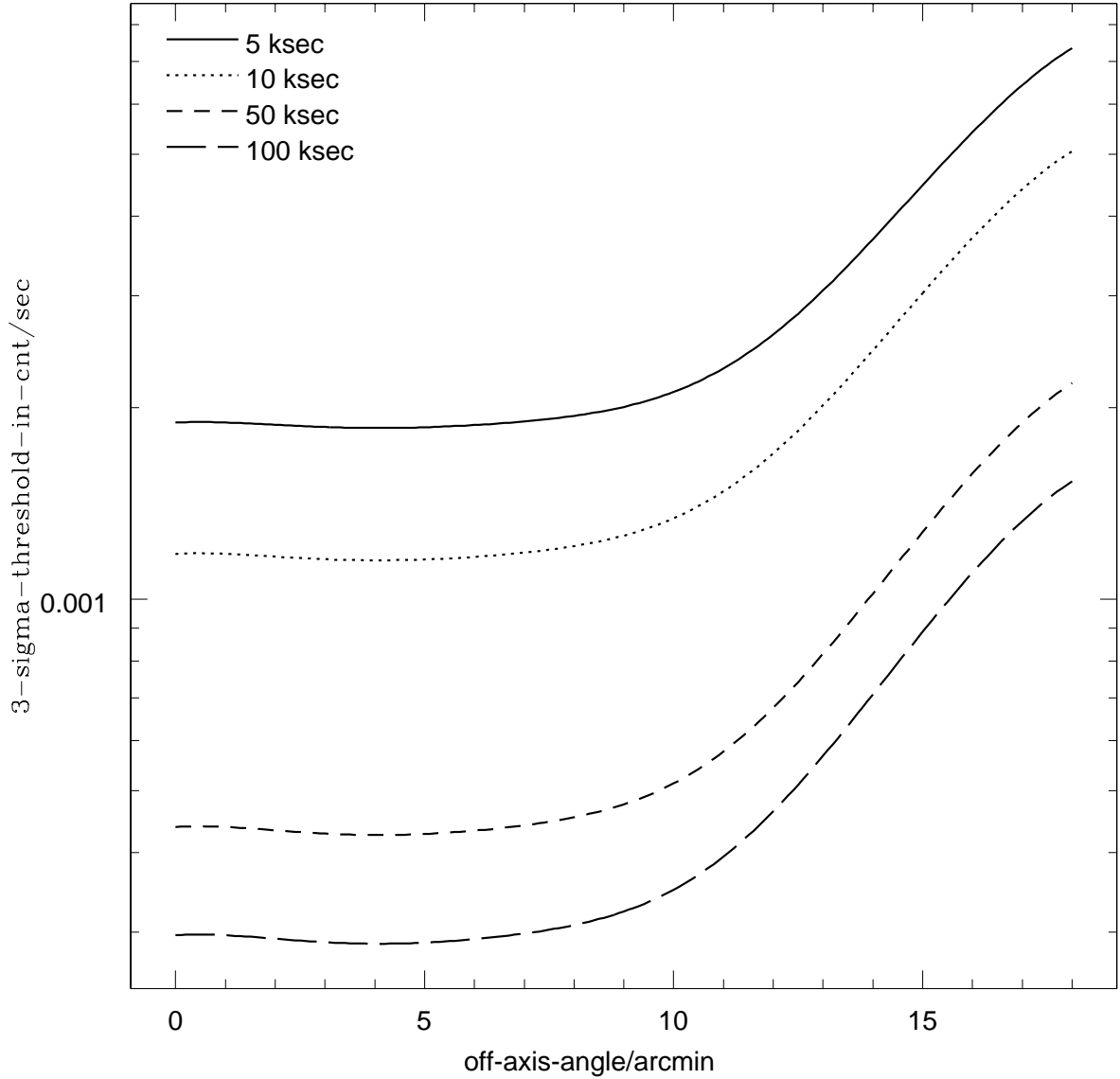}

\caption{The count rate thresholds for 3$\sigma$ detection computed for
different exposure times based on the best fit source size derived from the
simulations. For an exposure, the sensitivity is approximately the same within
$10^\prime$, and degrades outwards. At $\sim17^\prime$ it decreases by a factor
$\sim3.5$ for all the shown exposures. }

\end{figure}

\begin{figure}
\plotone{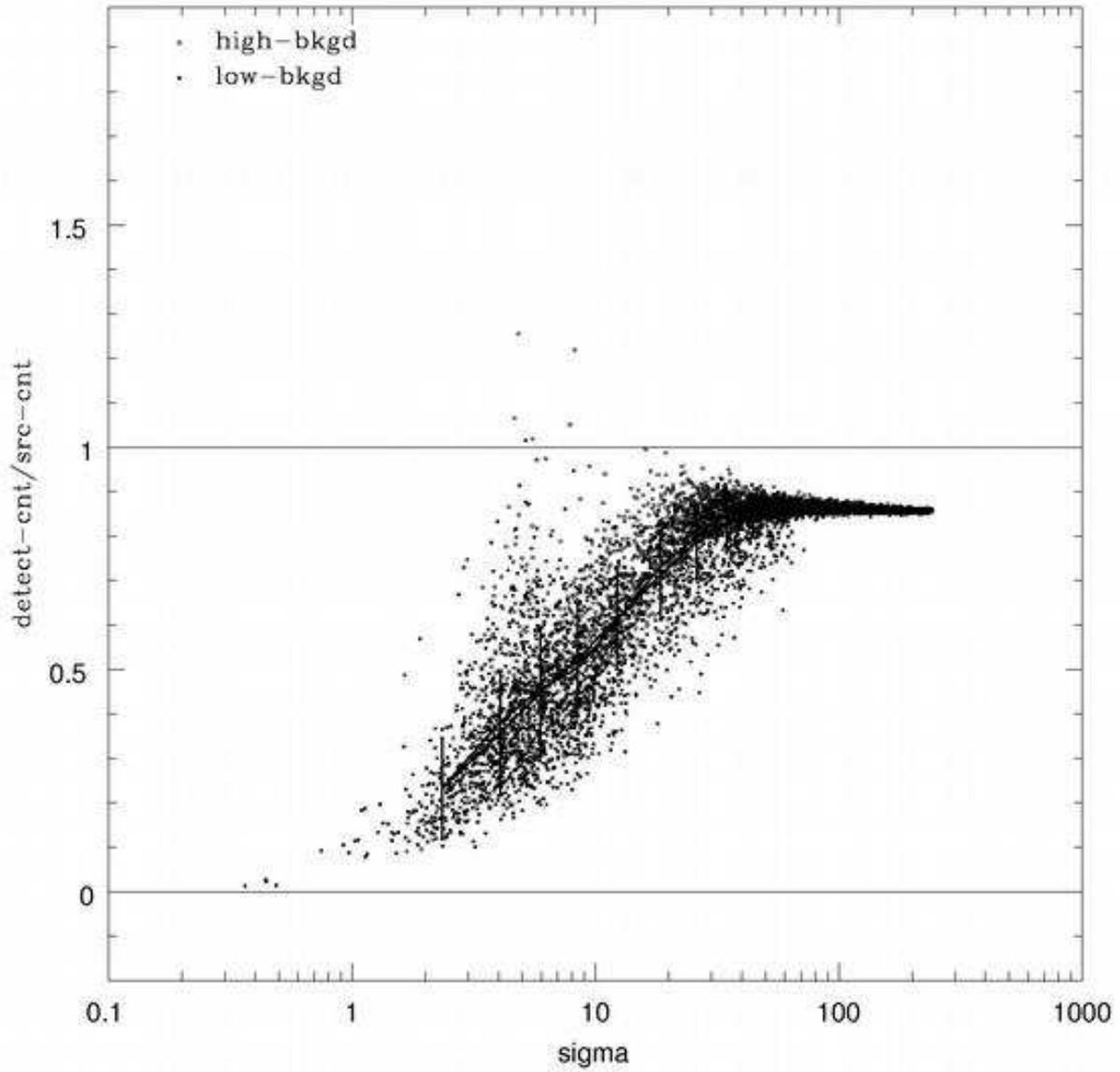}

\caption{Change in the correction factors with detection significance for
off-axis angles of 16.08'$\pm$0.1'. No significant dependence of correction
factors on the background levels is shown. }

\end{figure}

\begin{figure}
\plotone{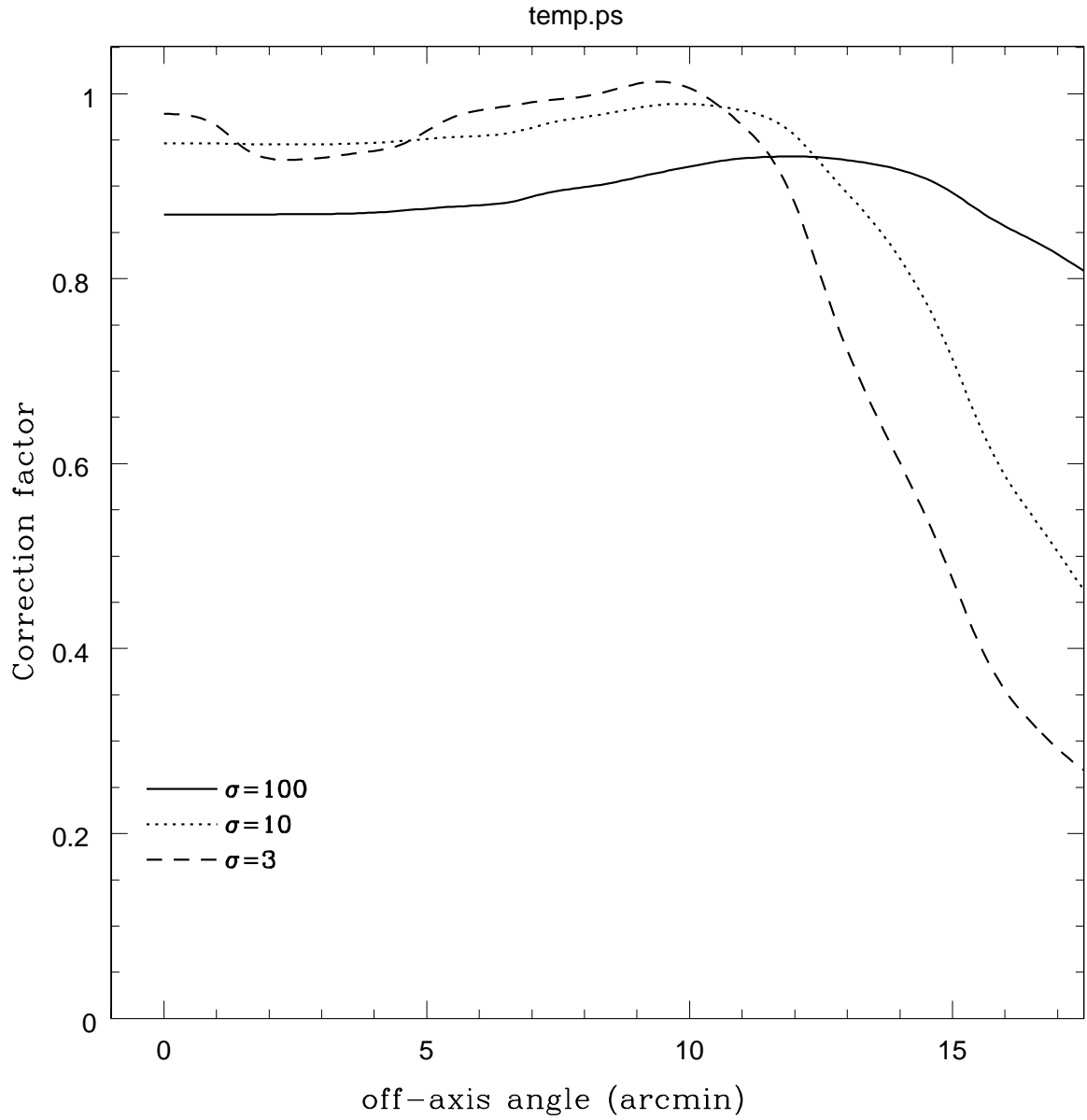}
\caption{The correction factors as a function of off-axis angles for different
detection significance. The decrease beyond $10^\prime$ can be attributed to the
increase in the point source sizes and the vignetting function with the off-axis angle.}

\end{figure}

\begin{figure}
\plotone{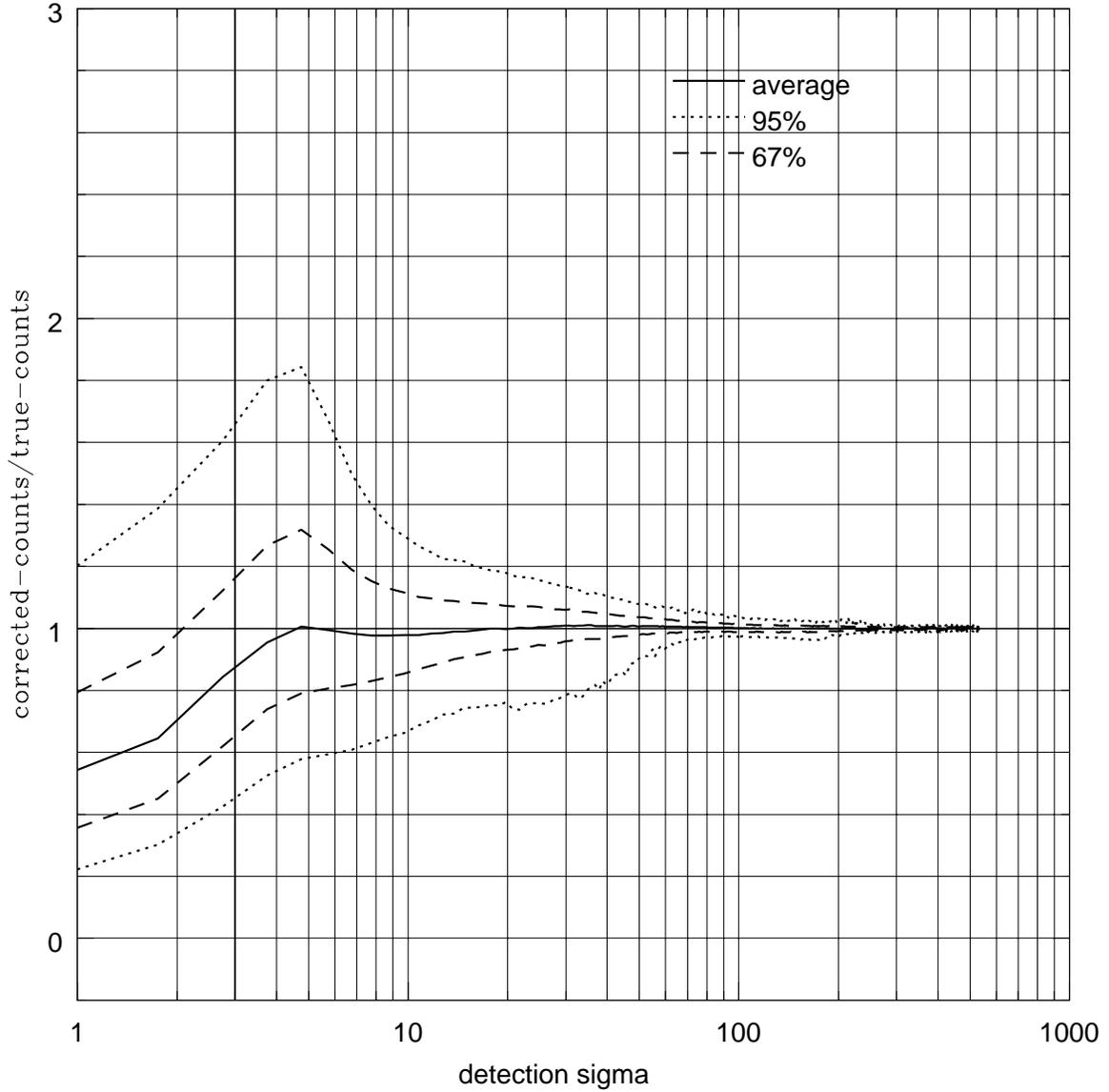}

\caption{The detection count correction factors from the simulations. The solid
line shows the  mean ratio of corrected counts to true counts, and the
dashed/dotted lines show the limits between which are 67\%/95\% of the
simulated sources.}

\end{figure}

\begin{figure}
\plotone{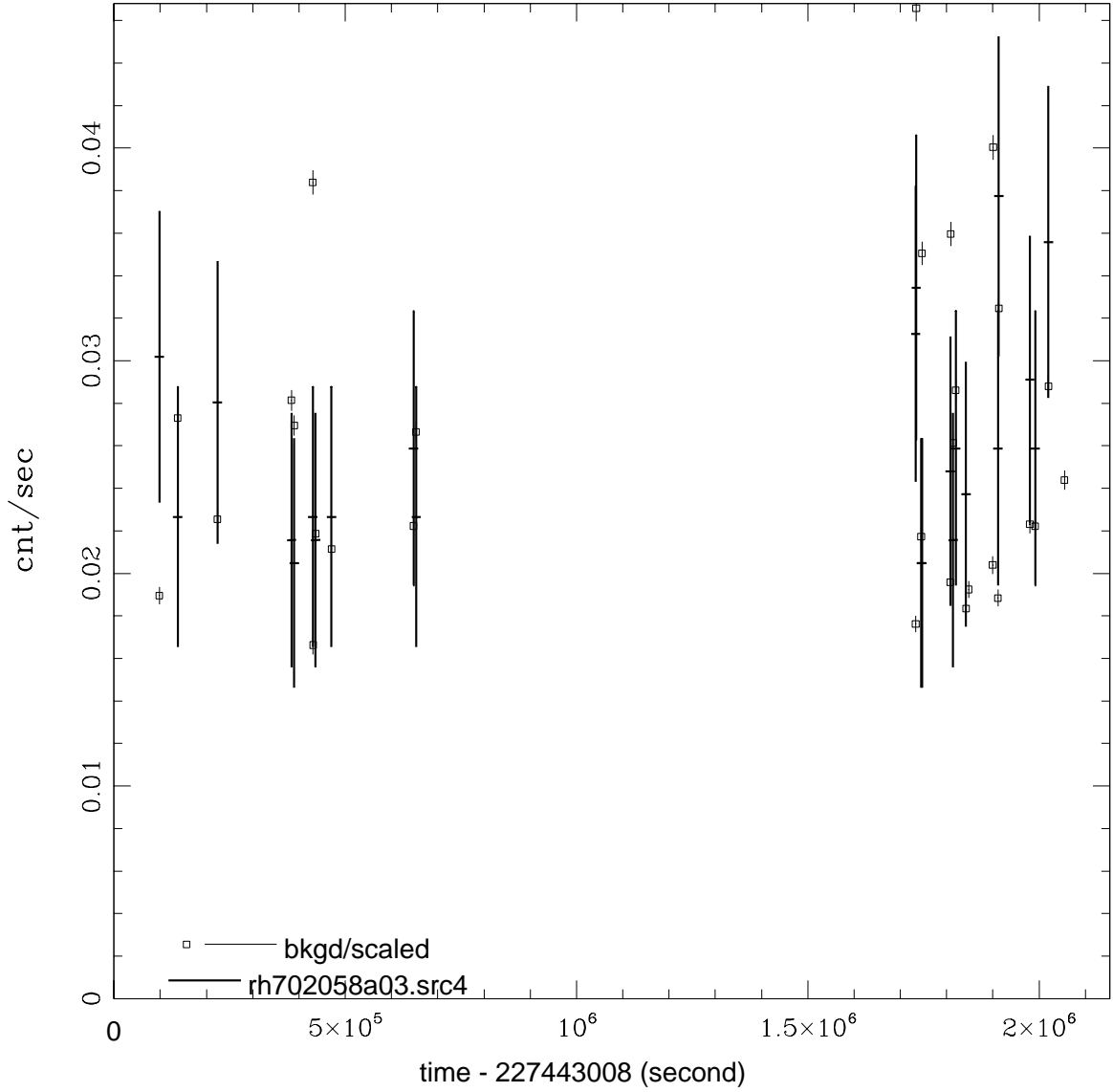}

\caption{ The HRI light curve for a ULX in Circinus galaxy which shows X-ray
eclipses with a 7.5 hour period in a Chandra observation. The light curve shows
clear temporal variations. However, the periodicity is not obvious due to the
wide separations between OBIs. }

\end{figure}

\begin{figure}
\plotone{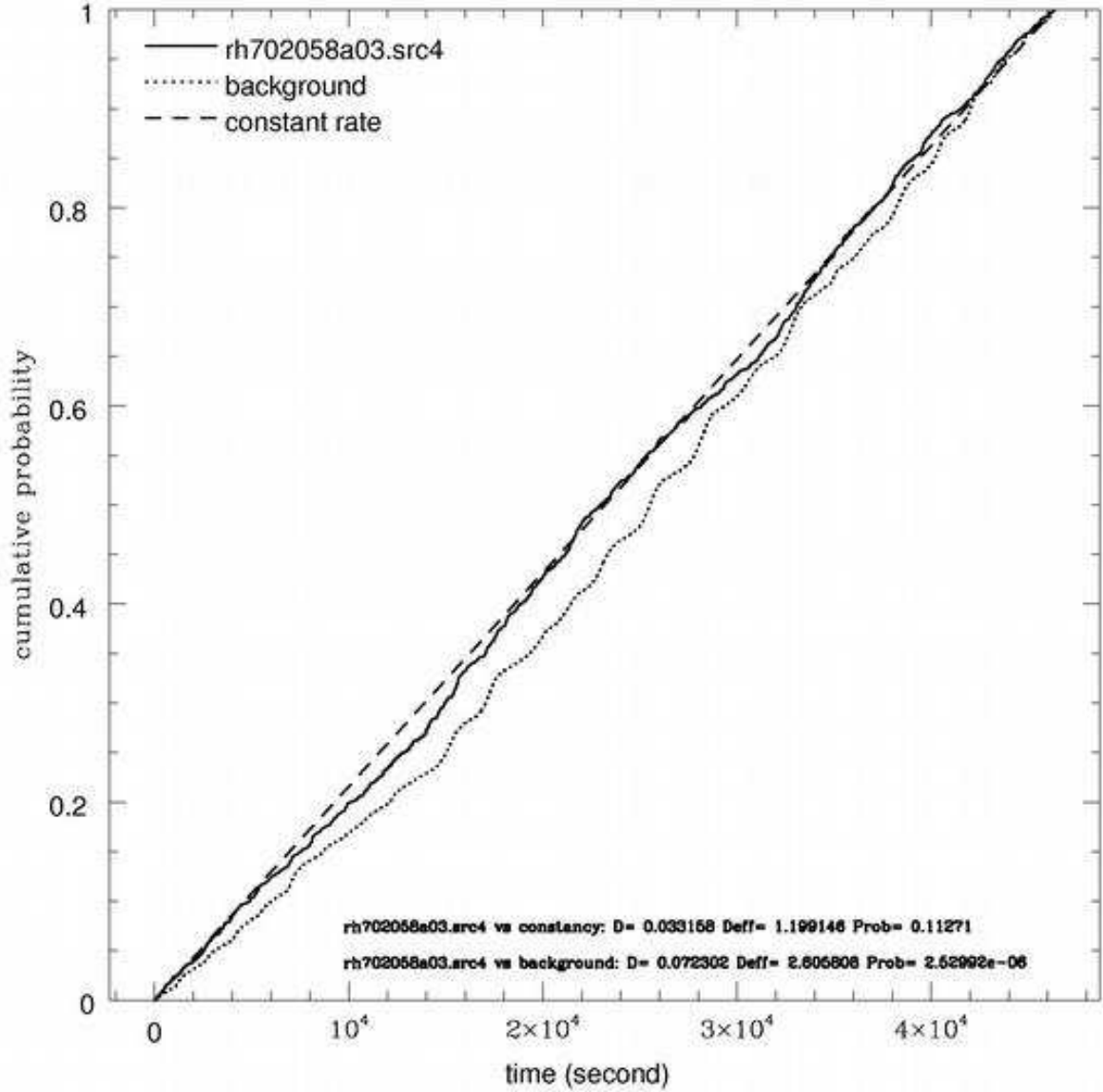}
\caption{ The cumulative probability curves for the Kolmogorov-Smirnov tests of
the ULX in Circinus galaxy.  The tests show that the source was variable with a
(insignificant) probability of 89\% during the HRI observation.  }

\end{figure}

\begin{figure}
\plotone{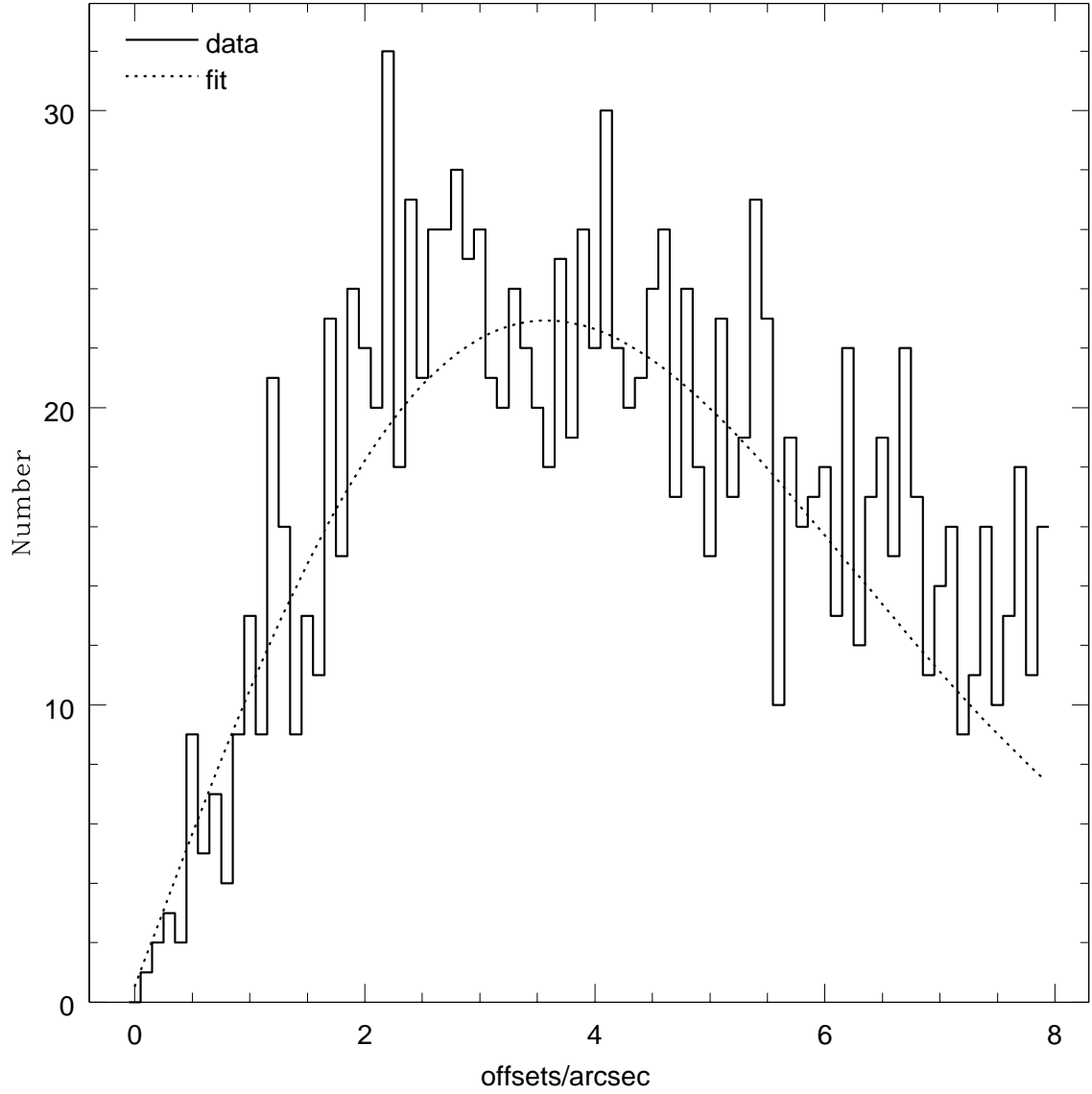}

\caption{Distribution of offsets between the HRI and catalogued positions for
1372 sources detected in HRI observations. This distribution can be fitted by a
2-D normal distribution with $\sigma=3\farcs62\pm0\farcs02$. }

\end{figure}


\clearpage

\begin{figure}
\plotone{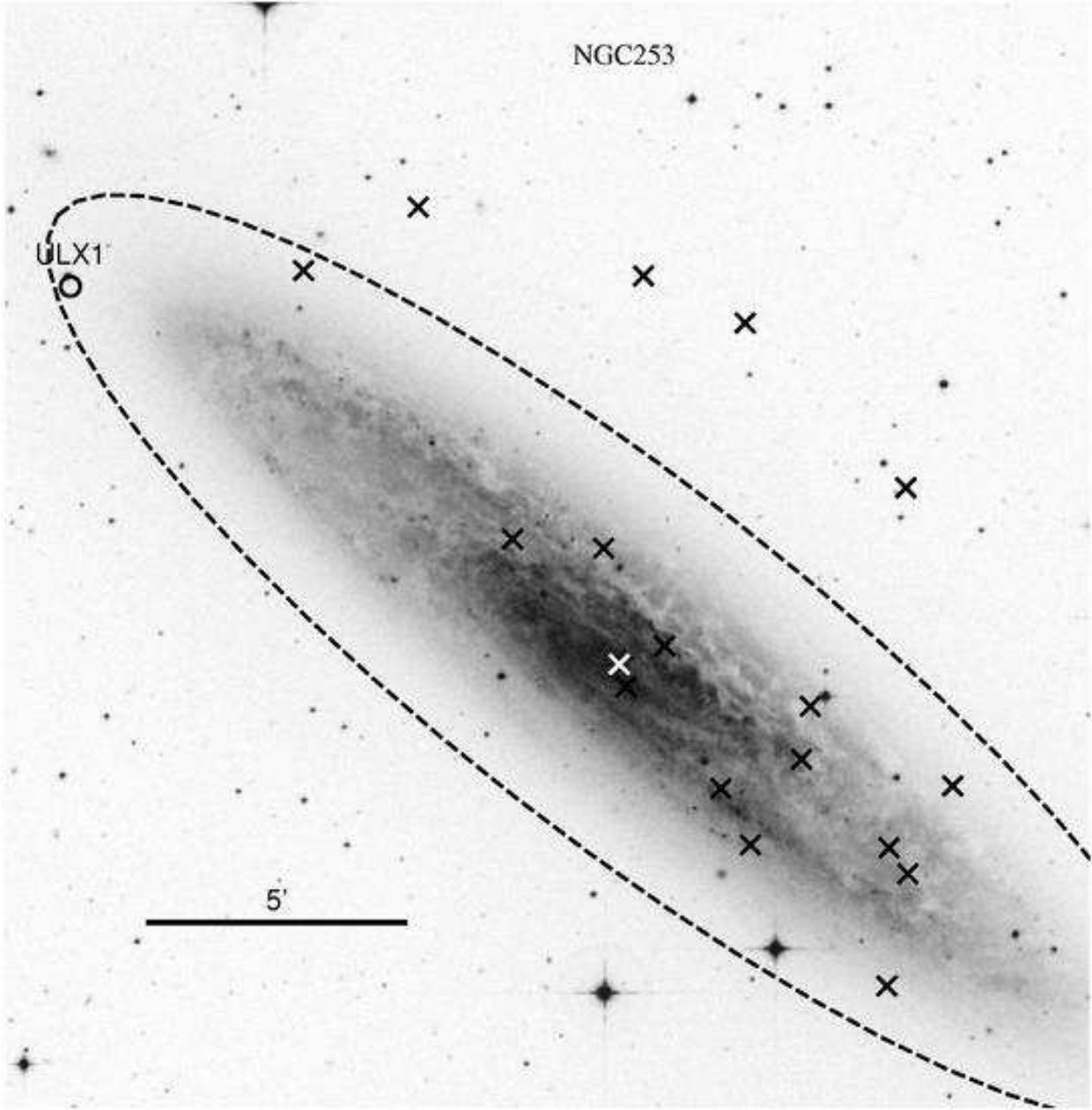}

\caption{The finding chart for the ULXs in NGC253. The ellipse is the $D_{25}$
isophote, and the galactic nucleus is marked by "+". ULXs in the clean sample
or ULX candidates are labelled as circles, and other X-ray sources are labelled
by "X". The DSS image is positioned with up as north and left as east.}

\end{figure}
\begin{figure}
\plotone{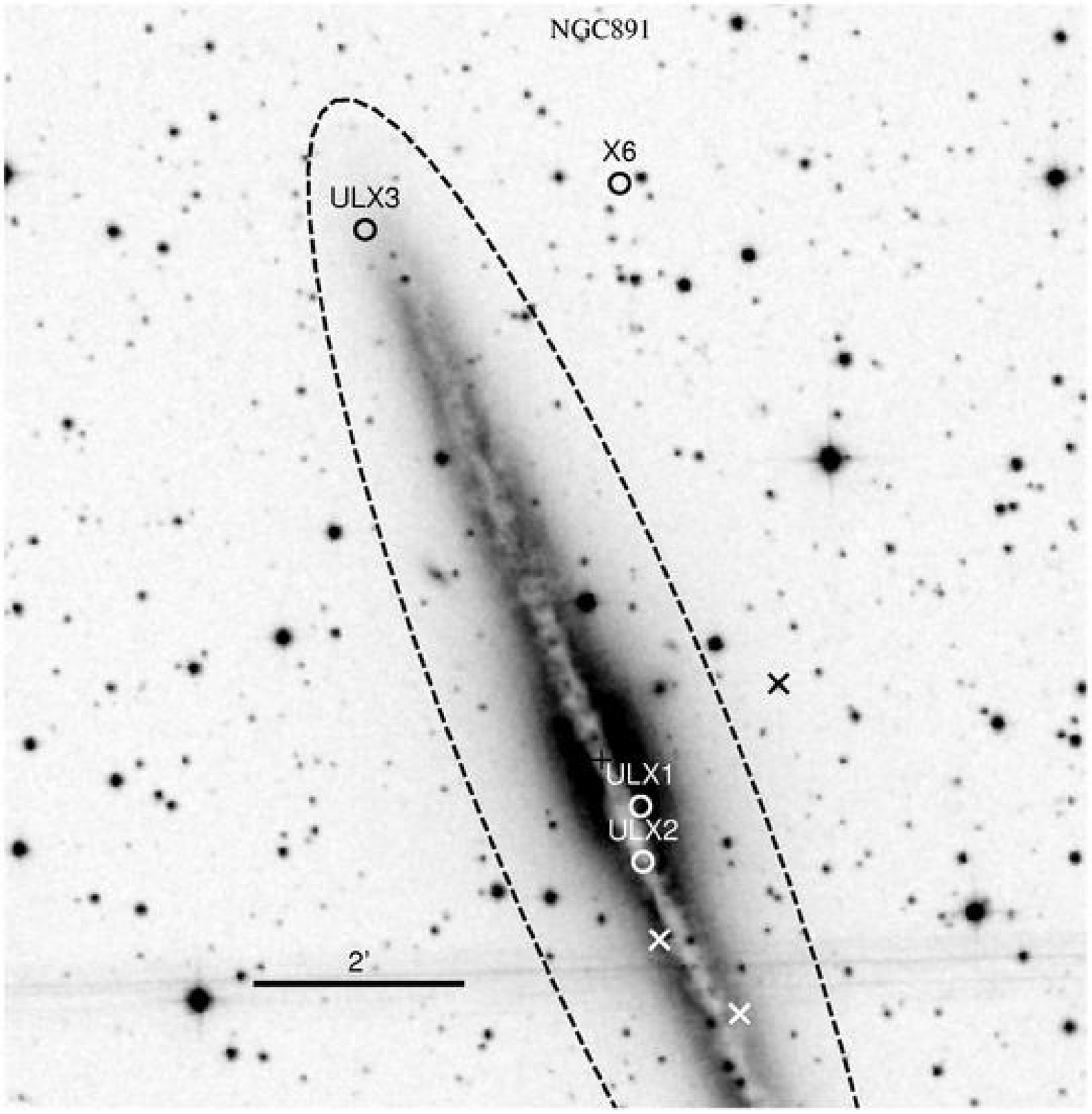}

\caption{The finding chart for the ULXs in NGC891.}

\end{figure}
\begin{figure}
\plotone{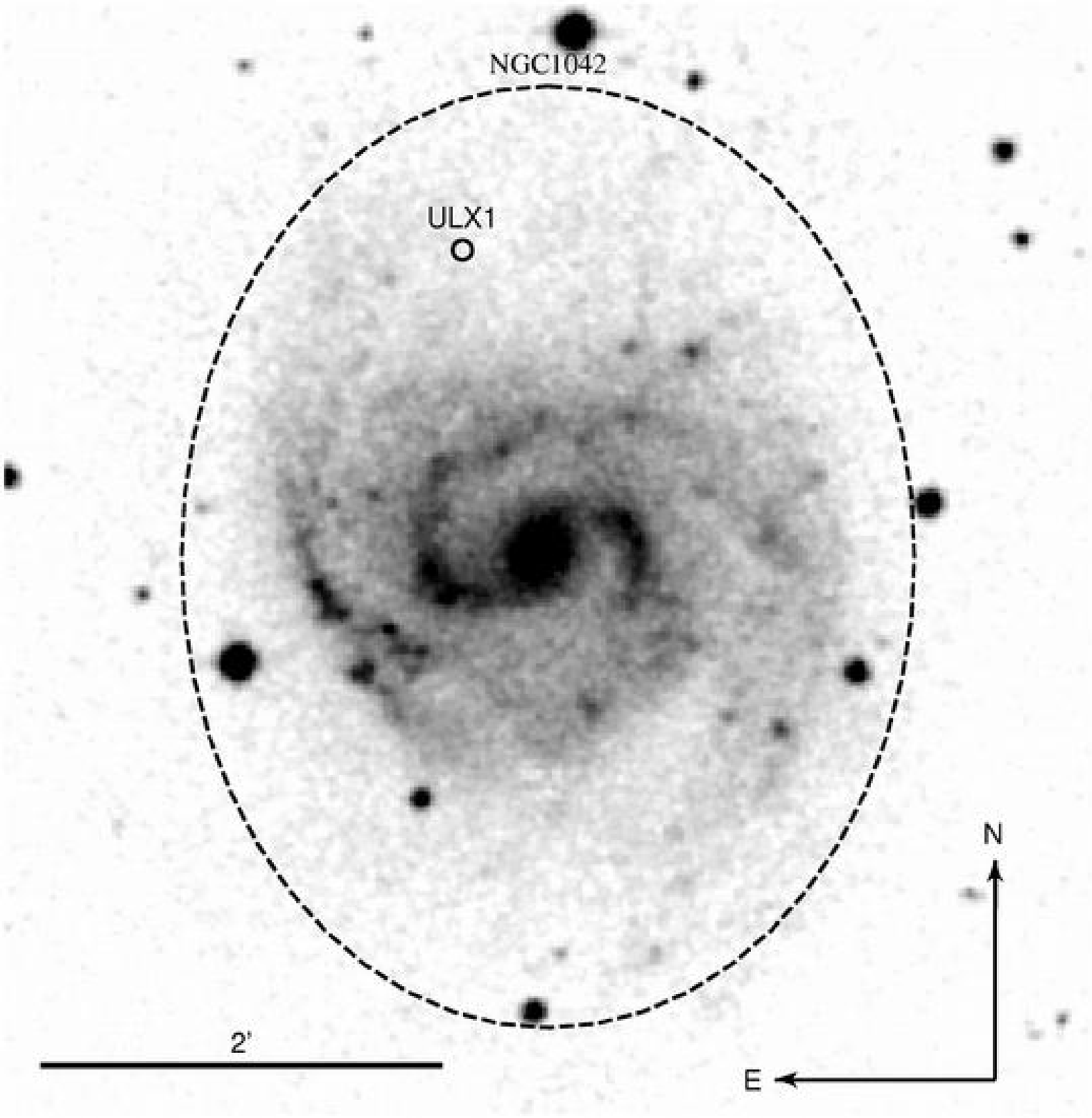}

\caption{The finding chart for the ULXs in NGC1042. }

\end{figure}
\begin{figure}
\plotone{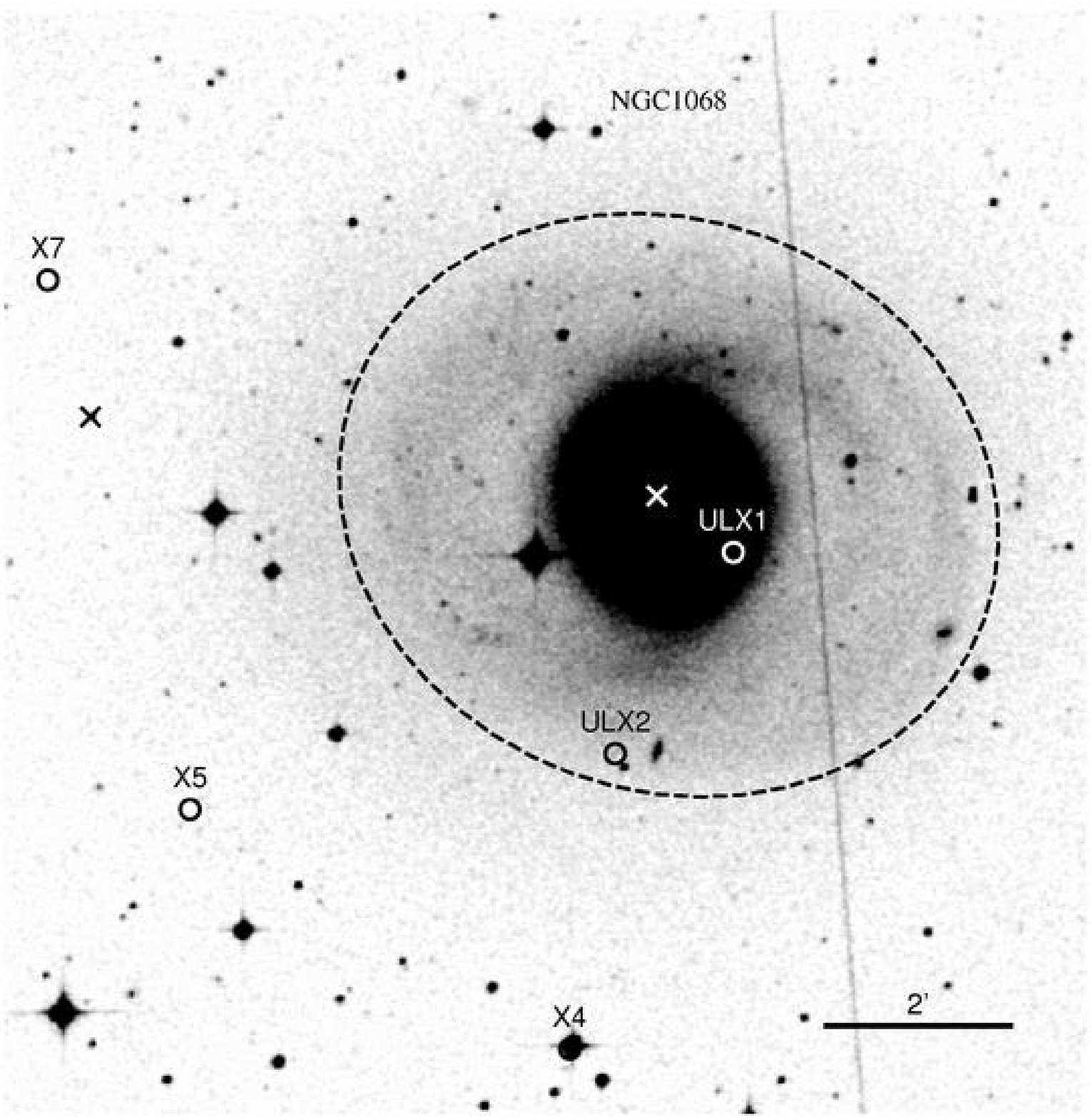}
\caption{The finding chart for the ULXs in NGC1068.}

\end{figure}
\begin{figure}
\plotone{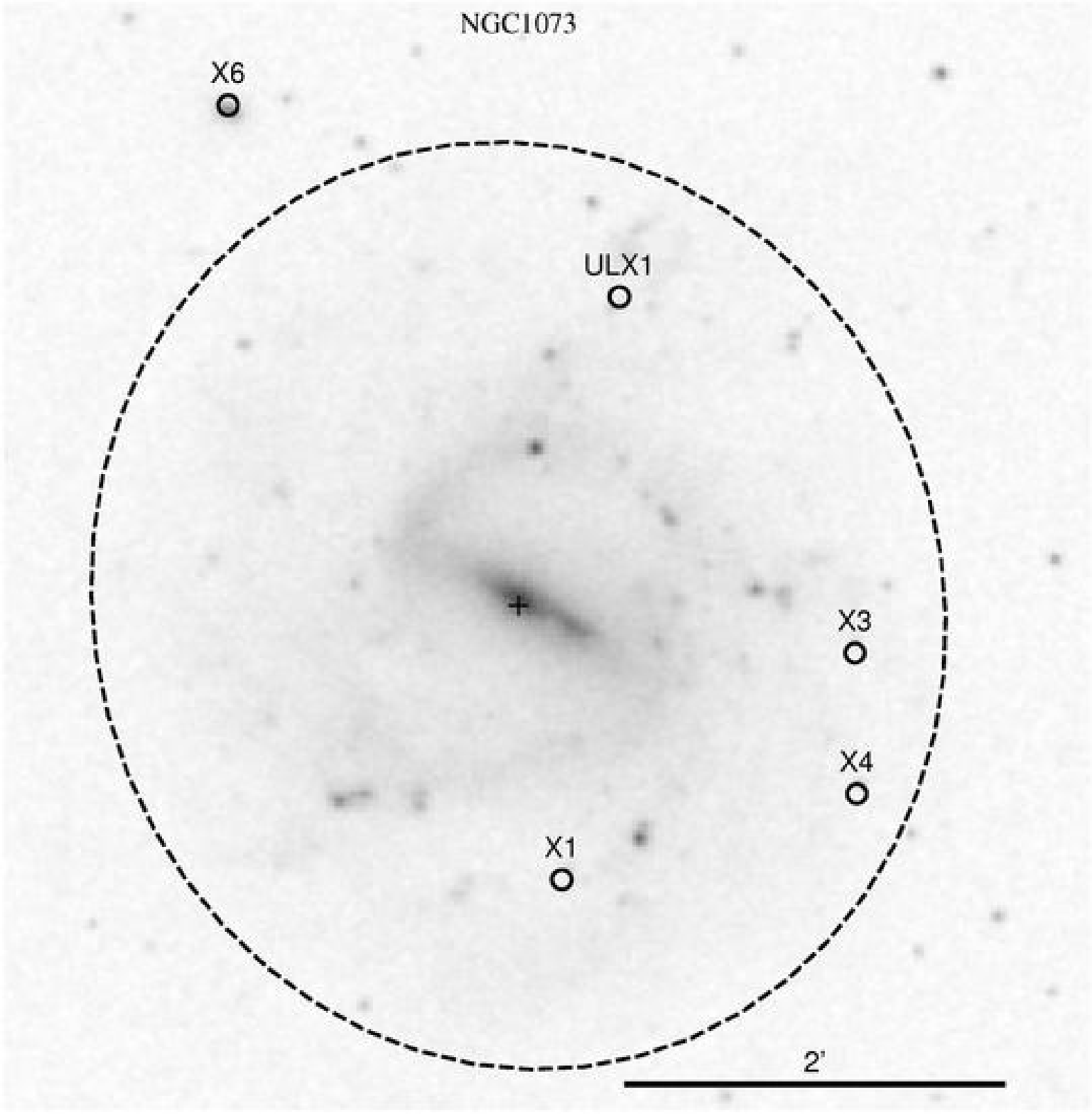}

\caption{The finding chart for the ULXs in NGC1073. }
\end{figure} 
\begin{figure}
\plotone{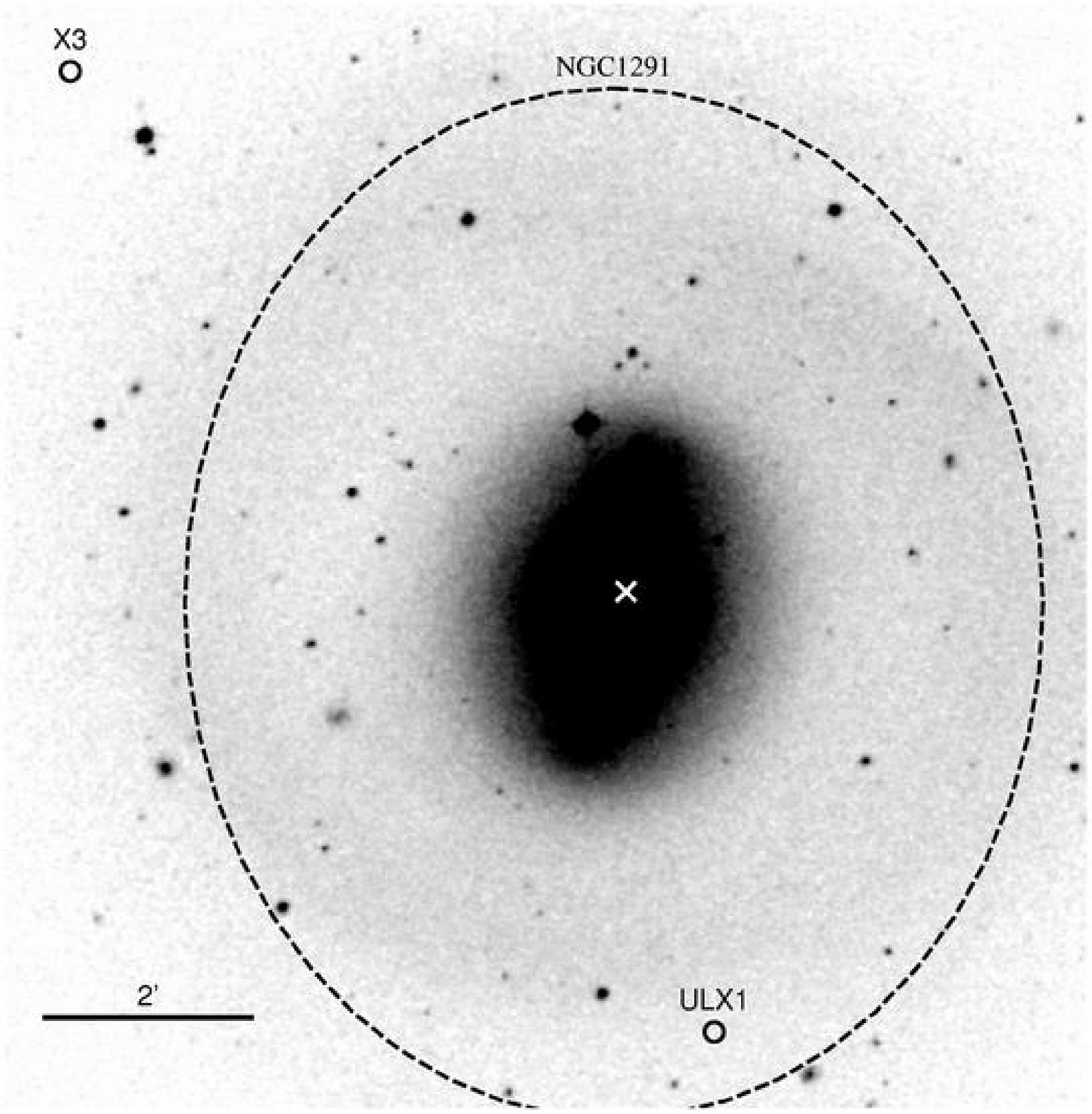}

\caption{The finding chart for the ULXs in NGC1291.}

\end{figure}
\begin{figure}
\plotone{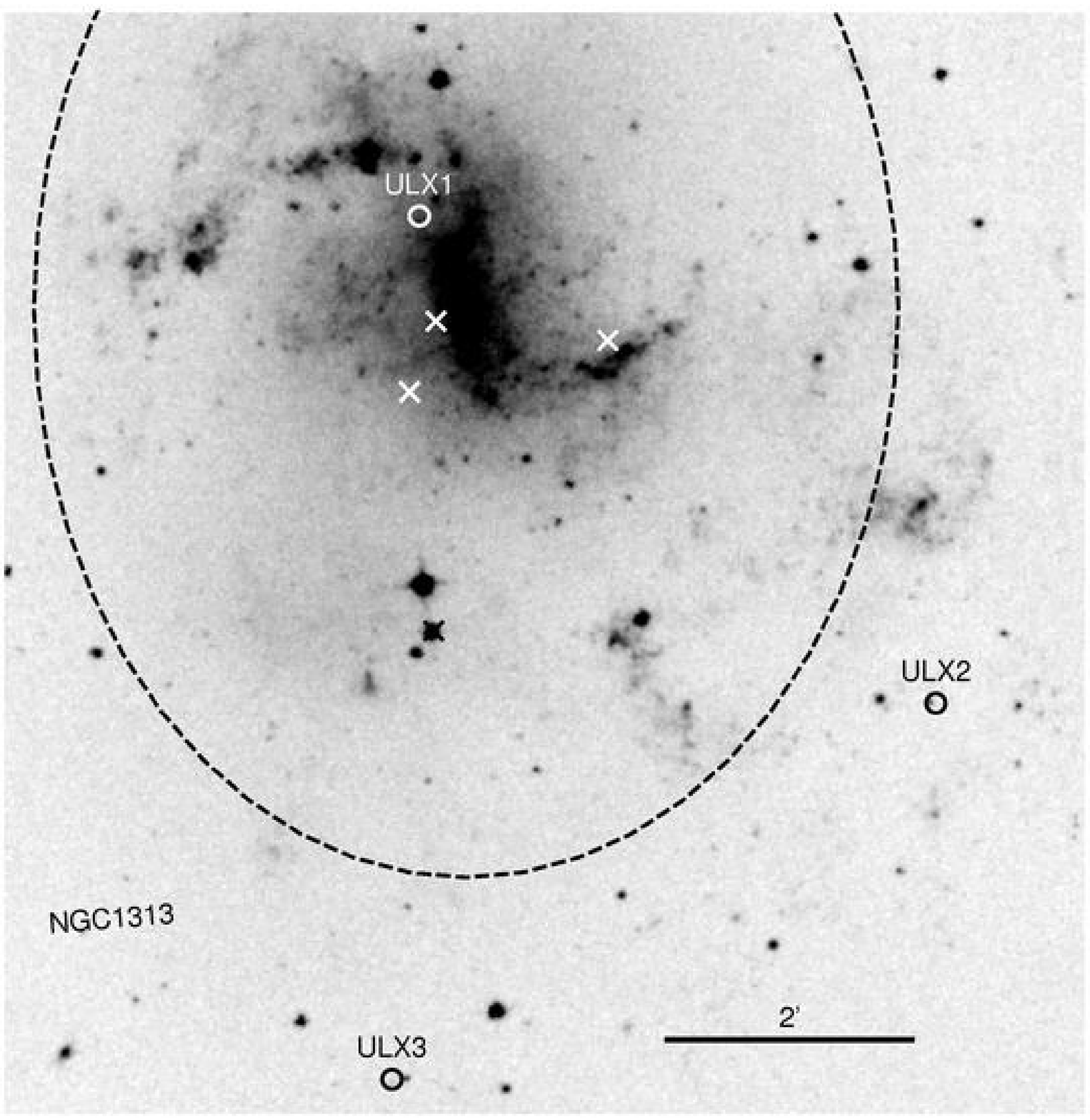}

\caption{The finding chart for the ULXs in NGC1313. }
\end{figure} 
\begin{figure}
\plotone{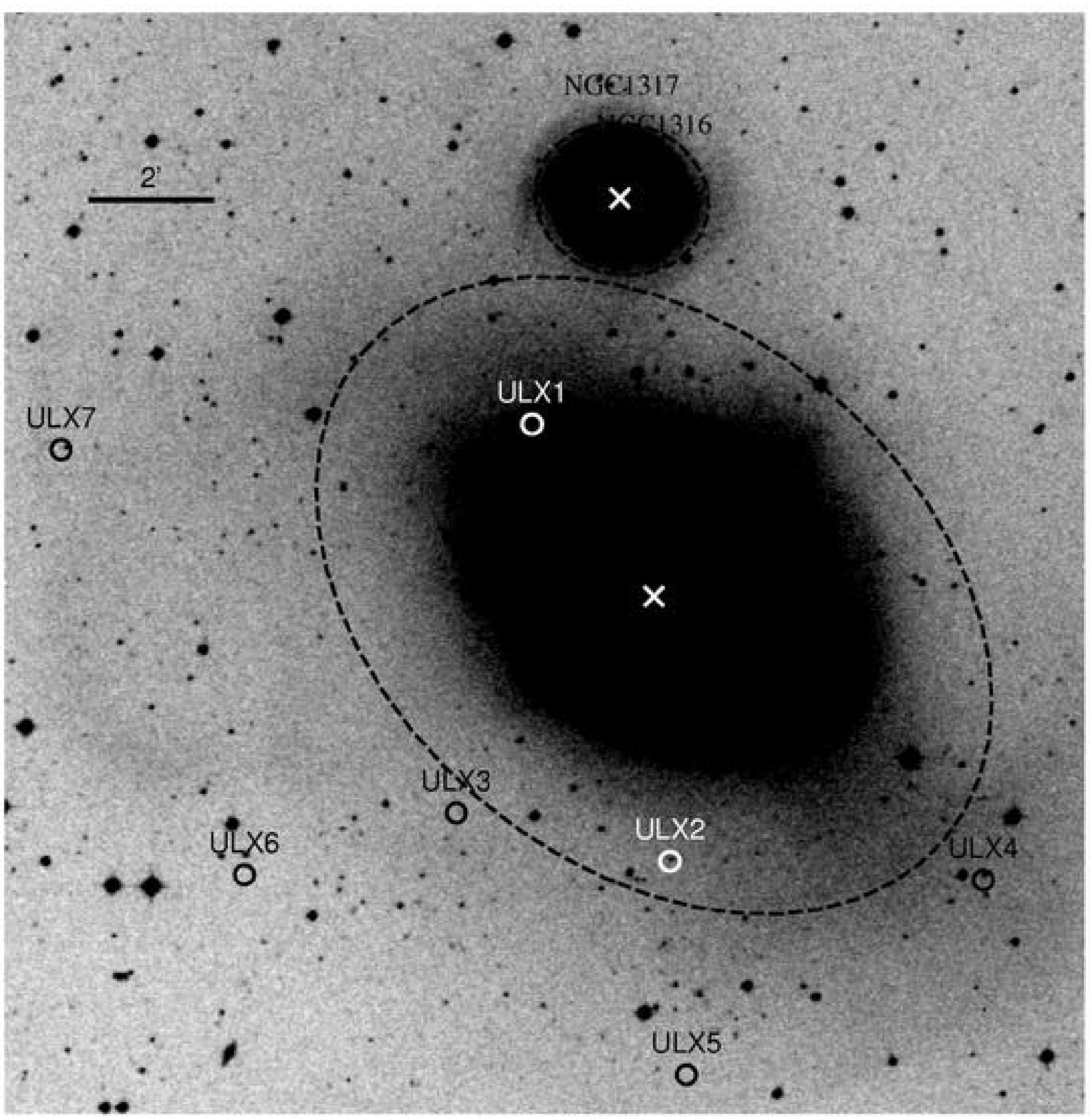}

\caption{The finding chart for the ULXs in NGC1316.}

\end{figure}
\clearpage

\begin{figure}
\plotone{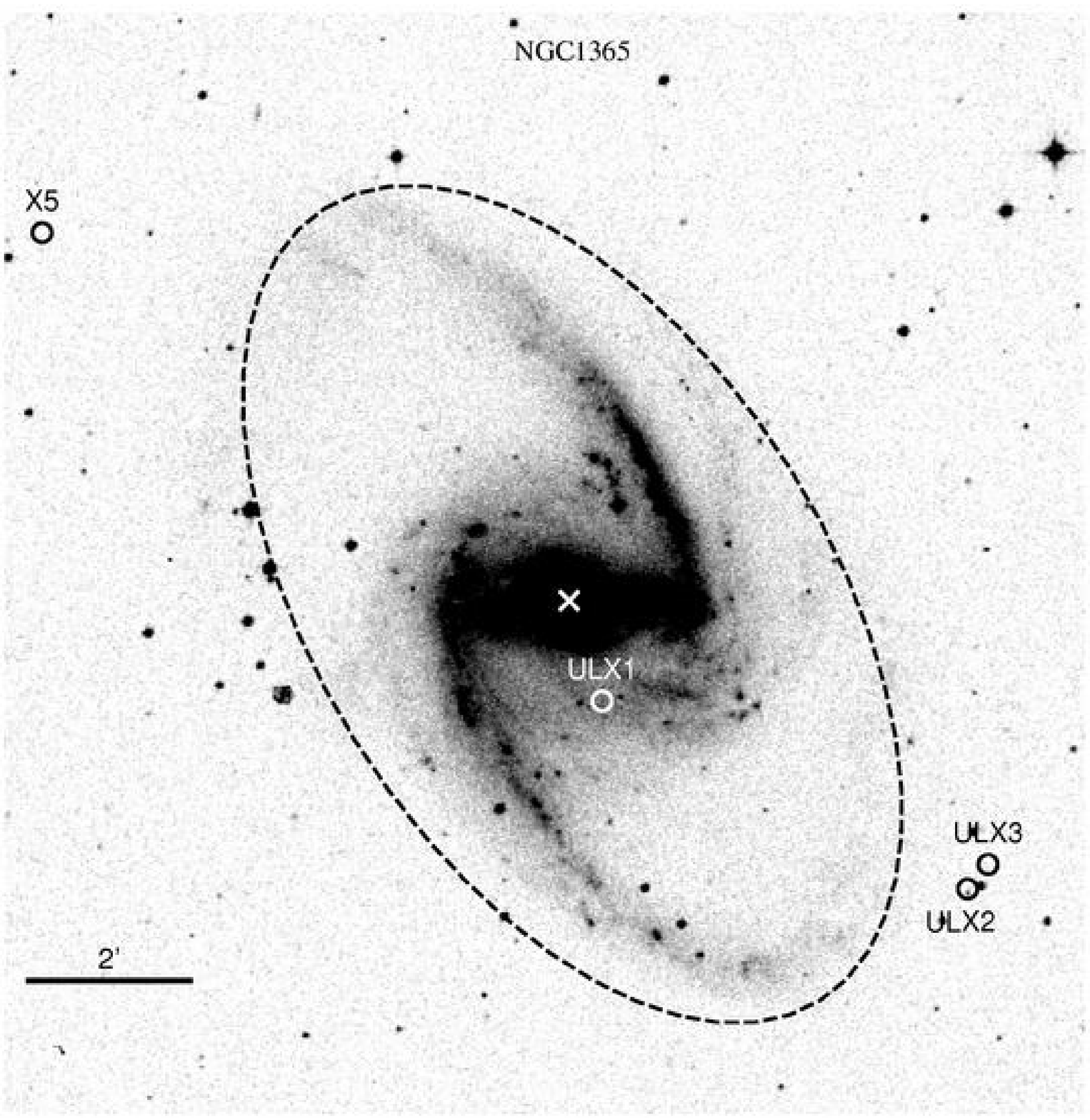}

\caption{The finding chart for the ULXs in NGC1365. }
\end{figure} 
\begin{figure}
\plotone{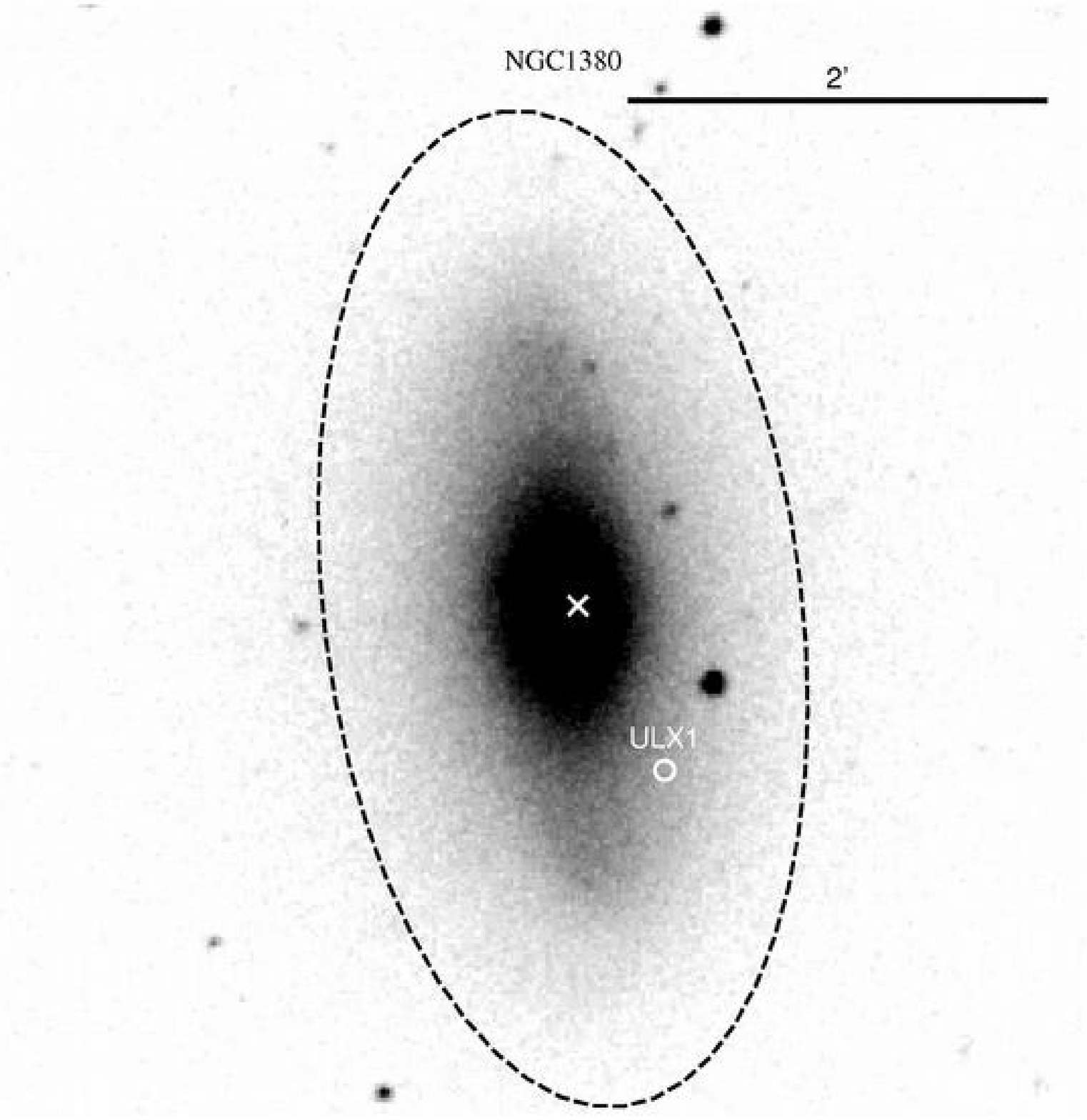}

\caption{The finding chart for the ULXs in NGC1380.}

\end{figure}
\begin{figure}
\plotone{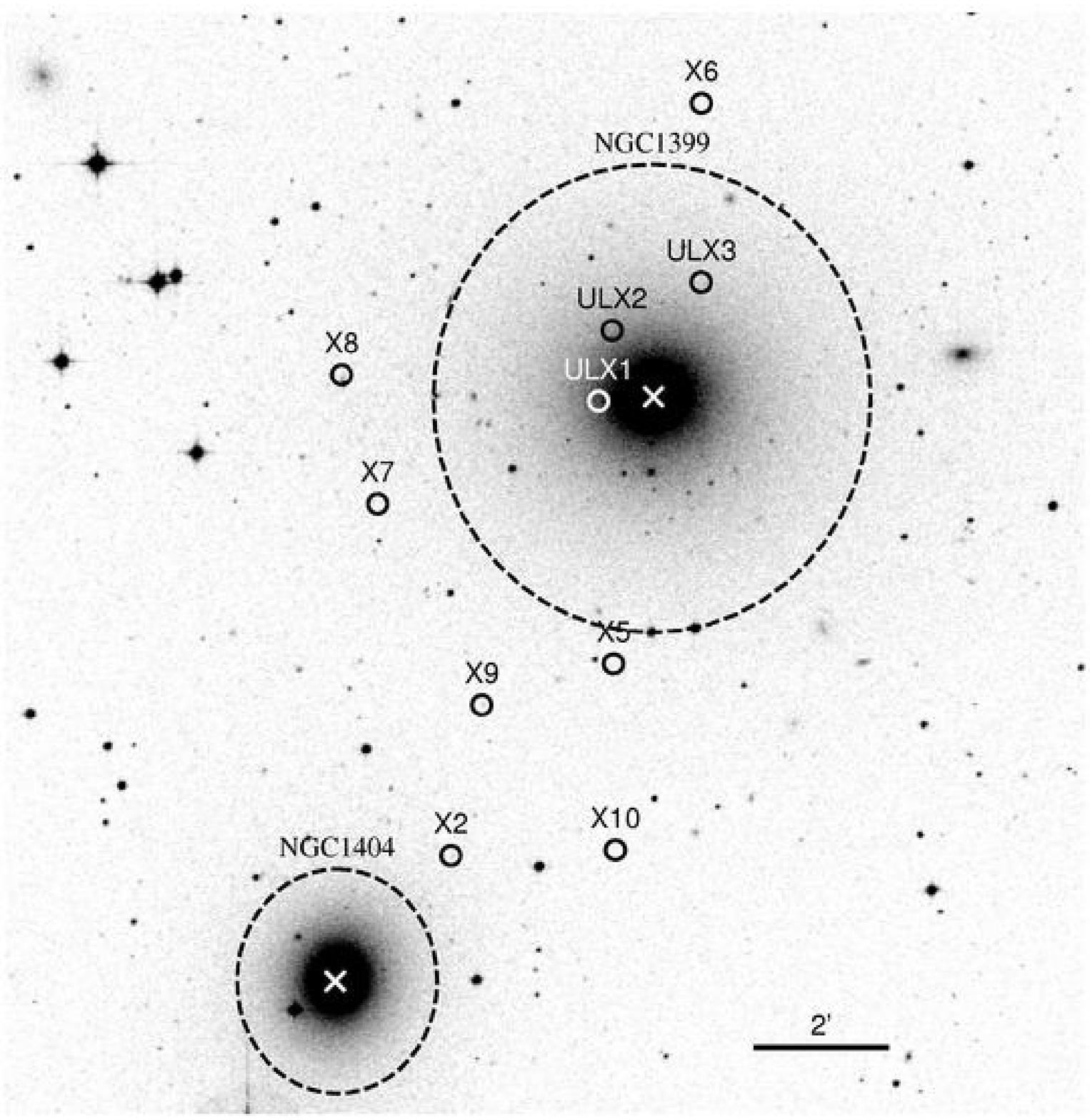}

\caption{The finding chart for the ULXs in NGC1399. }
\end{figure} 
\begin{figure}
\plotone{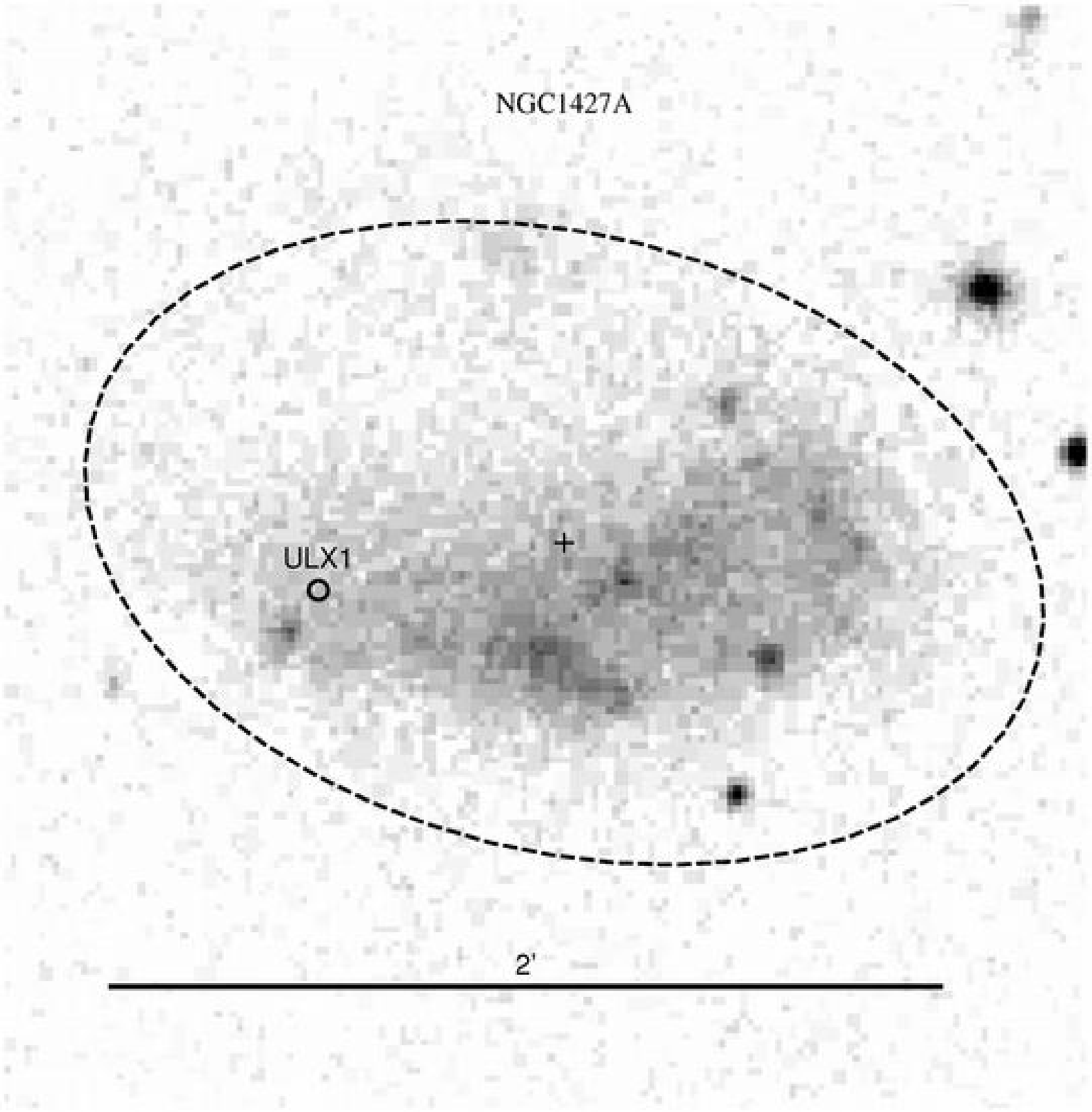}

\caption{The finding chart for the ULXs in NGC1427A.}

\end{figure}
\begin{figure}
\plotone{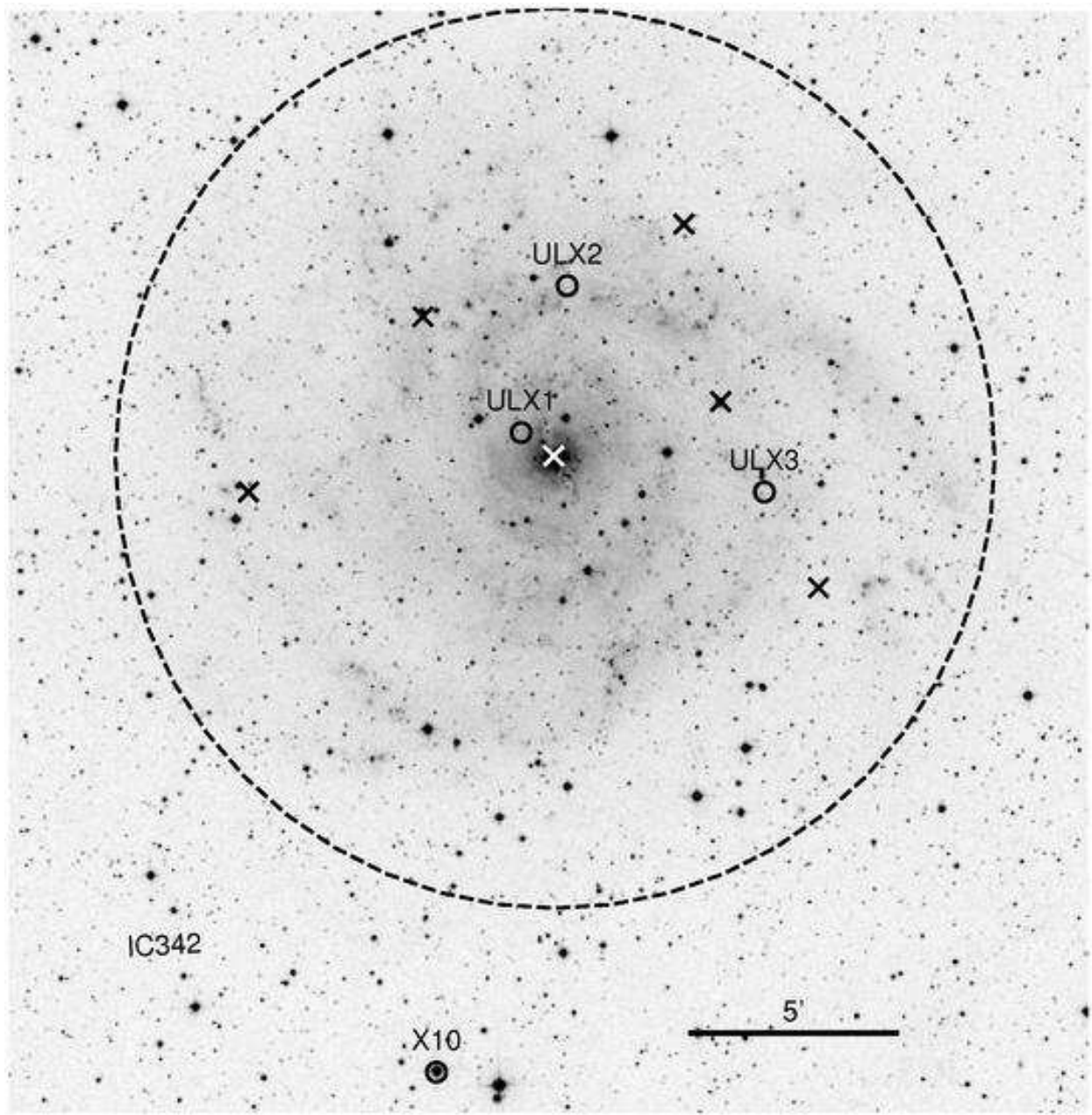}

\caption{The finding chart for the ULXs in PGC13826 (IC324). }
\end{figure} 
\begin{figure}
\plotone{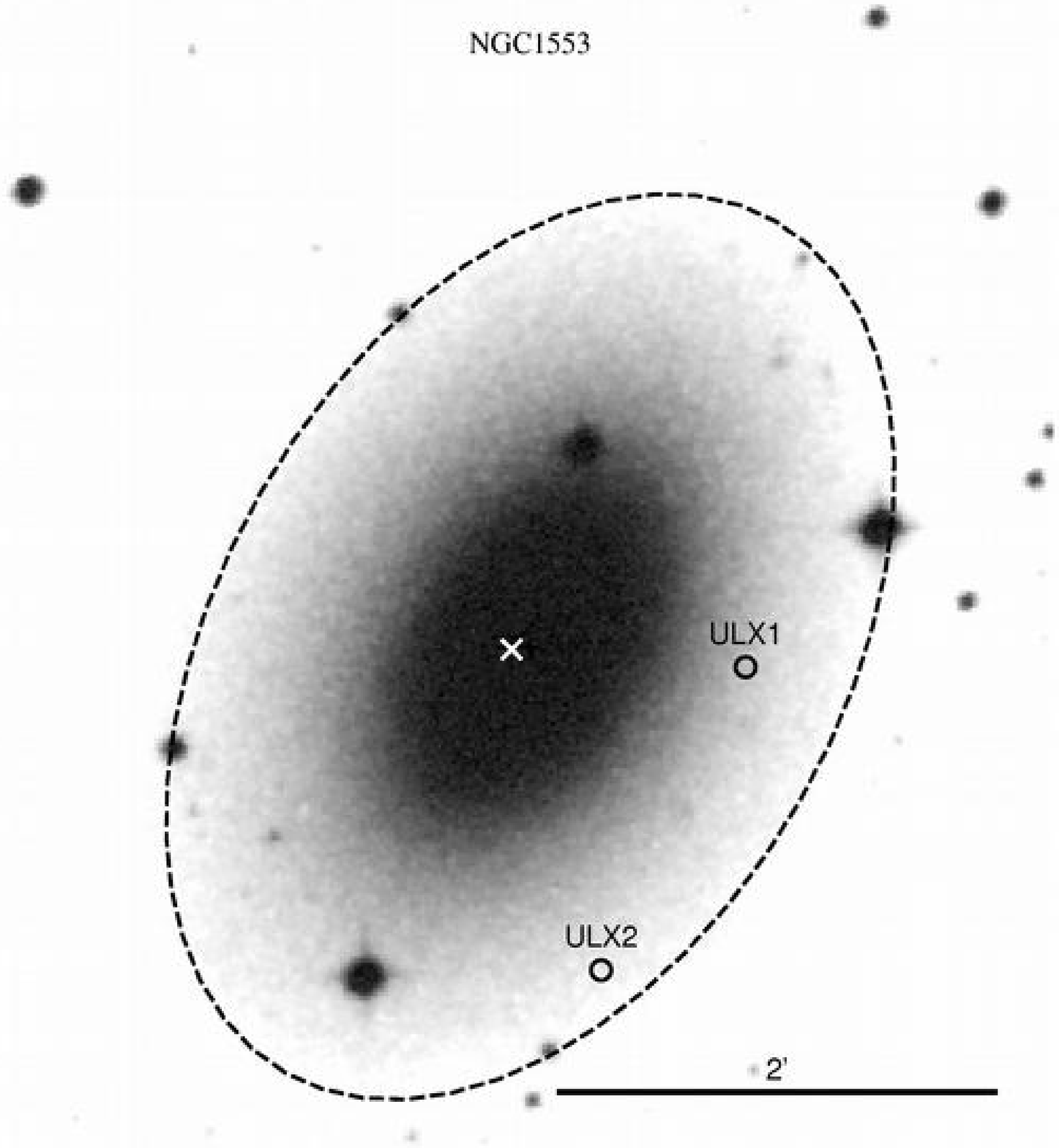}

\caption{The finding chart for the ULXs in NGC1553.}

\end{figure}
\begin{figure}
\plotone{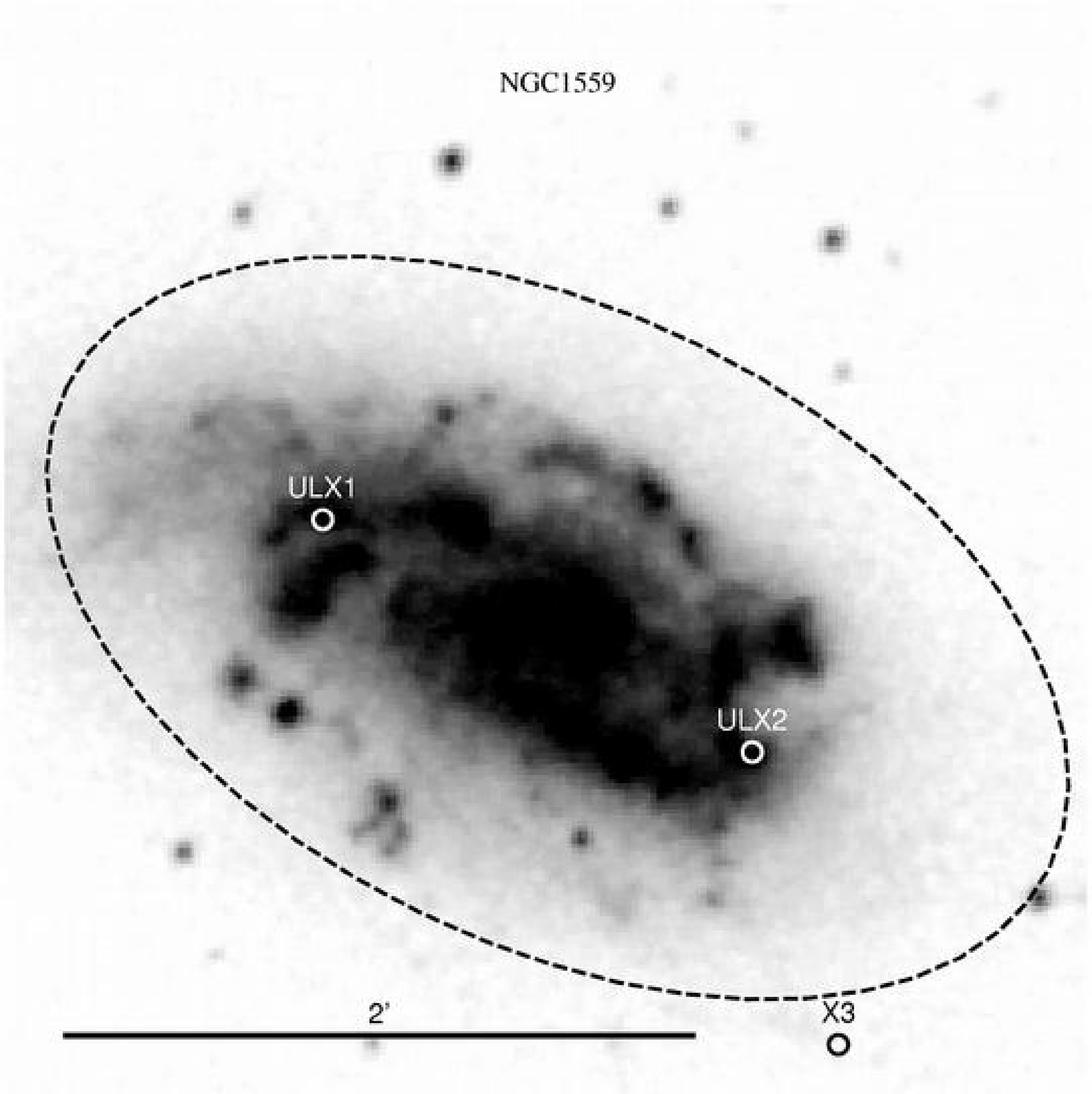}

\caption{The finding chart for the ULXs in NGC1559. }
\end{figure}
\begin{figure}
\plotone{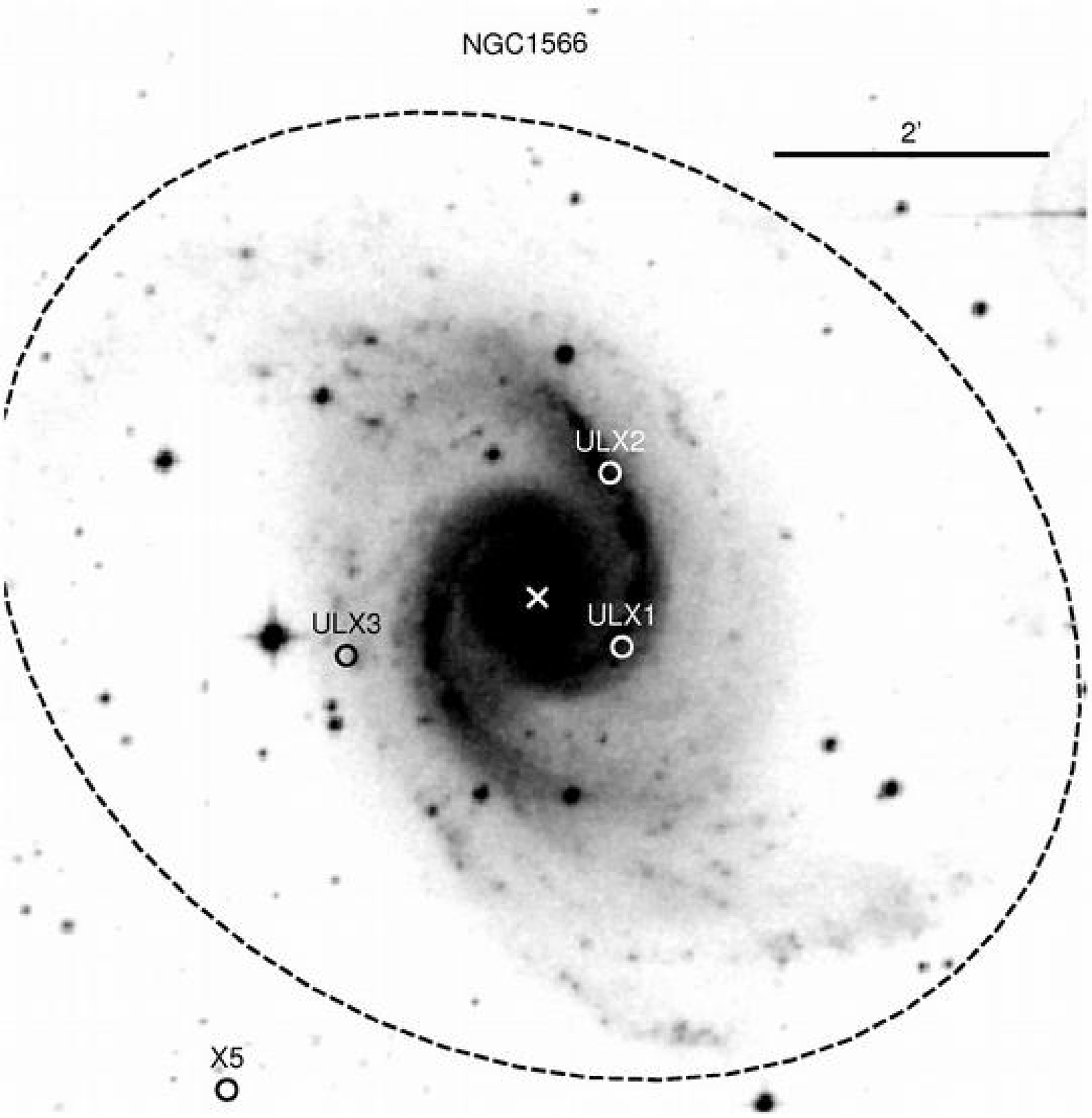}

\caption{The finding chart for the ULXs in NGC1566.}

\end{figure}
\clearpage

\begin{figure}
\plotone{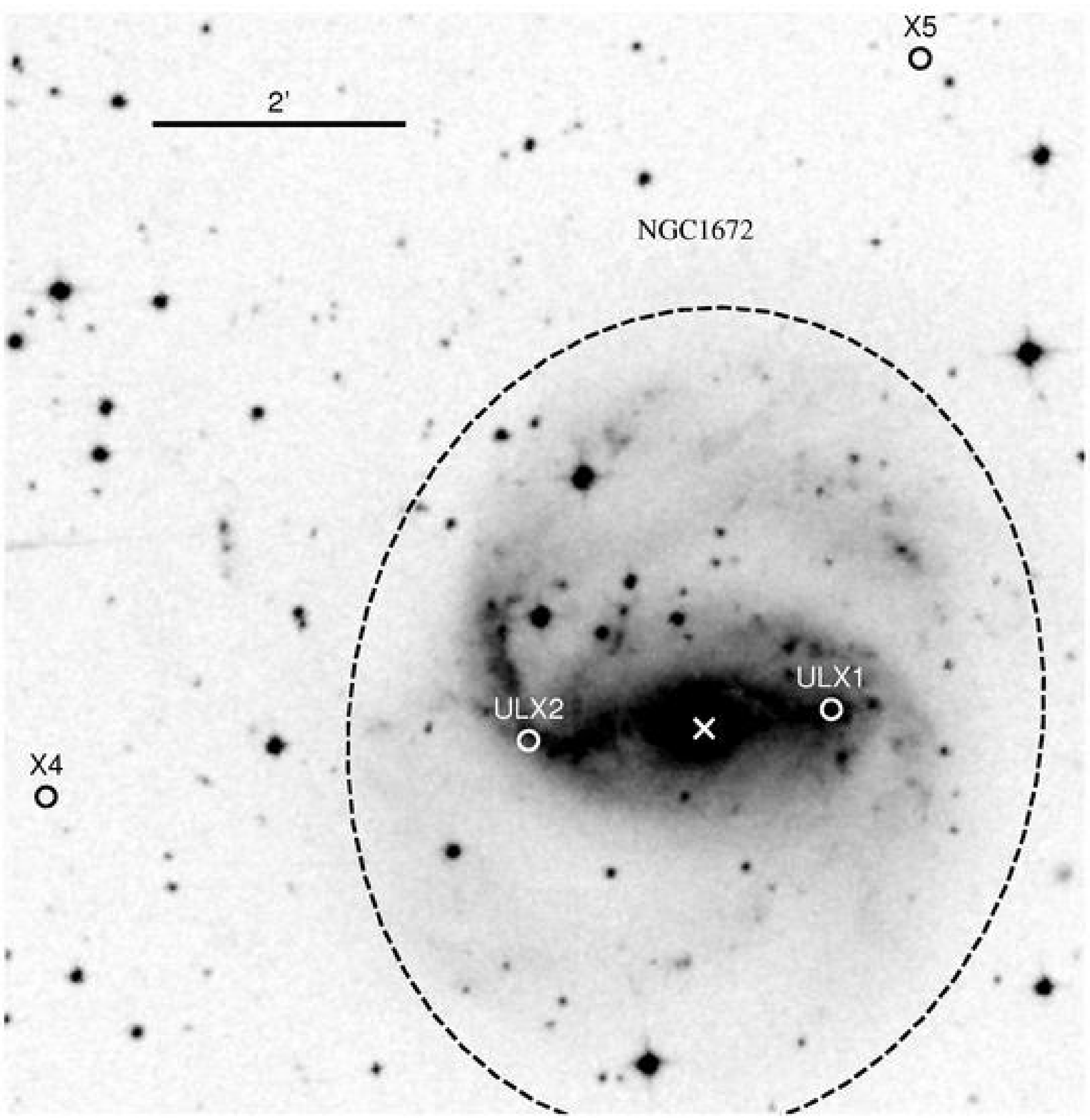}

\caption{The finding chart for the ULXs in NGC1672. }
\end{figure}
\begin{figure}
\plotone{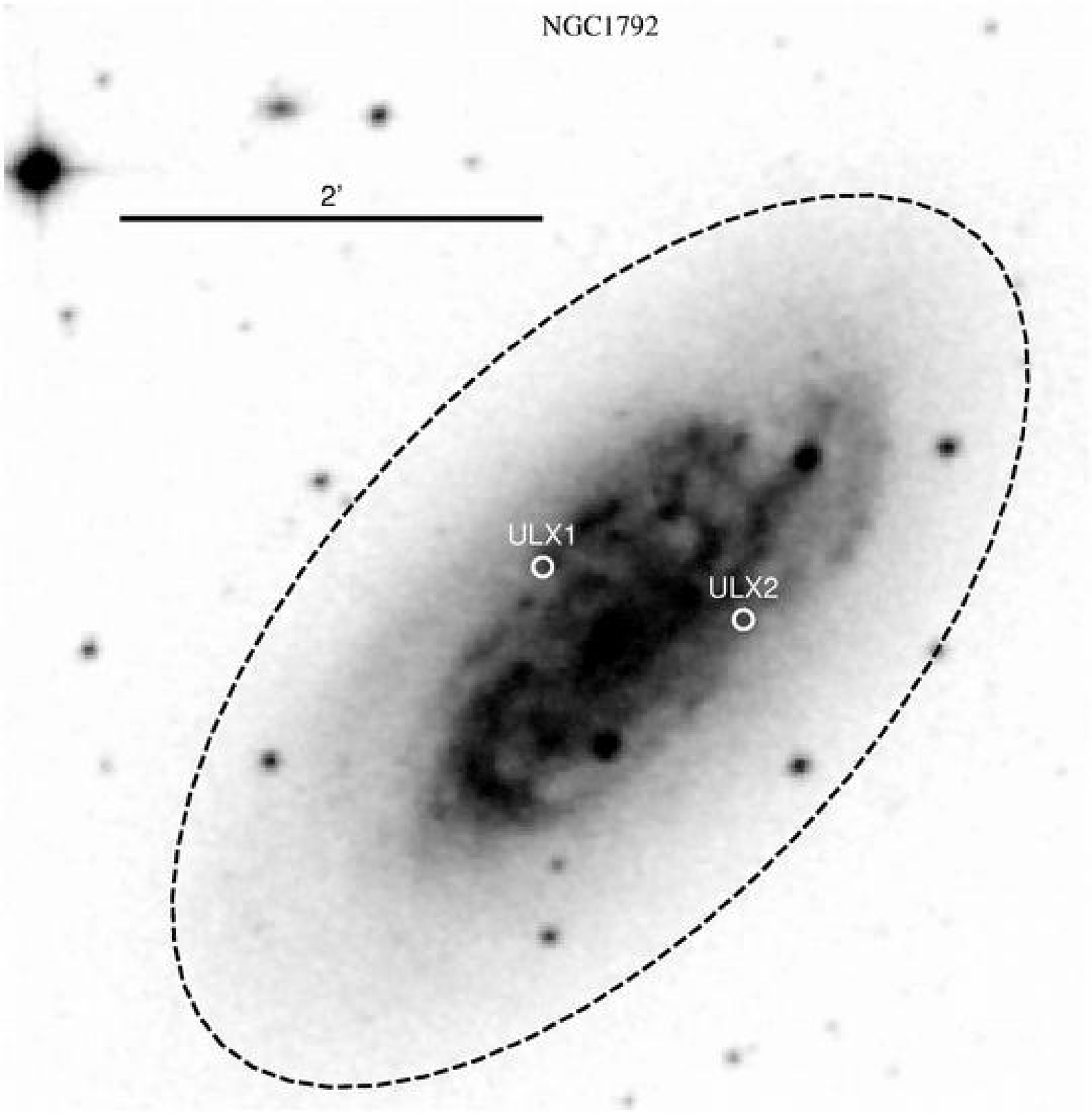}

\caption{The finding chart for the ULXs in NGC1792.}

\end{figure}
\begin{figure}
\plotone{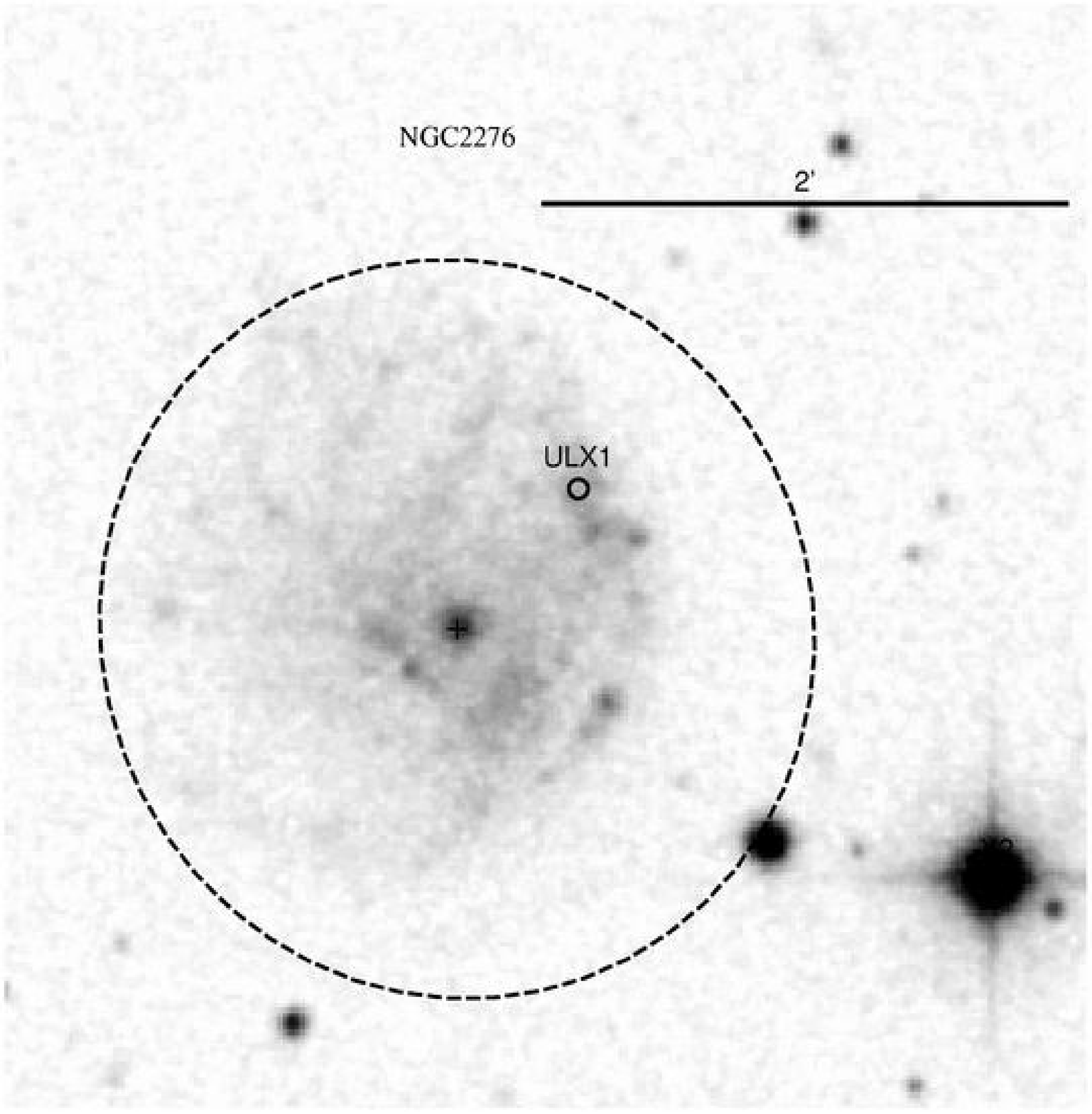}

\caption{The finding chart for the ULXs in NGC2276. }
\end{figure}
\begin{figure}
\plotone{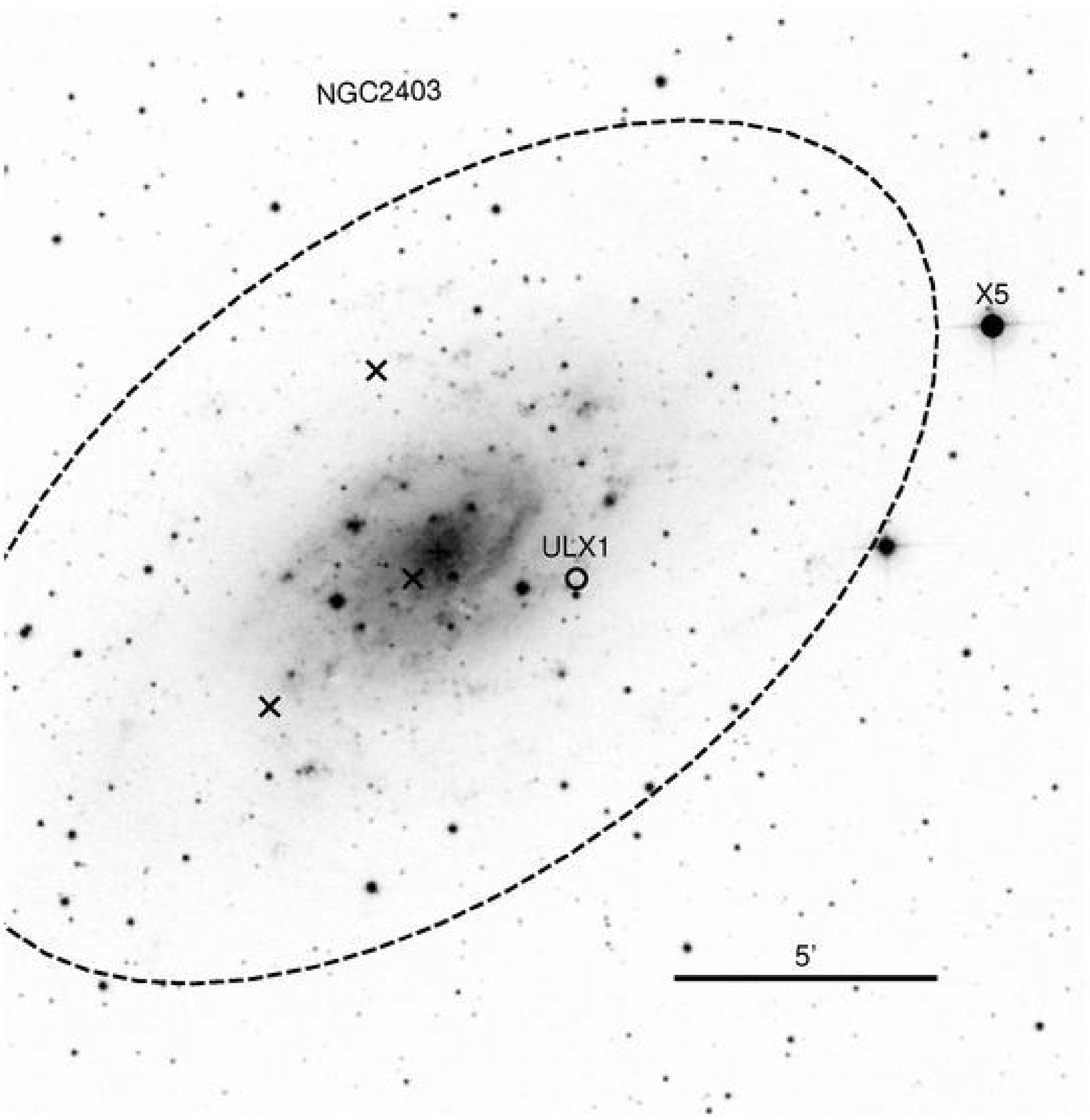}

\caption{The finding chart for the ULXs in NGC2403.}

\end{figure}
\begin{figure}
\plotone{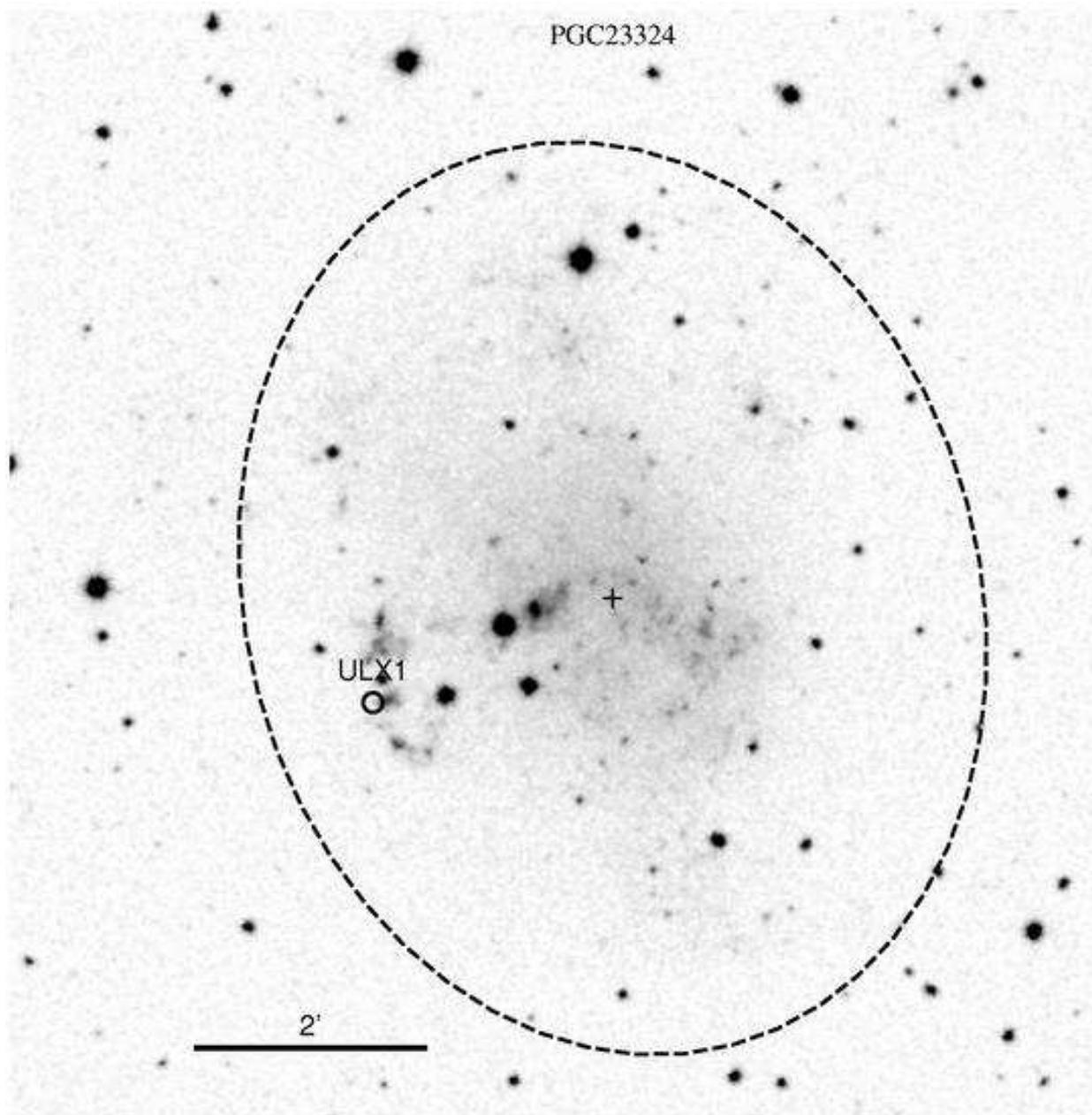}

\caption{The finding chart for the ULXs in PGC23324 (Holmberg II). }
\end{figure}
\begin{figure}
\plotone{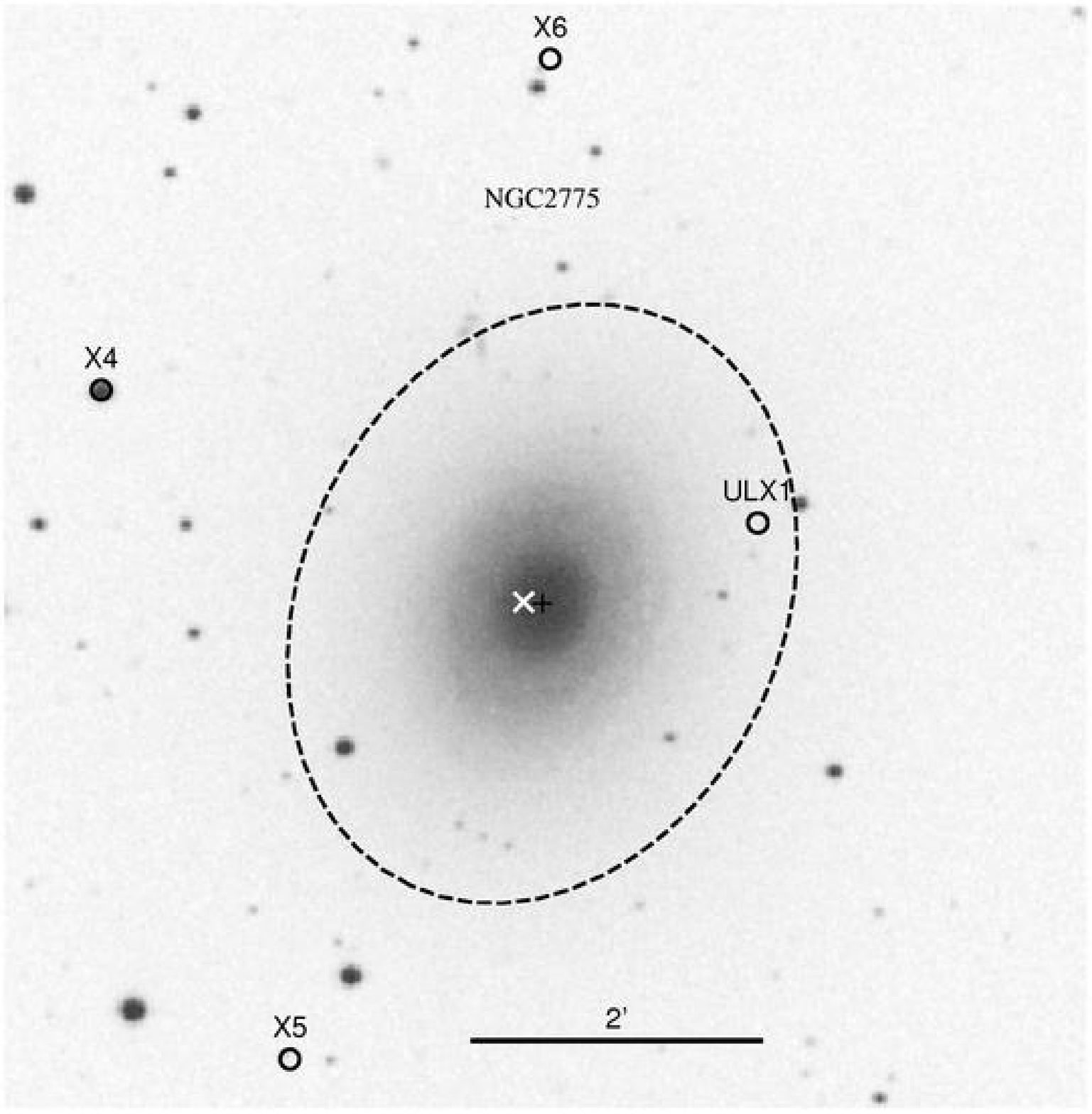}

\caption{The finding chart for the ULXs in NGC2775.}

\end{figure}
\begin{figure}
\plotone{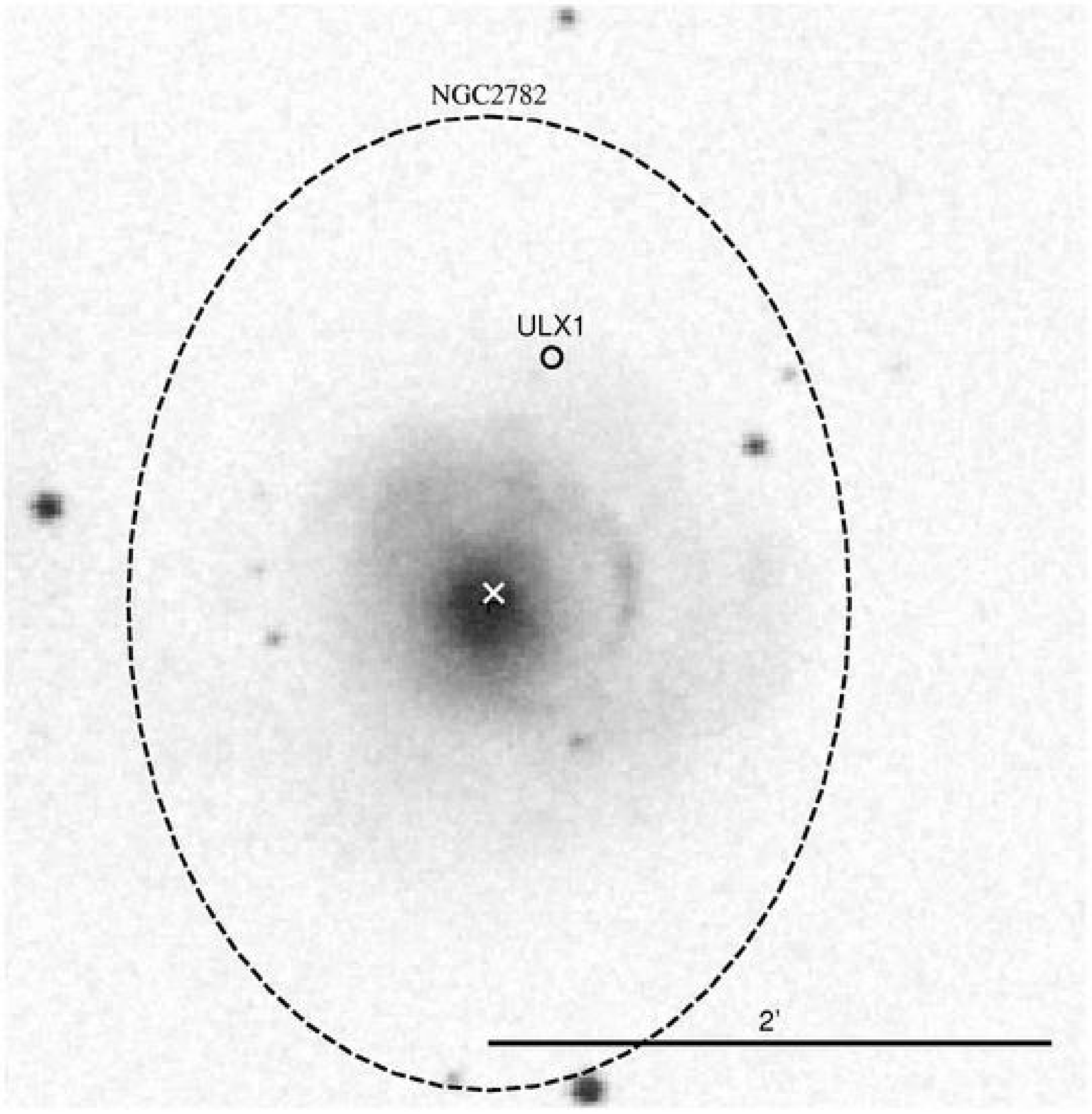}

\caption{The finding chart for the ULXs in NGC2782. }
\end{figure}
\begin{figure}
\plotone{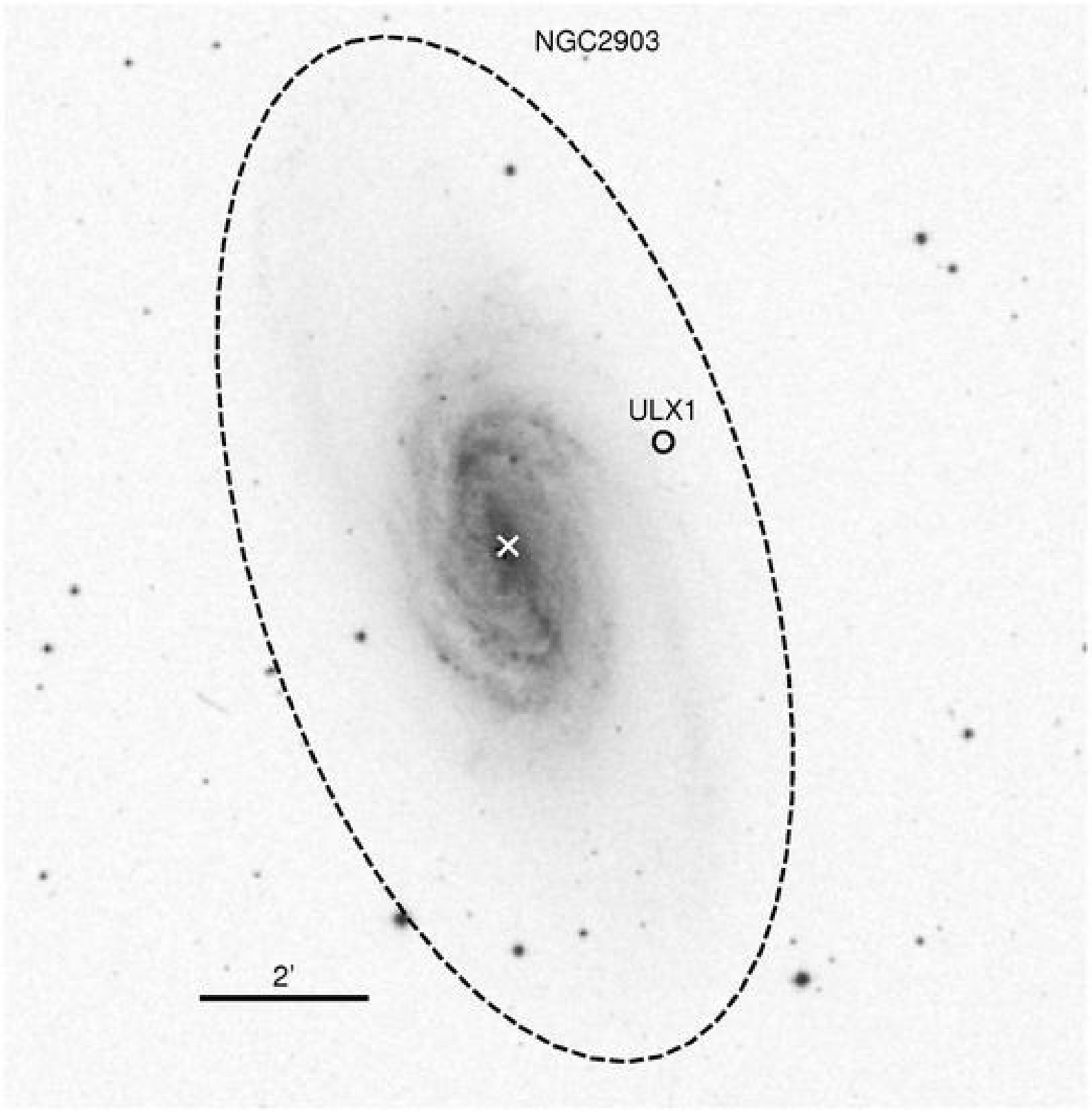}

\caption{The finding chart for the ULXs in NGC2903.}

\end{figure}
\clearpage

\begin{figure}
\plotone{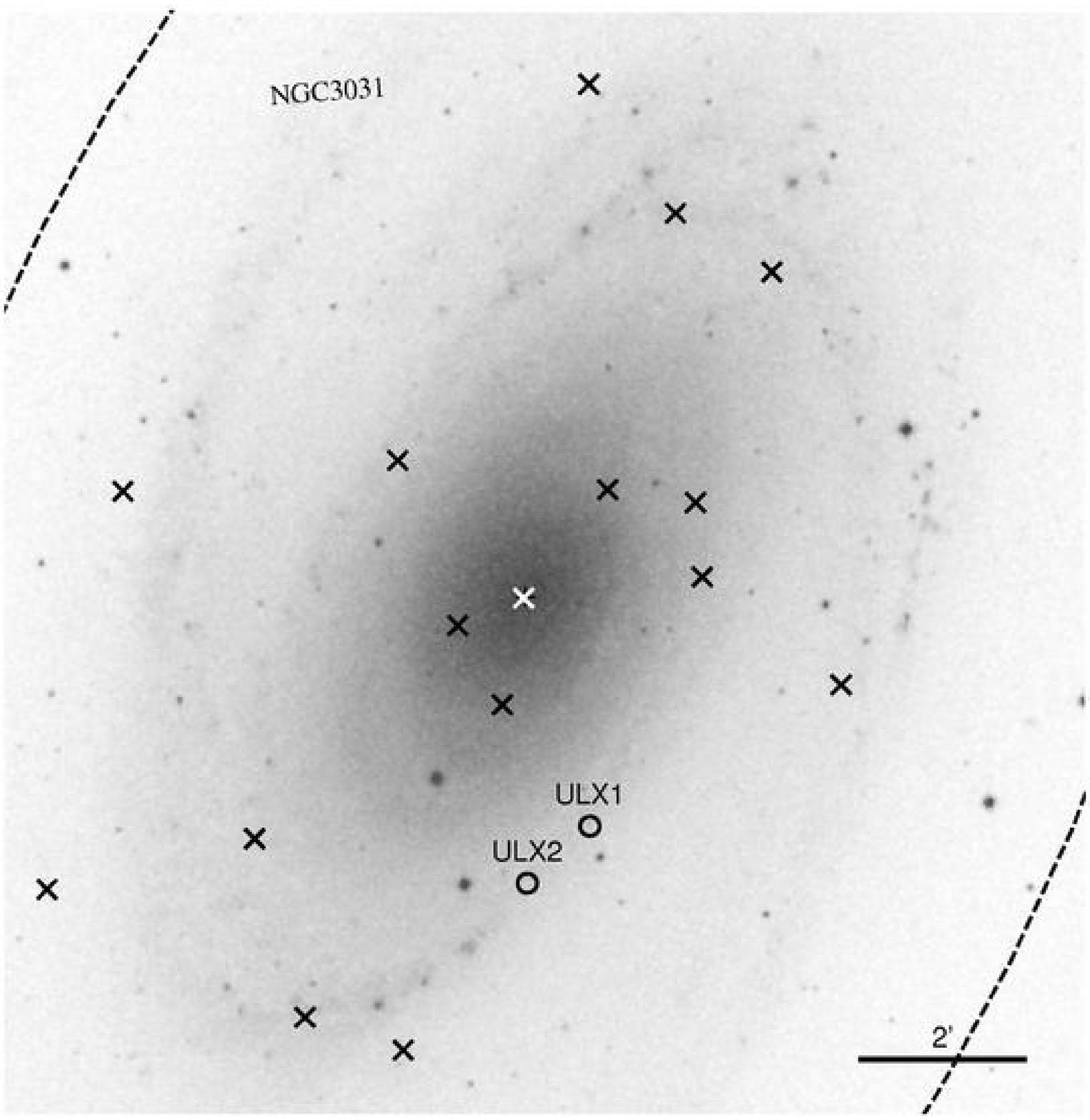}

\caption{The finding chart for the ULXs in NGC3031. }
\end{figure}
\begin{figure}
\plotone{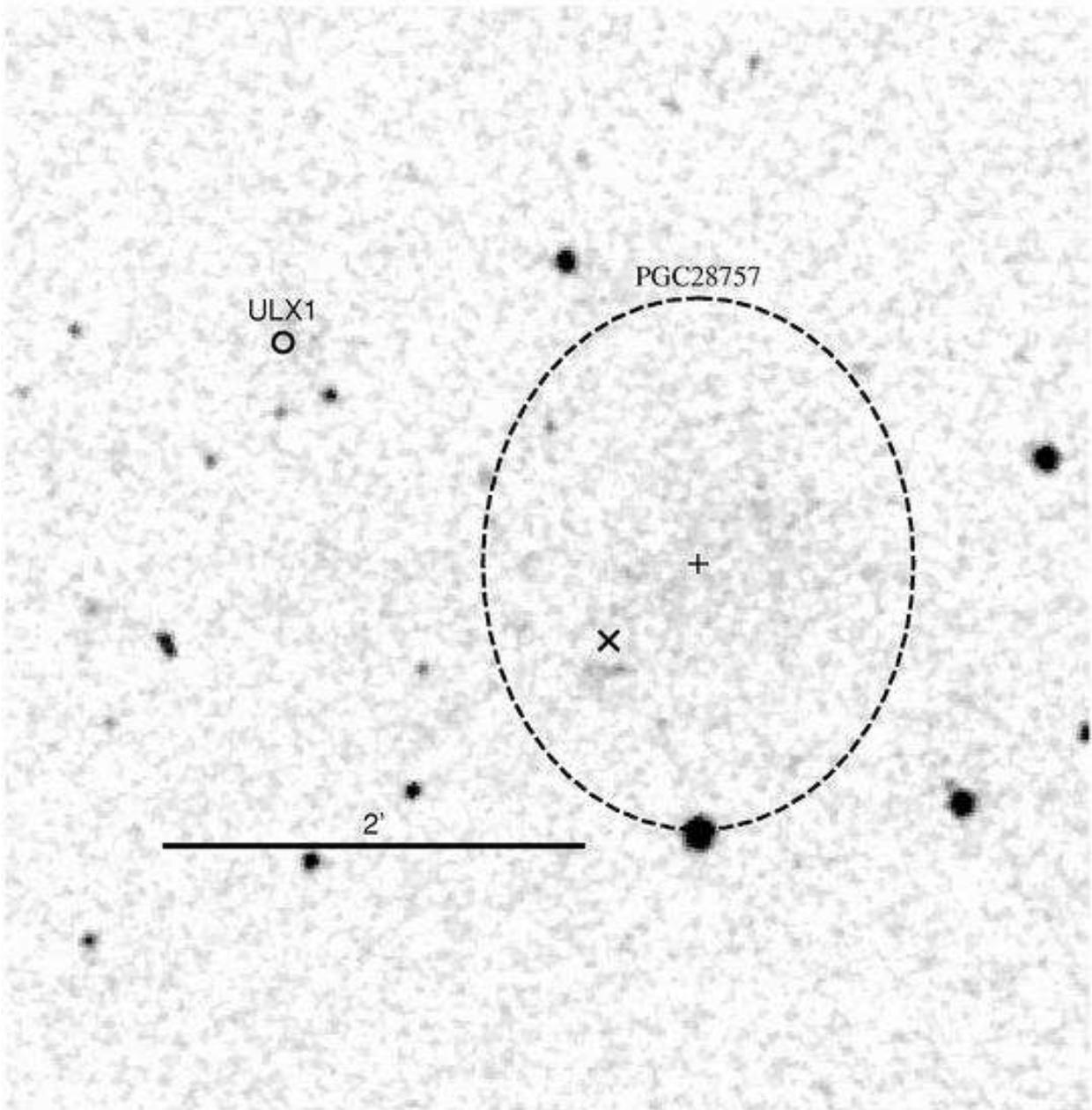}

\caption{The finding chart for the ULXs in PGC28757 (Holmberg IX).}

\end{figure}
\begin{figure}
\plotone{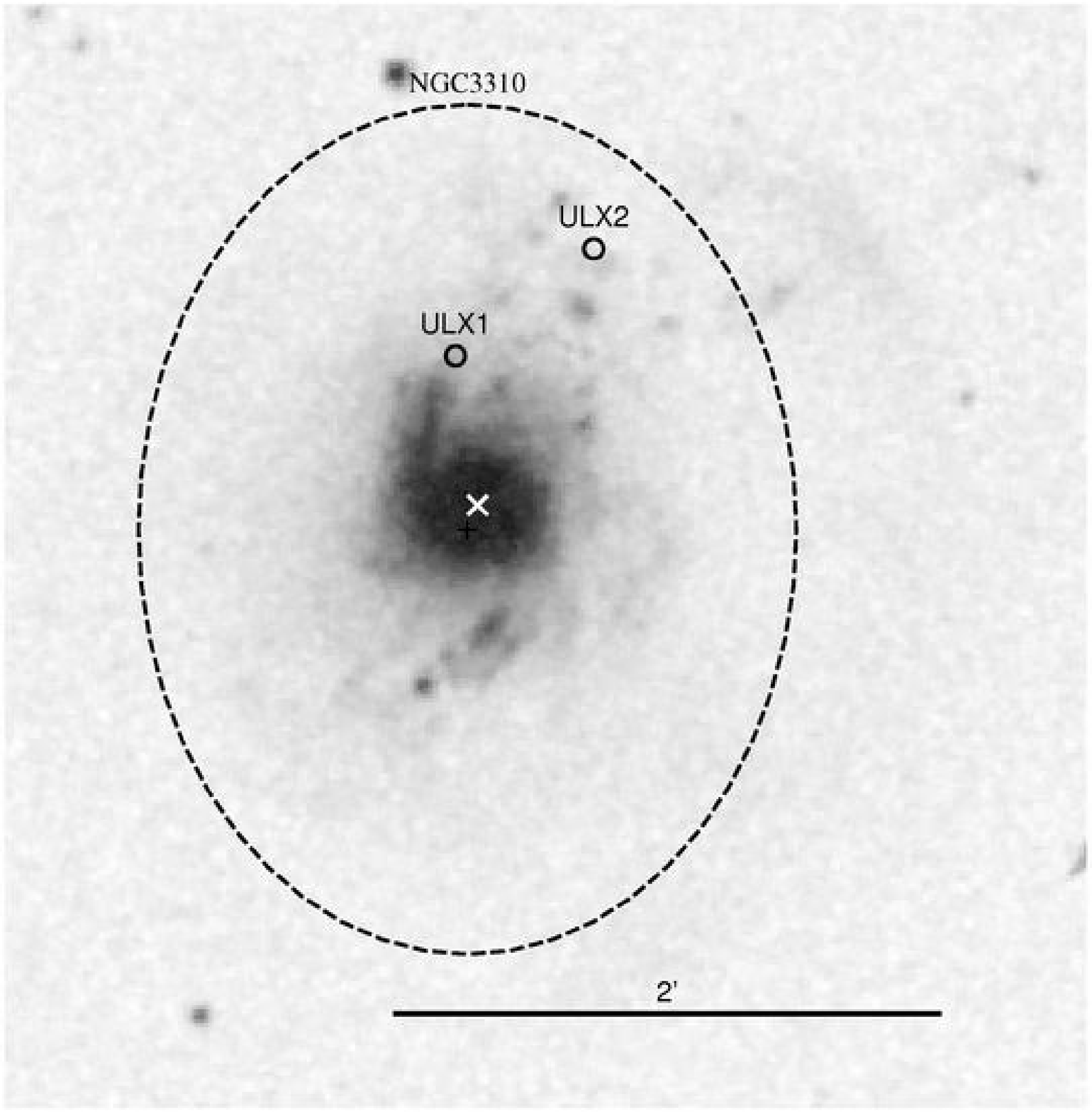}

\caption{The finding chart for the ULXs in NGC3310. }
\end{figure}
\begin{figure}
\plotone{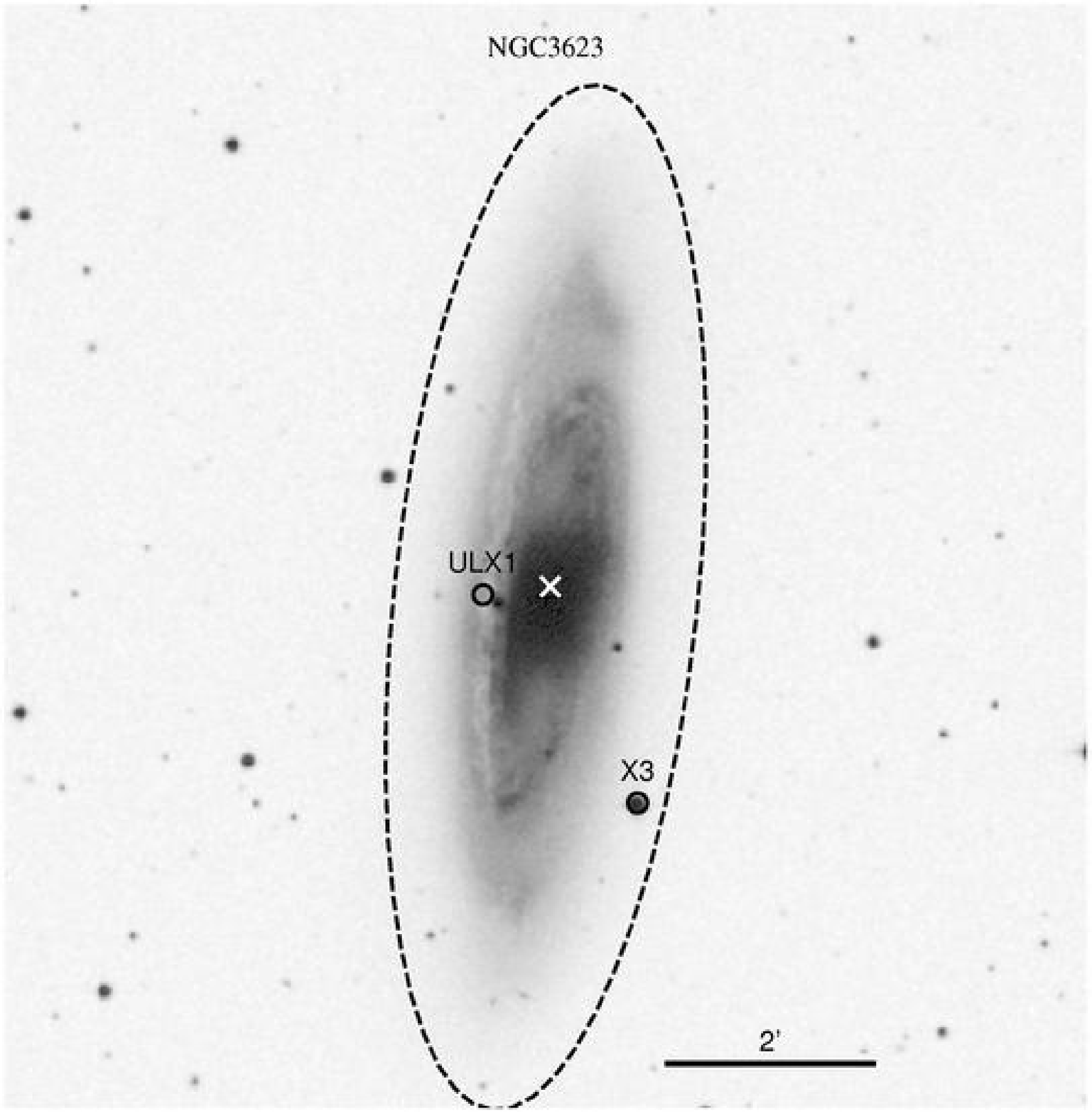}

\caption{The finding chart for the ULXs in NGC3623.}

\end{figure}
\begin{figure}
\plotone{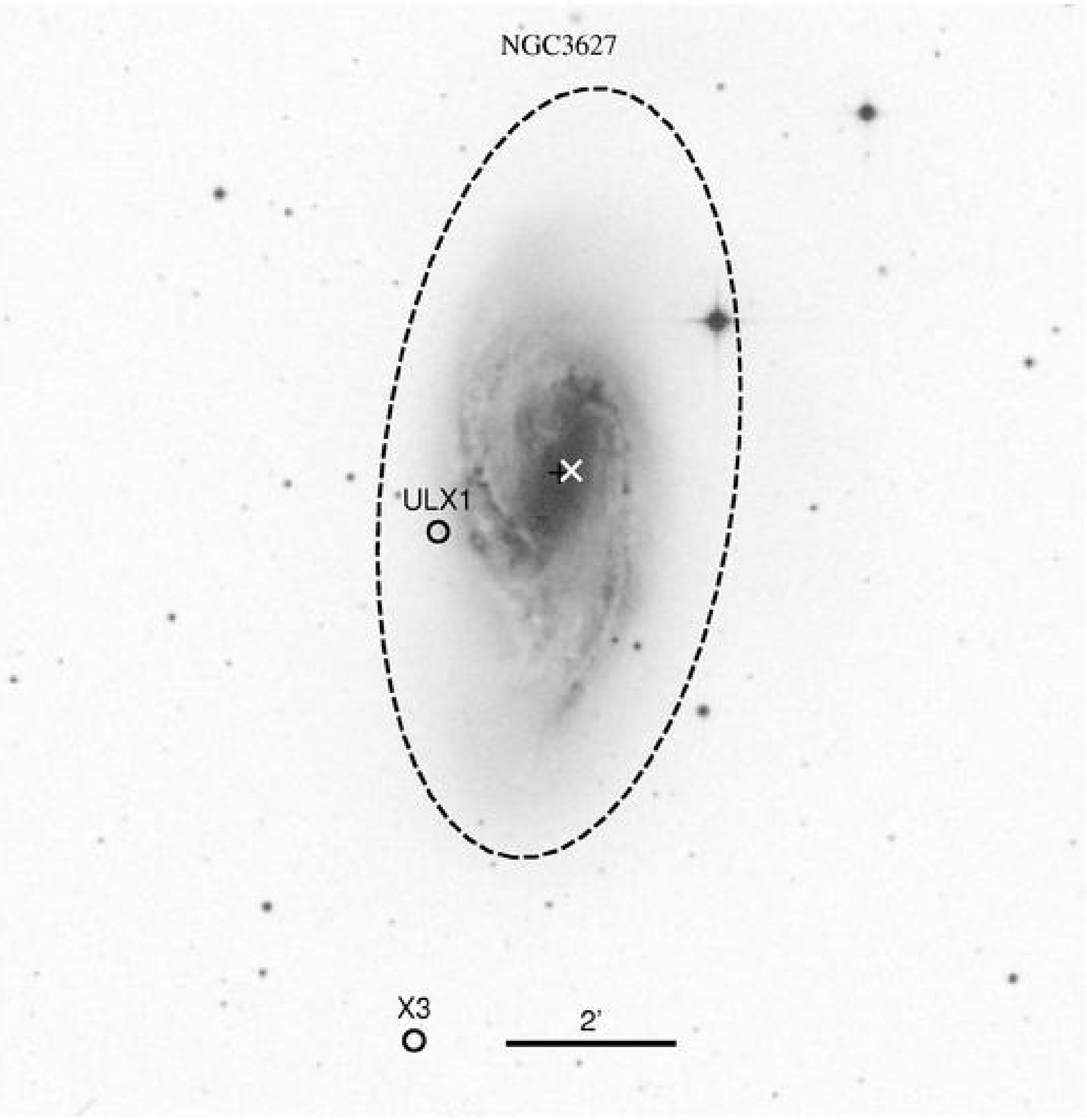}

\caption{The finding chart for the ULXs in NGC3627. }
\end{figure}
\begin{figure}
\plotone{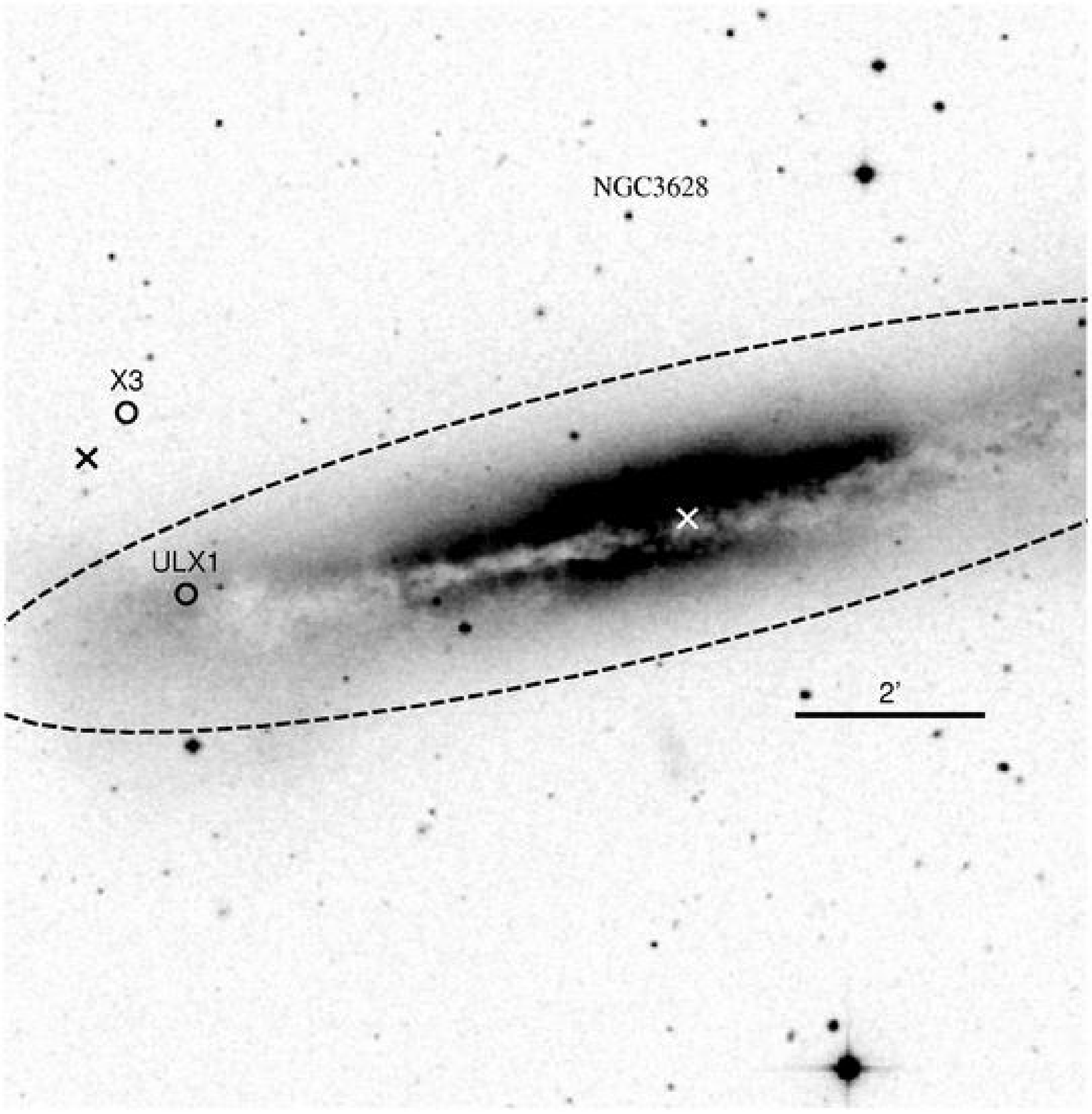}

\caption{The finding chart for the ULXs in NGC3628.}

\end{figure}
\begin{figure}
\plotone{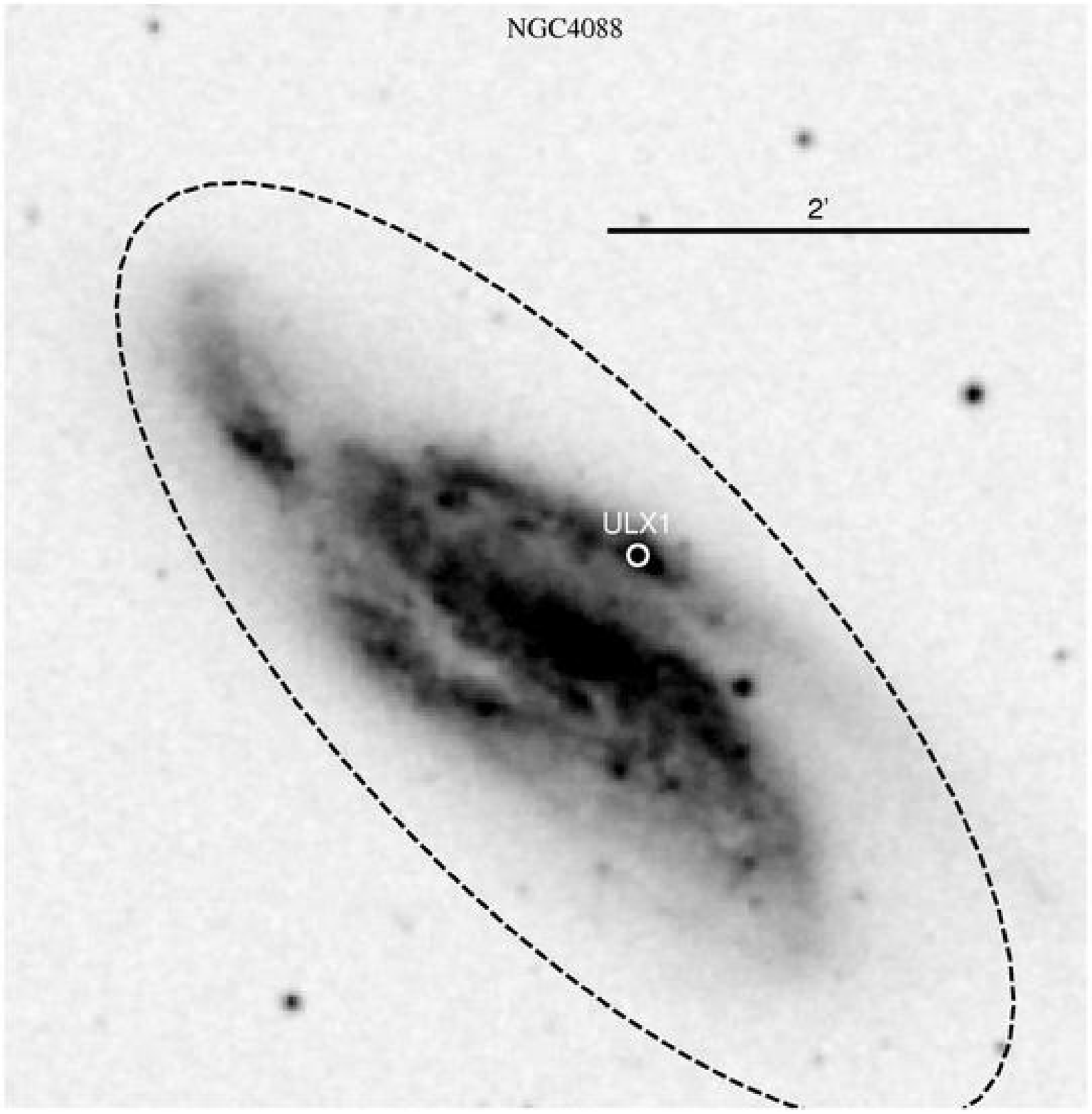}

\caption{The finding chart for the ULXs in NGC4088. }
\end{figure}
\begin{figure}
\plotone{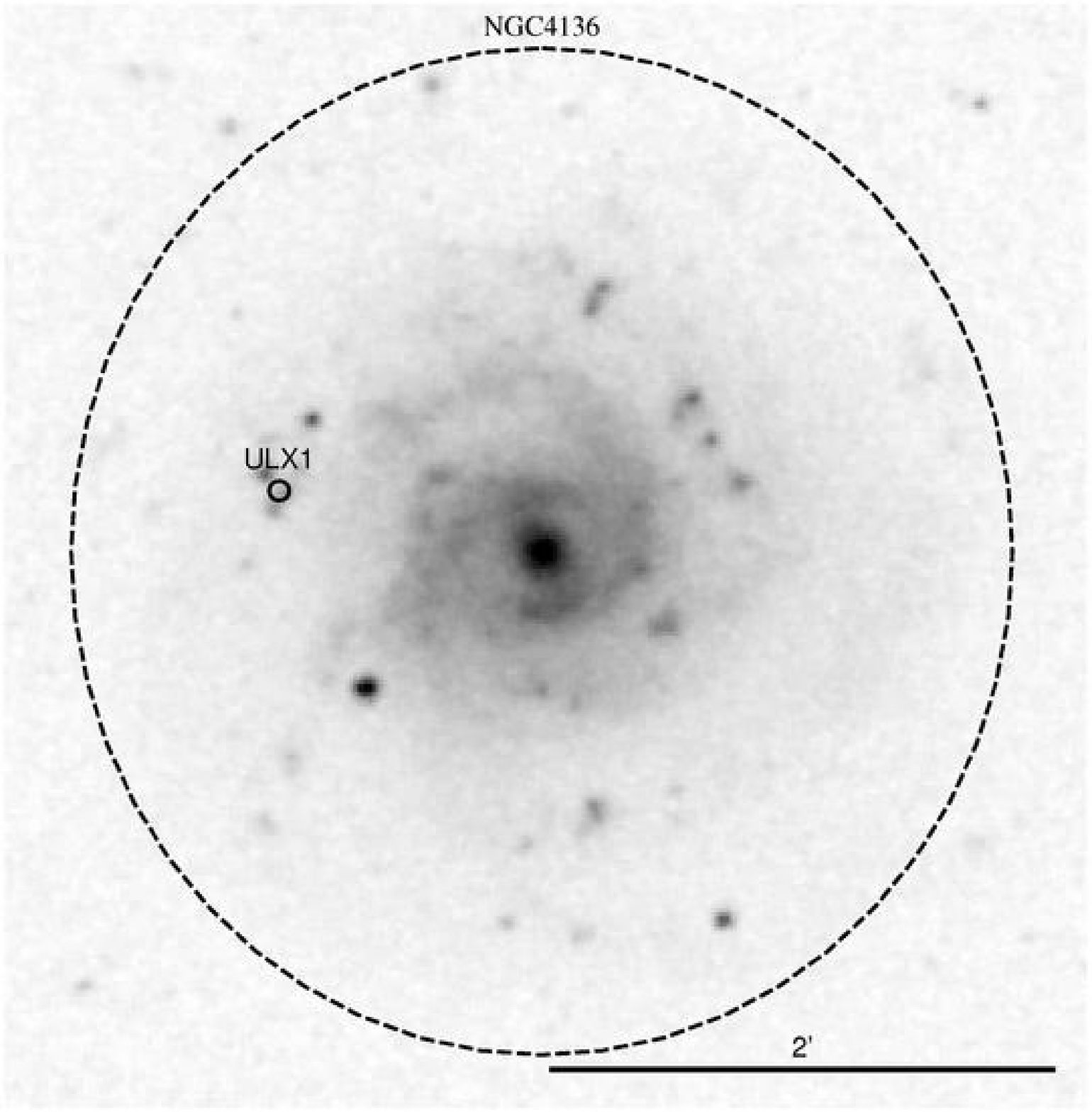}

\caption{The finding chart for the ULXs in NGC4136.}

\end{figure}
\clearpage

\begin{figure}
\plotone{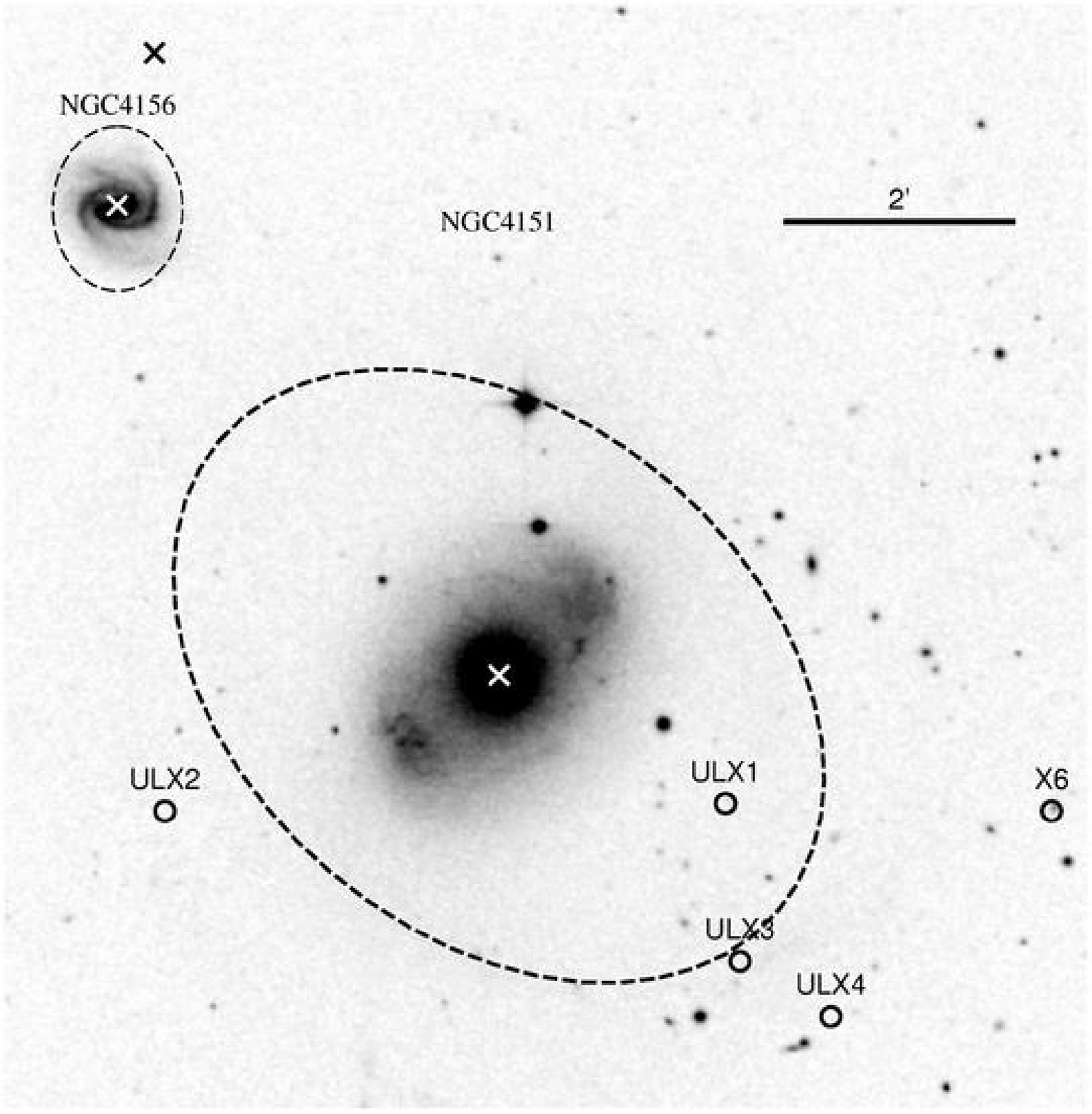}

\caption{The finding chart for the ULXs in NGC4151. }
\end{figure}
\begin{figure}
\plotone{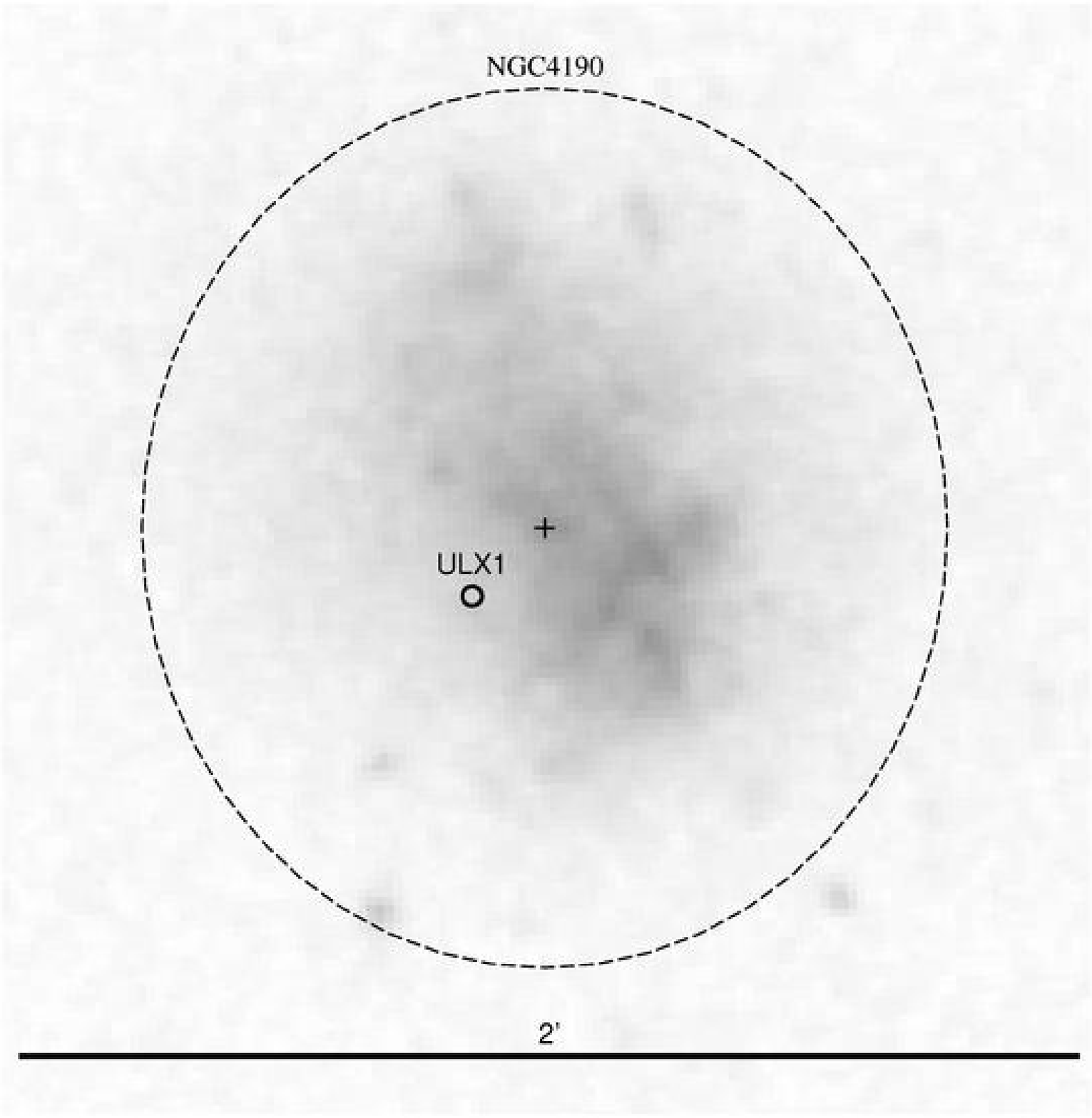}

\caption{The finding chart for the ULXs in NGC4190.}

\end{figure}
\begin{figure}
\plotone{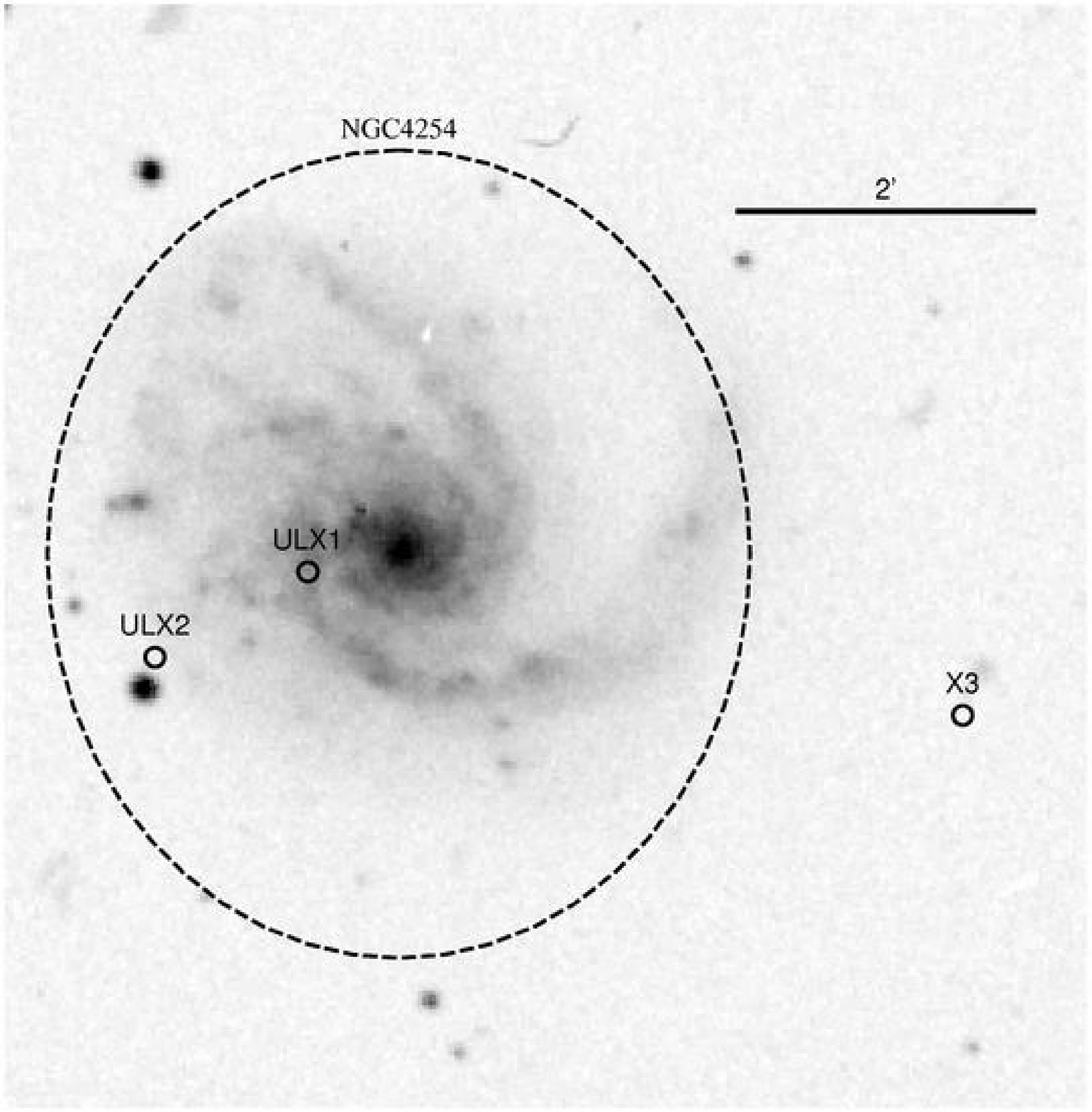}

\caption{The finding chart for the ULXs in NGC4254. }
\end{figure}
\begin{figure}
\plotone{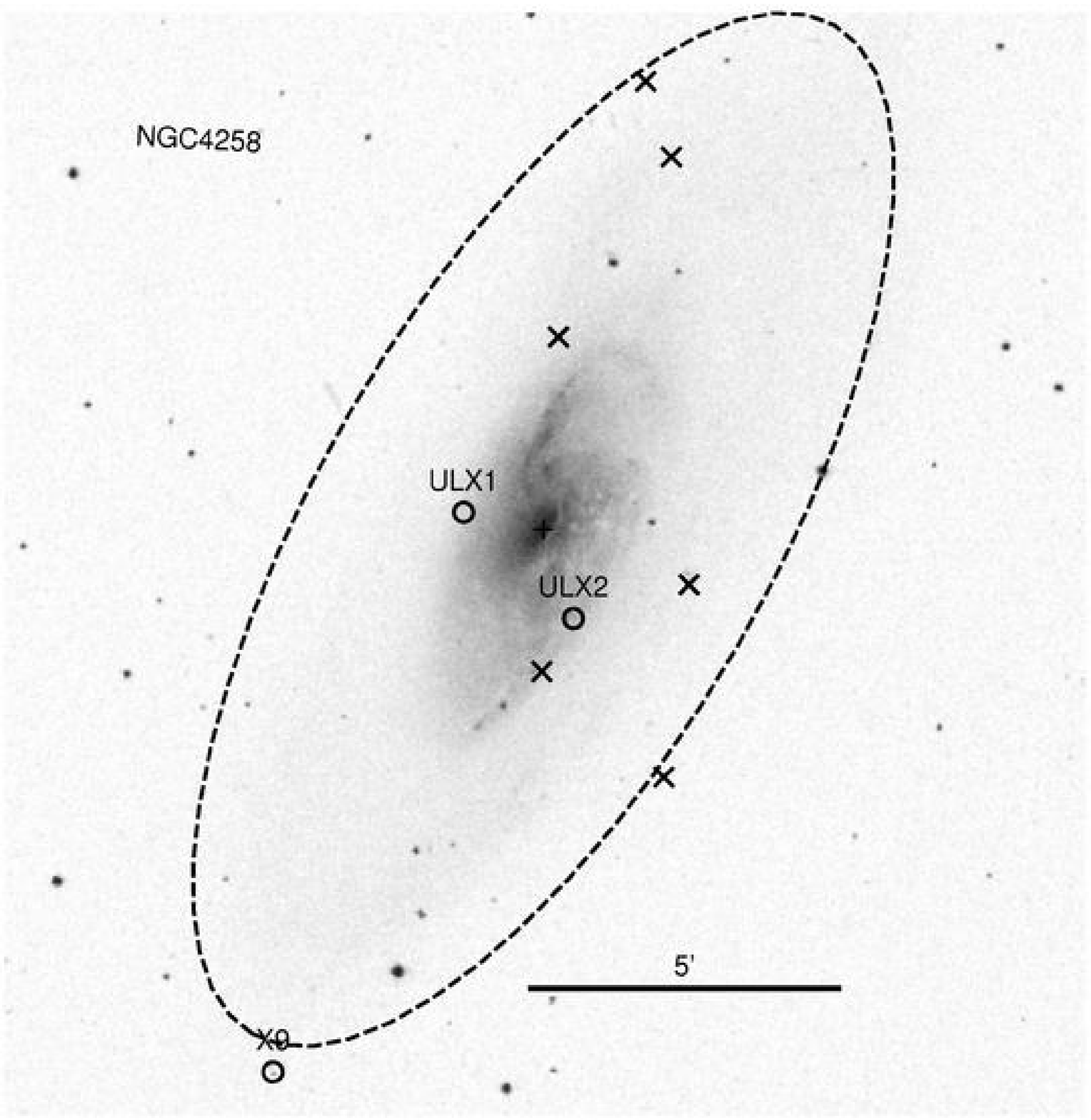}

\caption{The finding chart for the ULXs in NGC4258.}

\end{figure}
\begin{figure}
\plotone{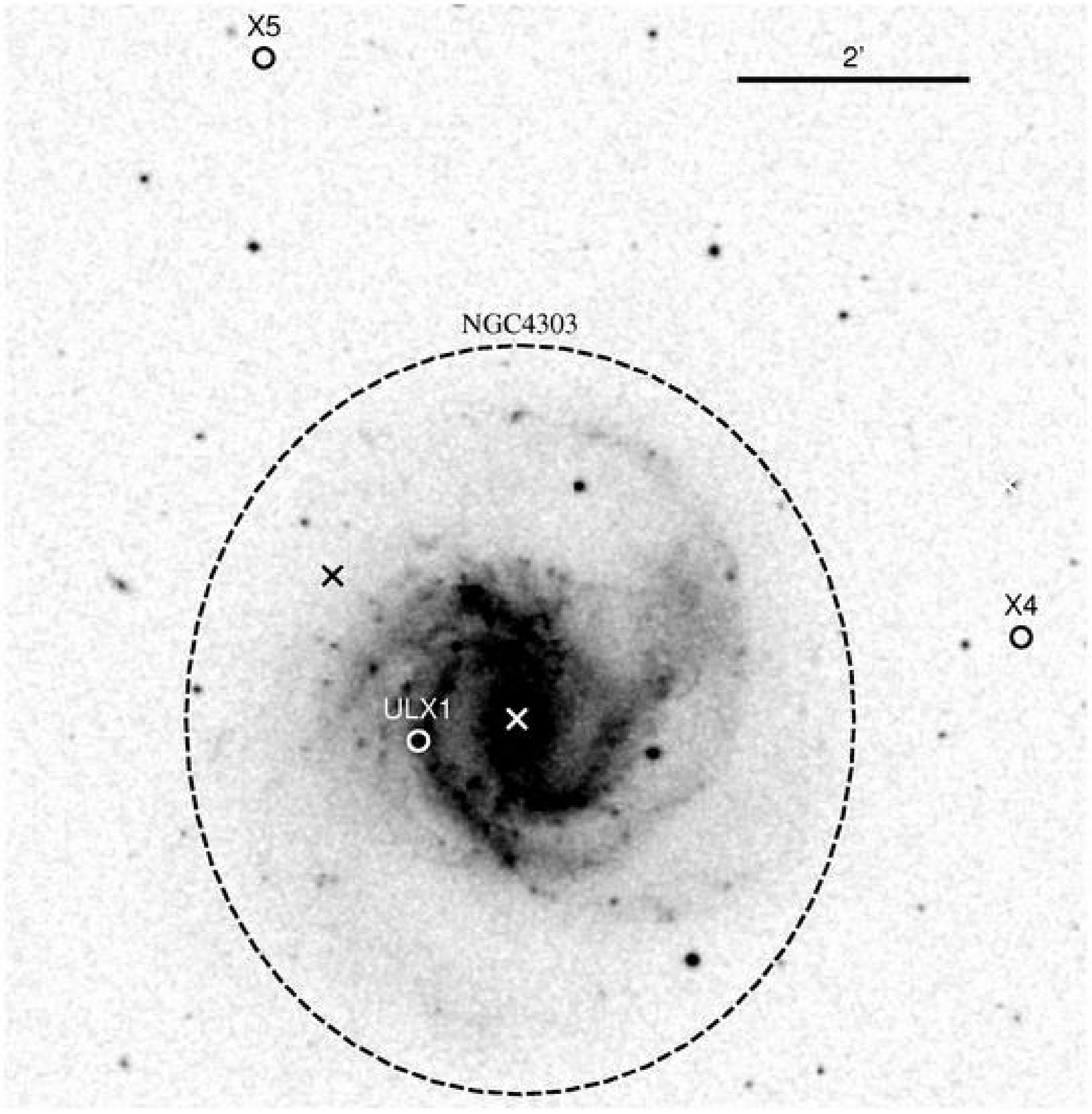}

\caption{The finding chart for the ULXs in NGC4303. }
\end{figure}
\begin{figure}
\plotone{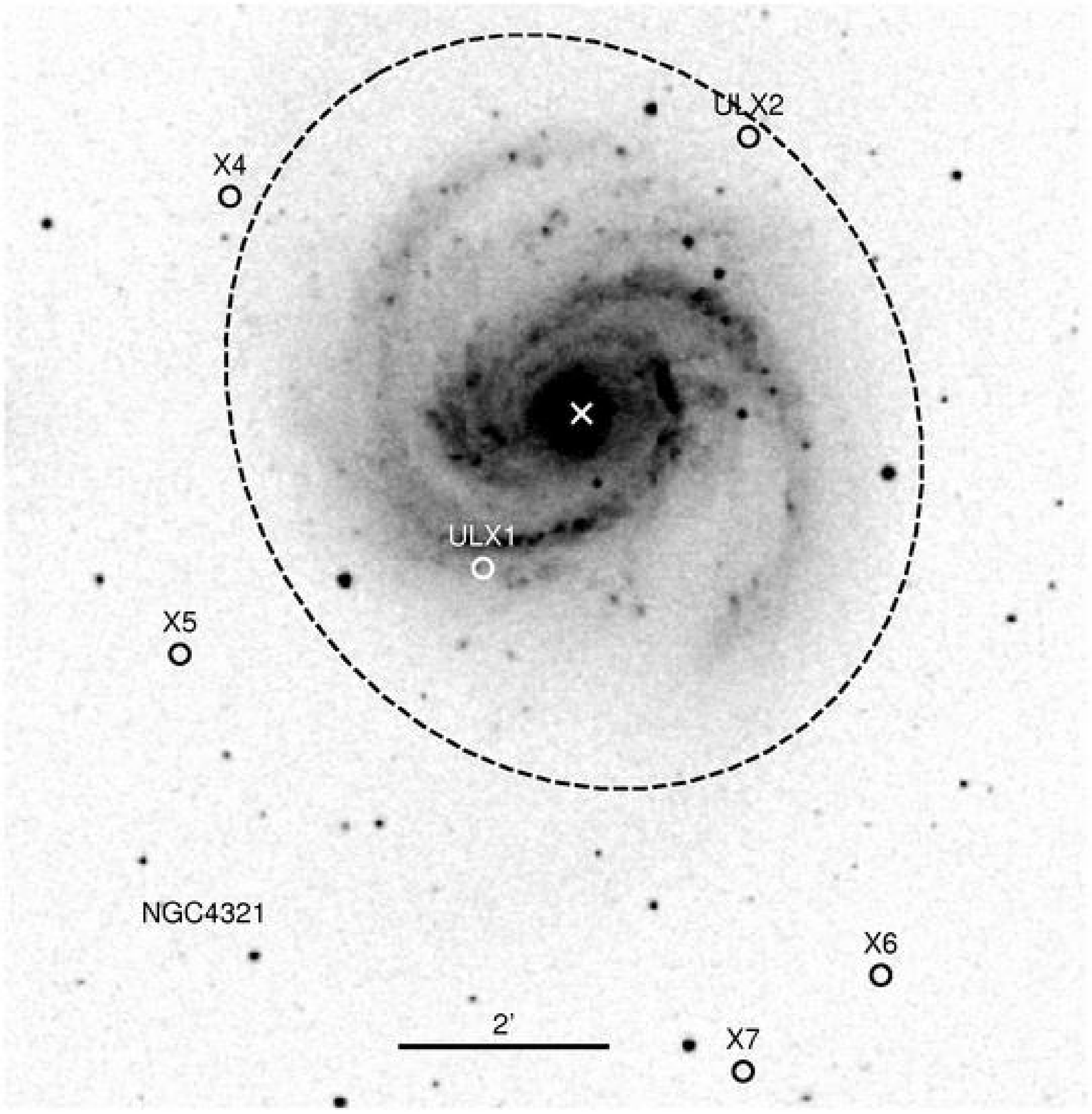}

\caption{The finding chart for the ULXs in NGC4321.}

\end{figure}
\begin{figure}
\plotone{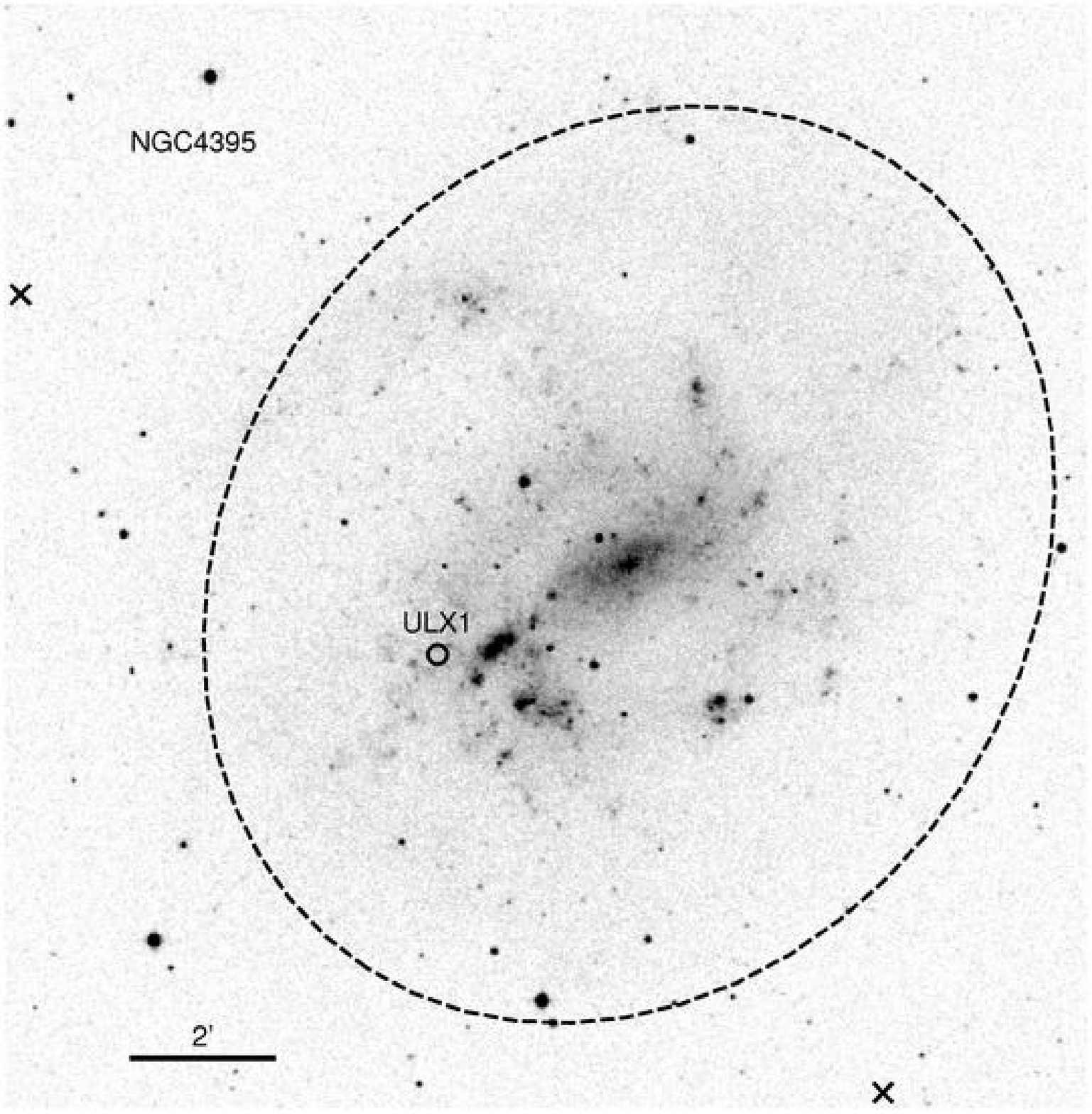}

\caption{The finding chart for the ULXs in NGC4395. }
\end{figure}
\begin{figure}
\plotone{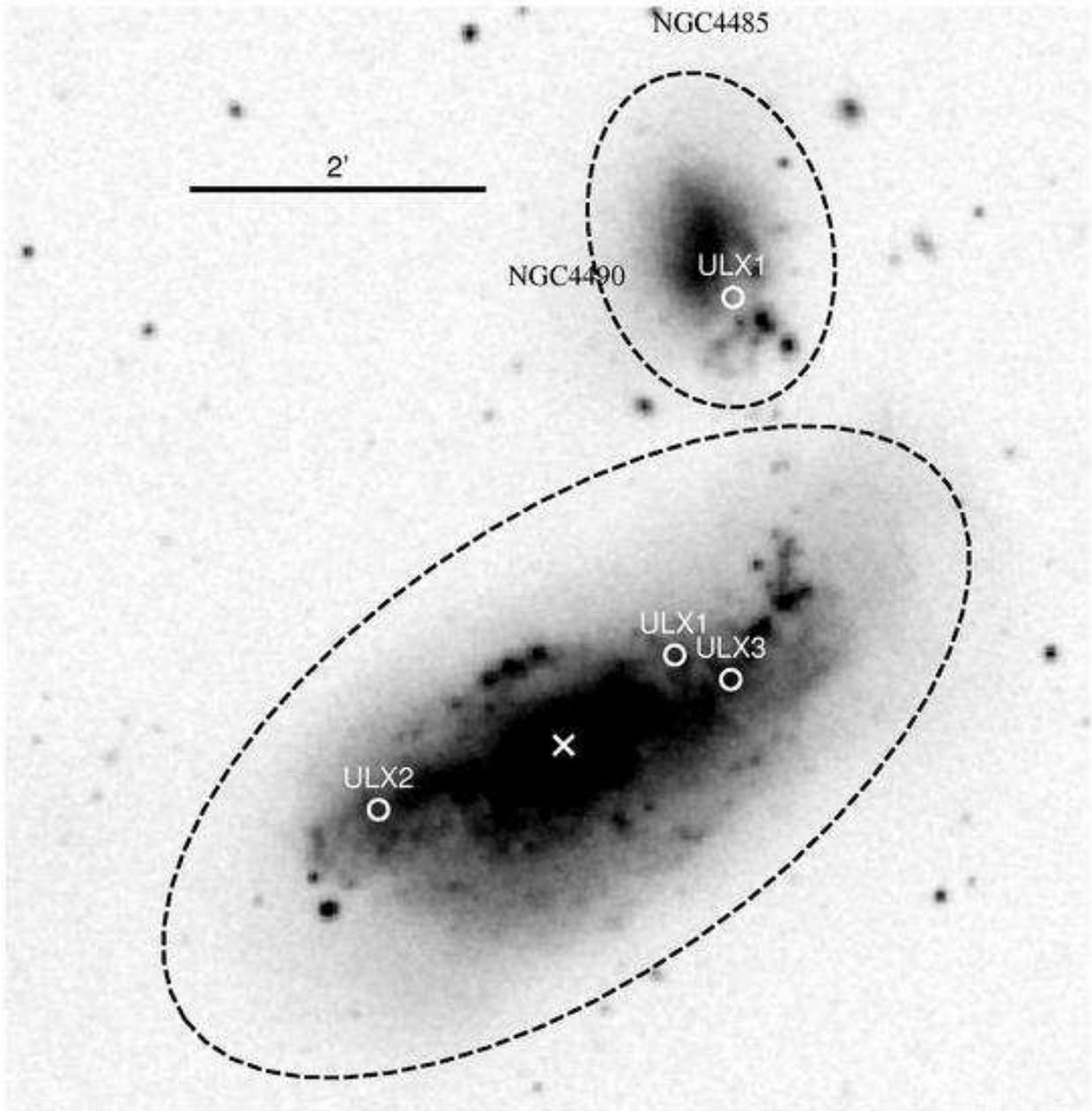}

\caption{The finding chart for the ULXs in NGC4490 and NGC4485.}

\end{figure}
\clearpage

\begin{figure}
\plotone{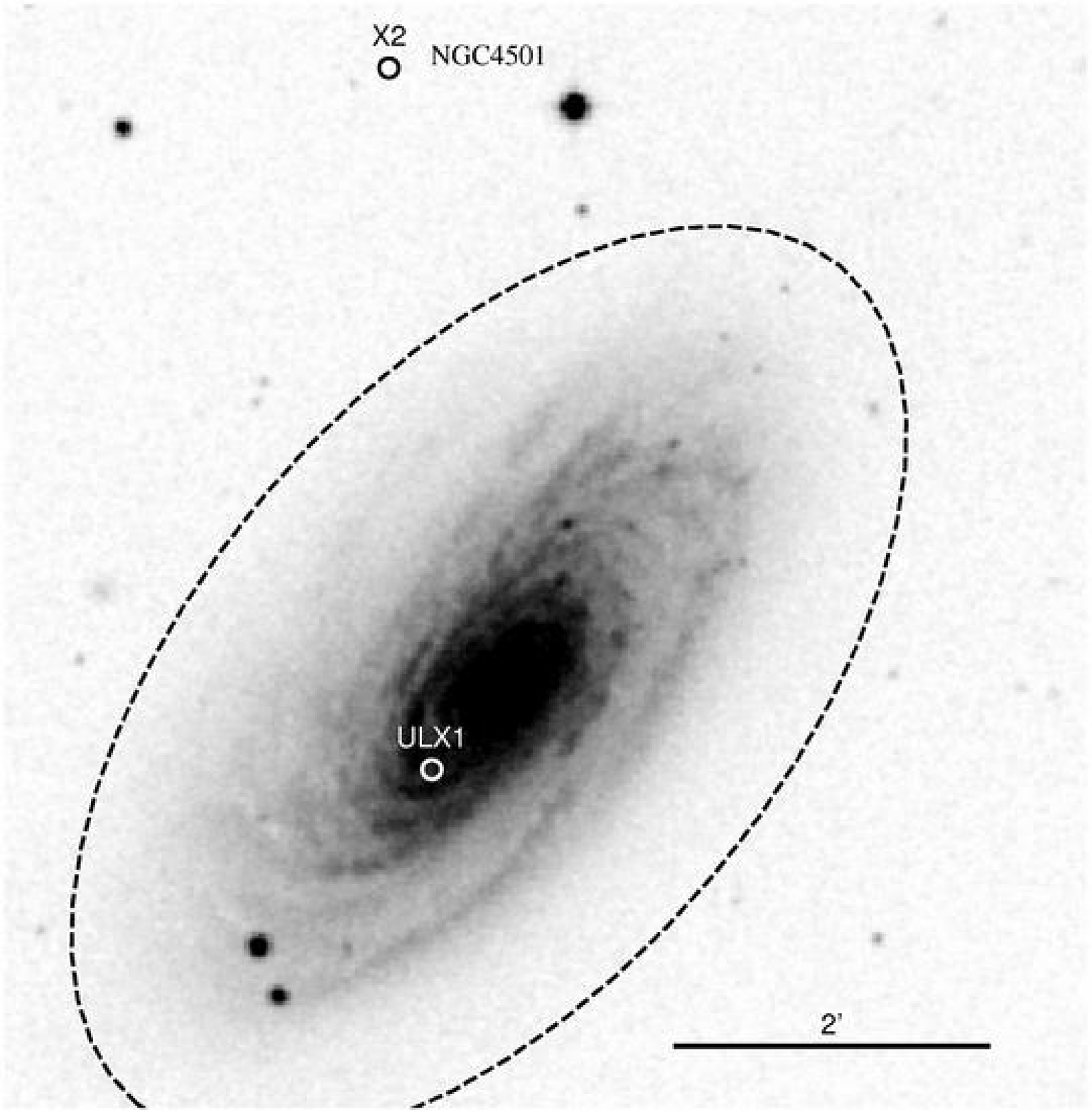}

\caption{The finding chart for the ULXs in NGC4501. }
\end{figure}
\begin{figure}
\plotone{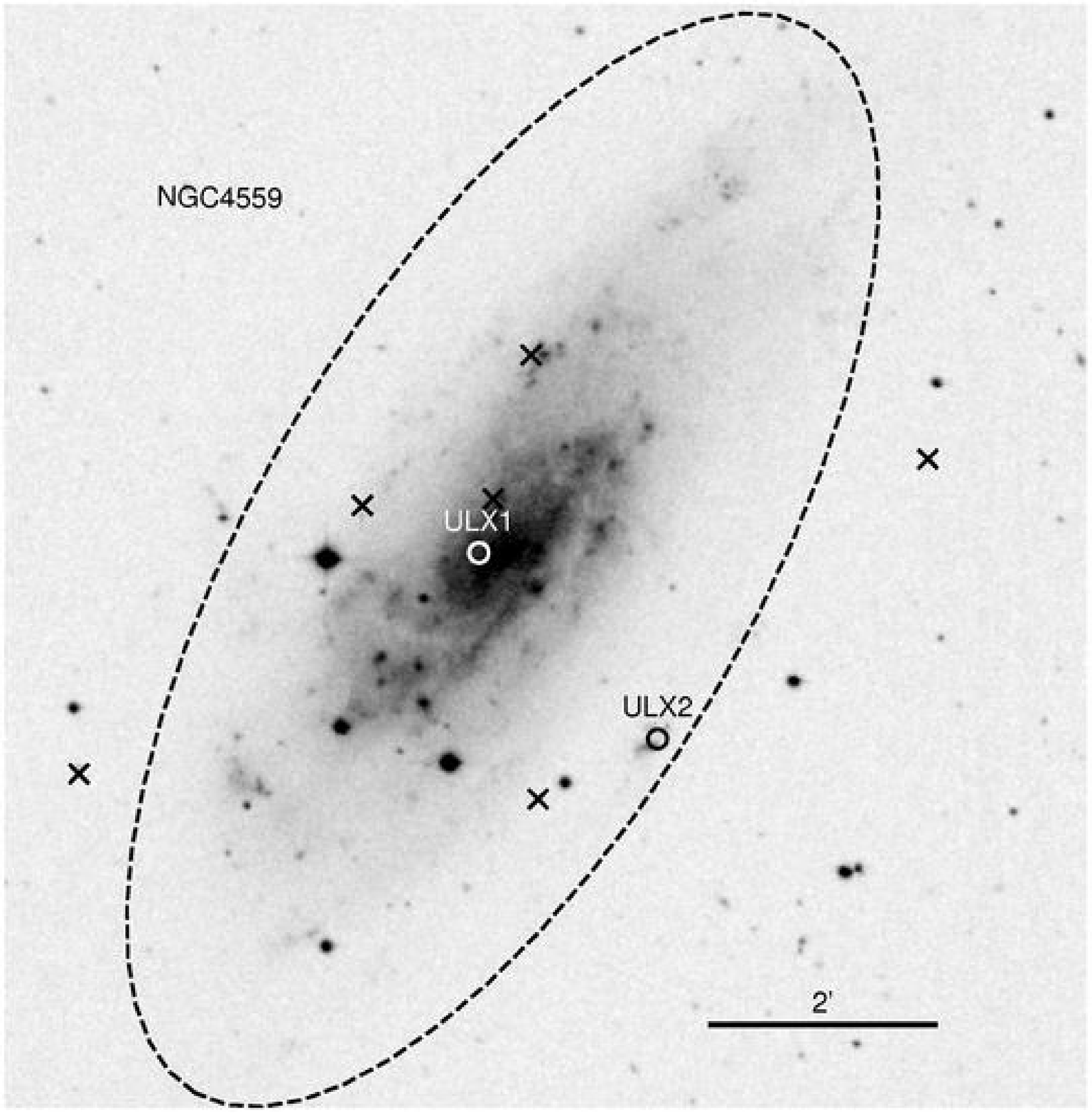}

\caption{The finding chart for the ULXs in NGC4559.}

\end{figure}
\begin{figure}
\plotone{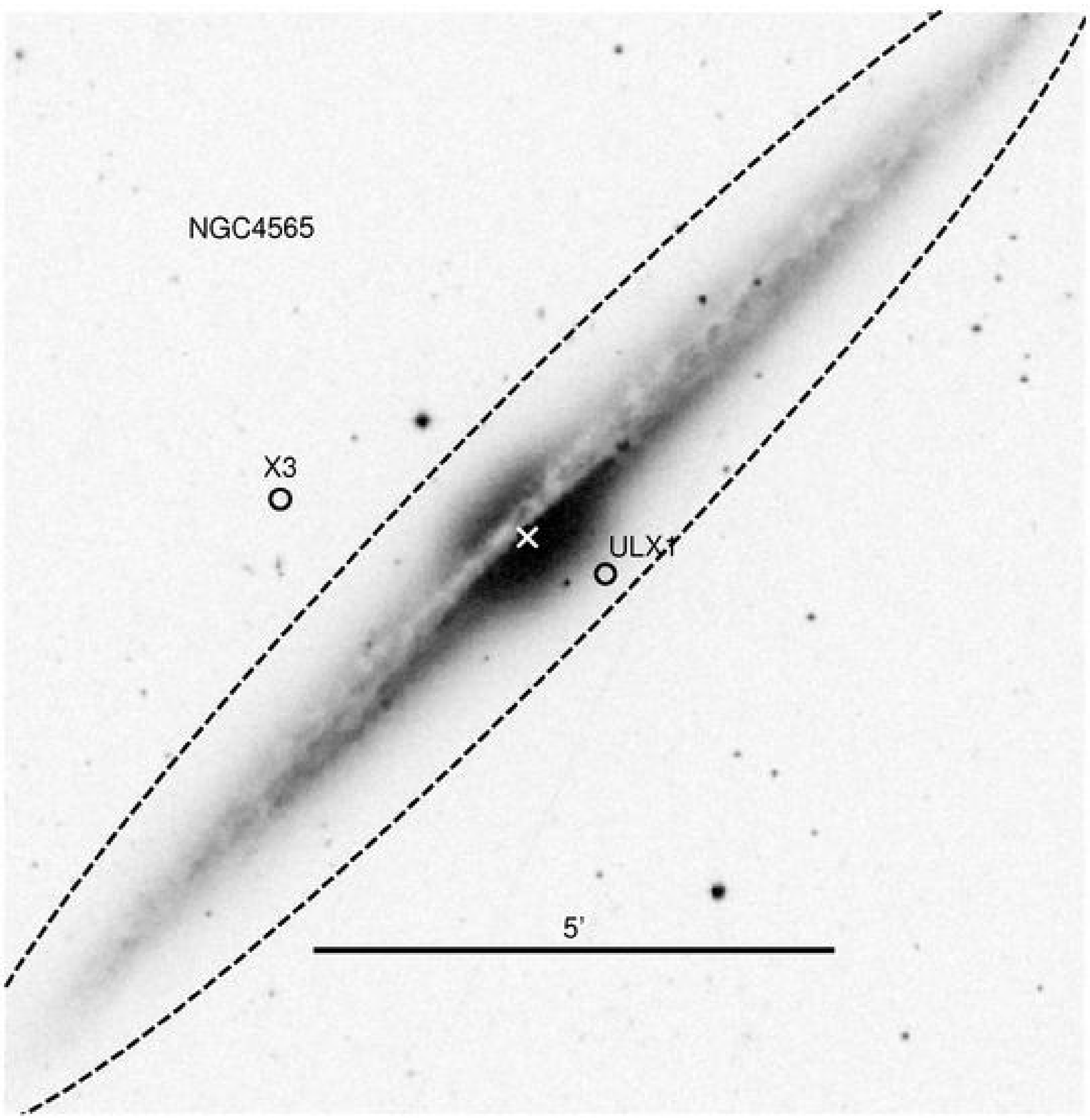}

\caption{The finding chart for the ULXs in NGC4565. }
\end{figure}
\begin{figure}
\plotone{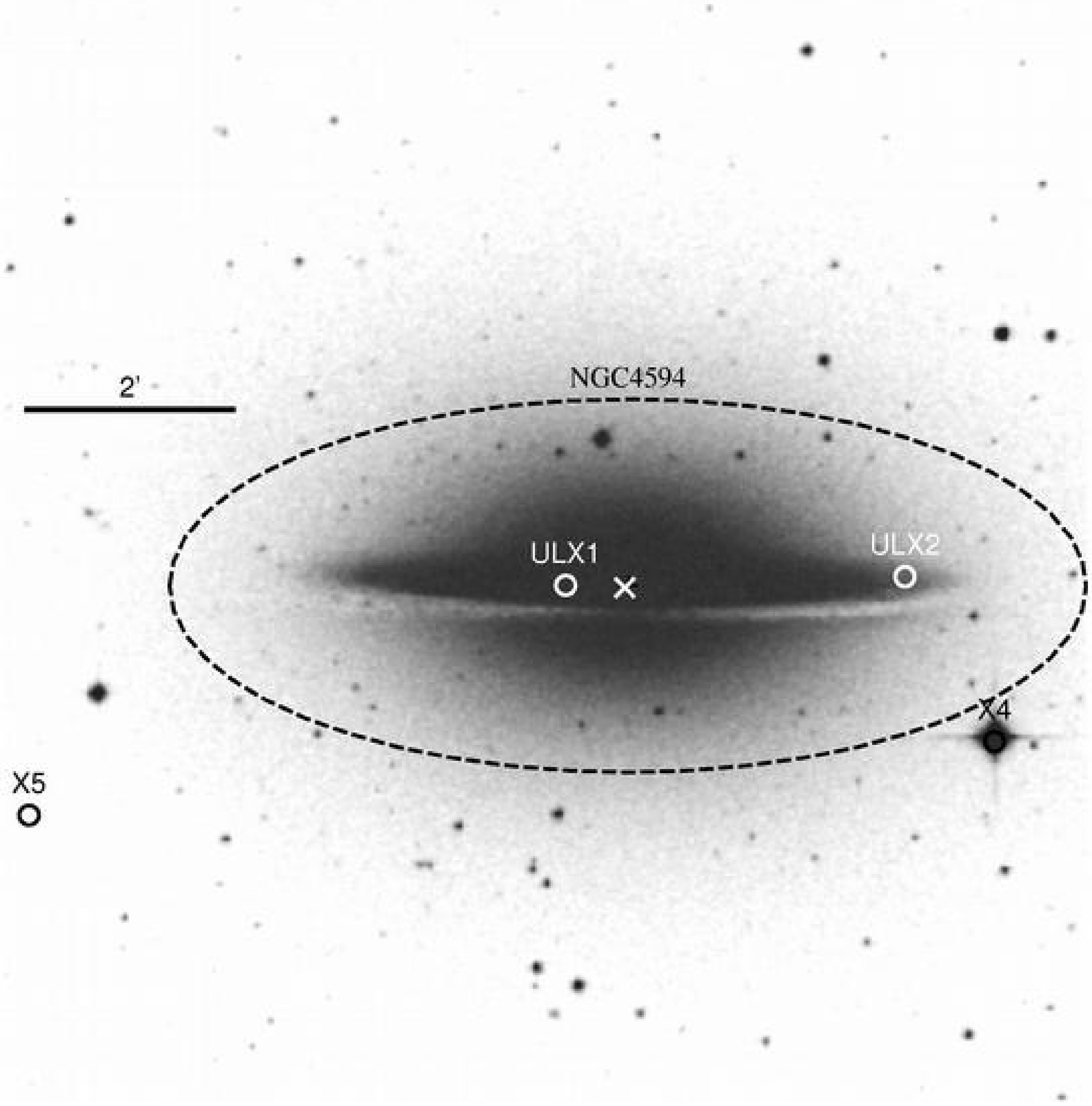}

\caption{The finding chart for the ULXs in NGC4594.}

\end{figure}
\begin{figure}
\plotone{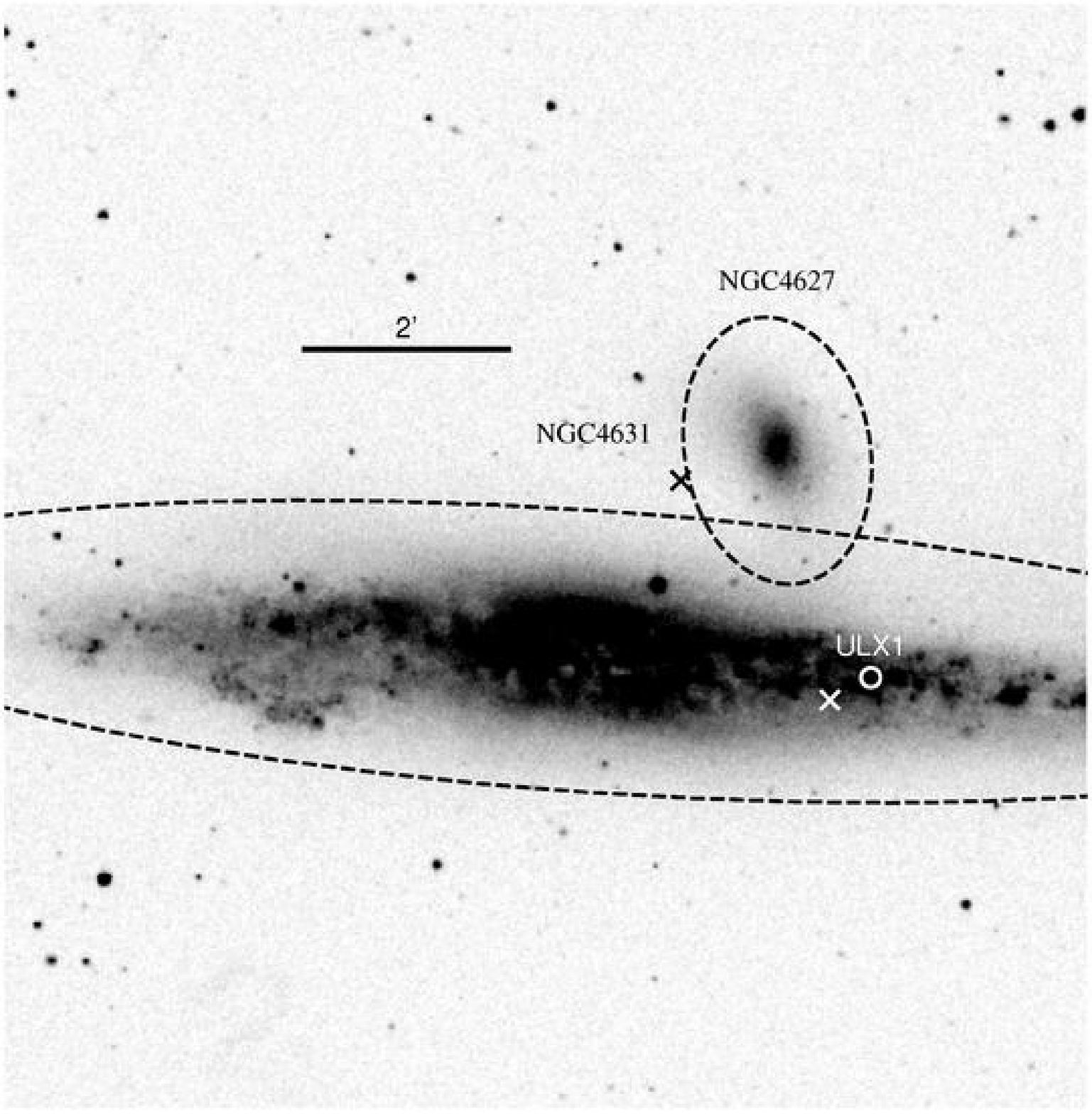}

\caption{The finding chart for the ULXs in NGC4631. }
\end{figure}
\begin{figure}
\plotone{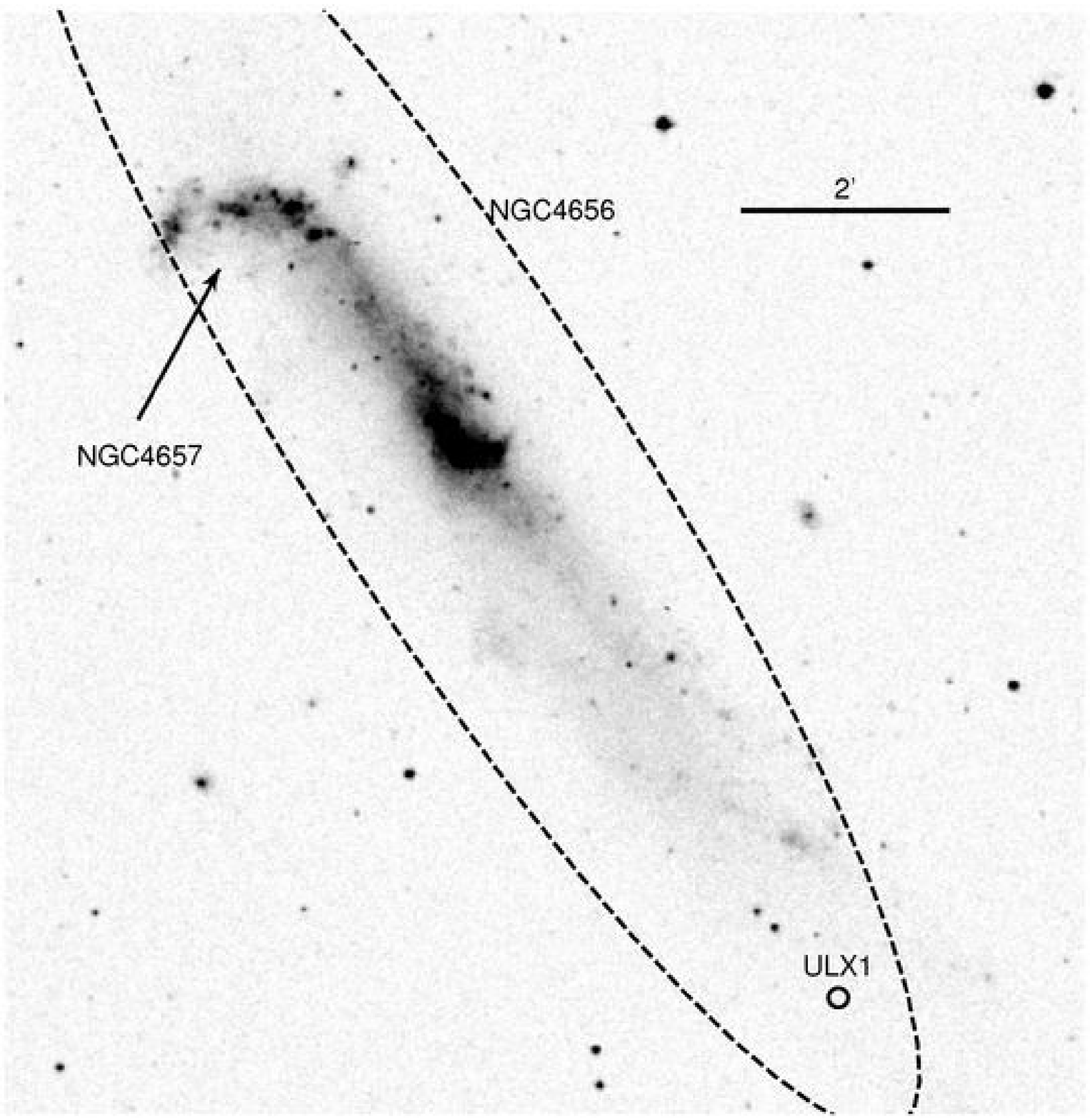}

\caption{The finding chart for the ULXs in NGC4656.}

\end{figure}
\begin{figure}
\plotone{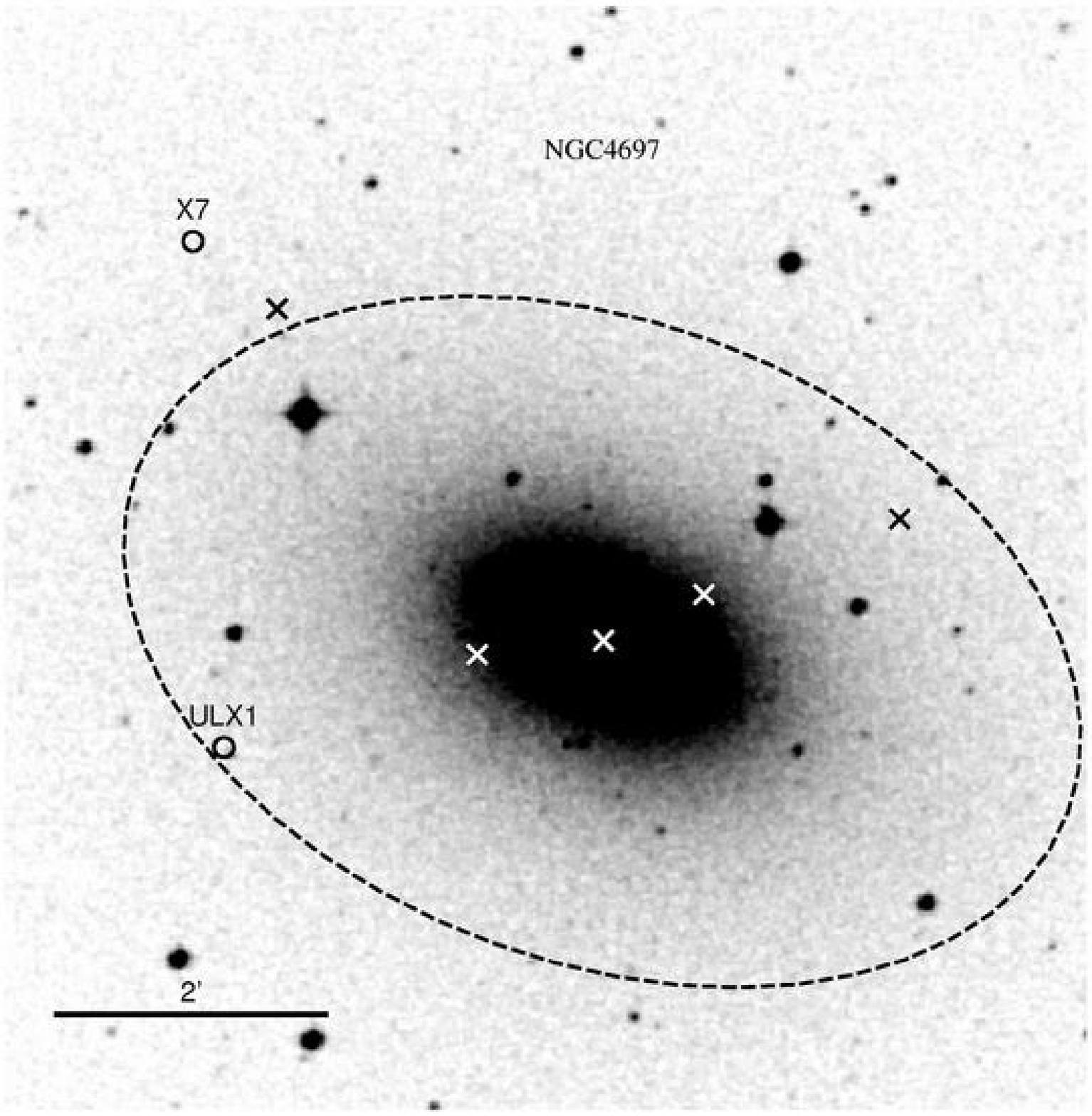}

\caption{The finding chart for the ULXs in NGC4697. }
\end{figure}
\begin{figure}
\plotone{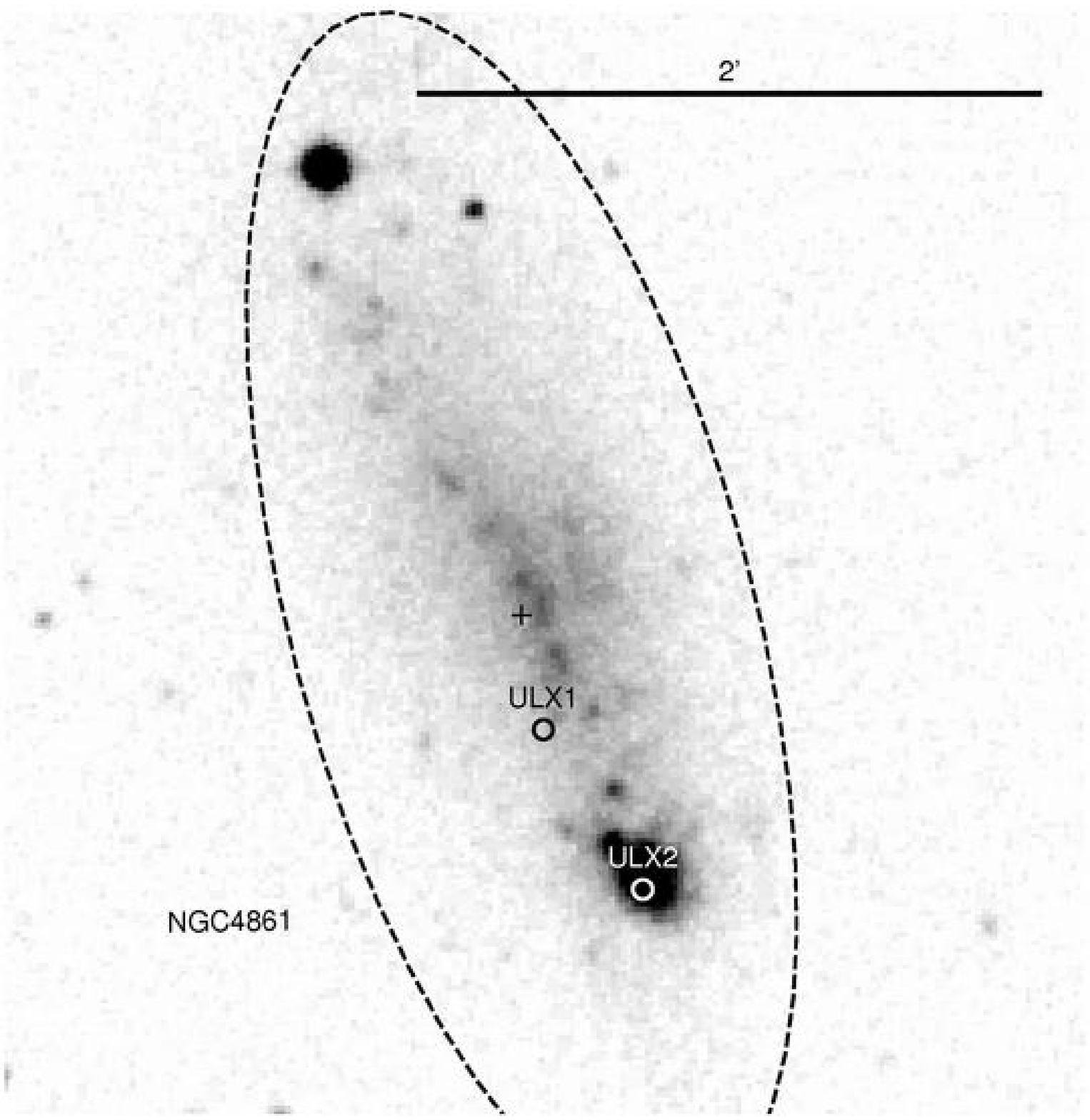}

\caption{The finding chart for the ULXs in NGC4861.}

\end{figure}
\clearpage

\begin{figure}
\plotone{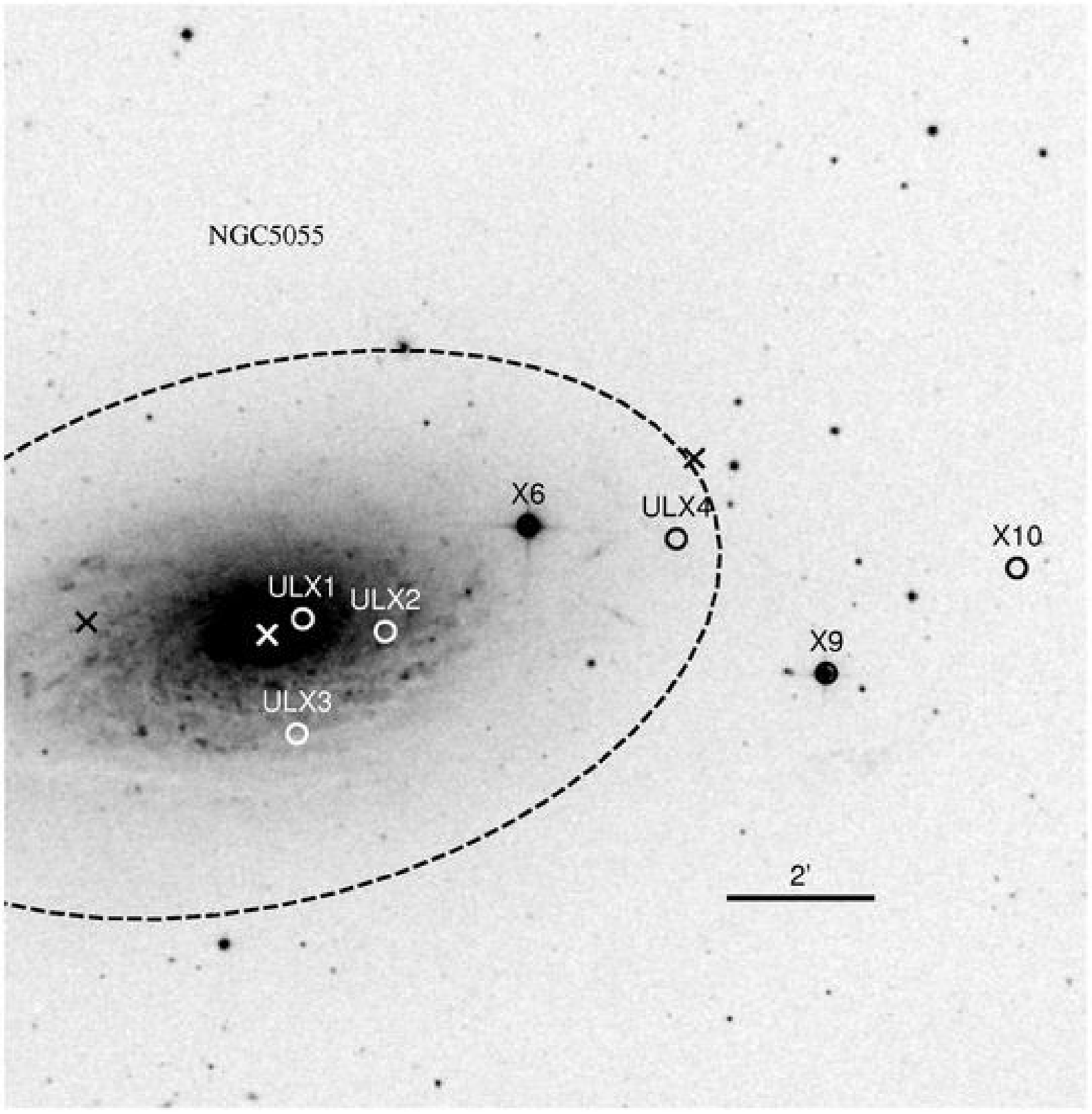}

\caption{The finding chart for the ULXs in NGC5055. }
\end{figure}
\begin{figure}
\plotone{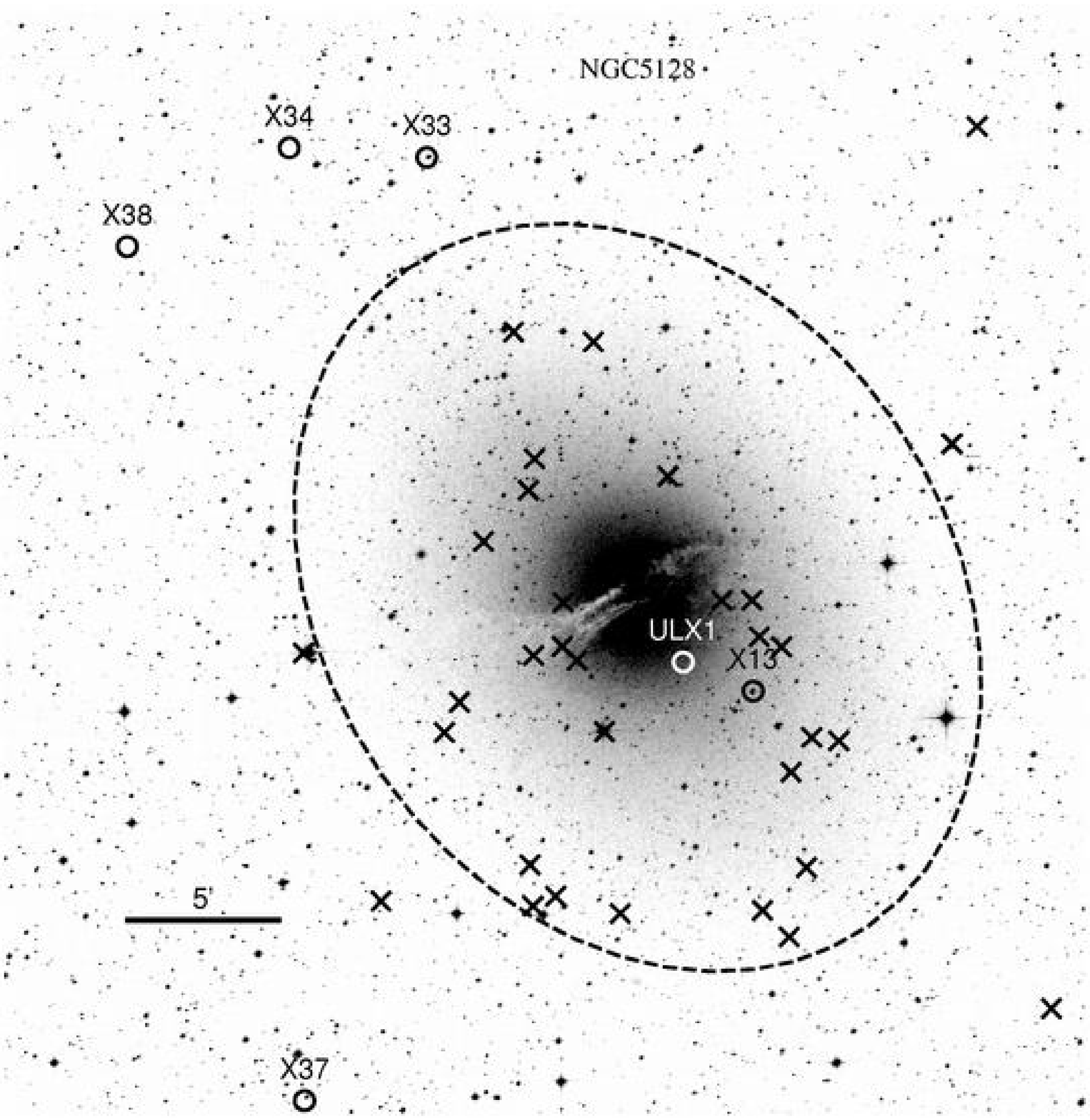}

\caption{The finding chart for the ULXs in NGC5128.}

\end{figure}
\begin{figure}
\plotone{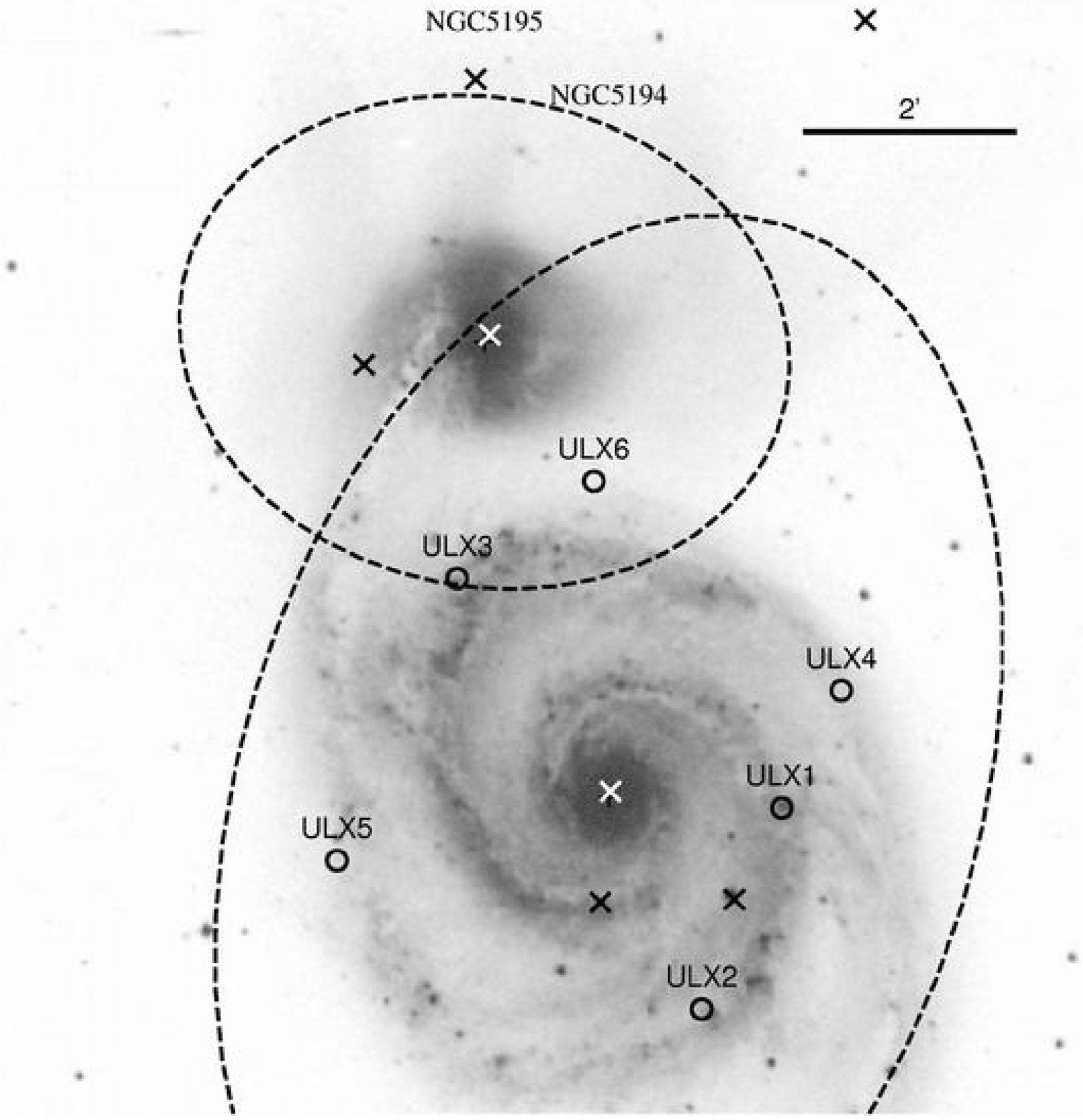}

\caption{The finding chart for the ULXs in NGC5194. }
\end{figure}
\begin{figure}
\plotone{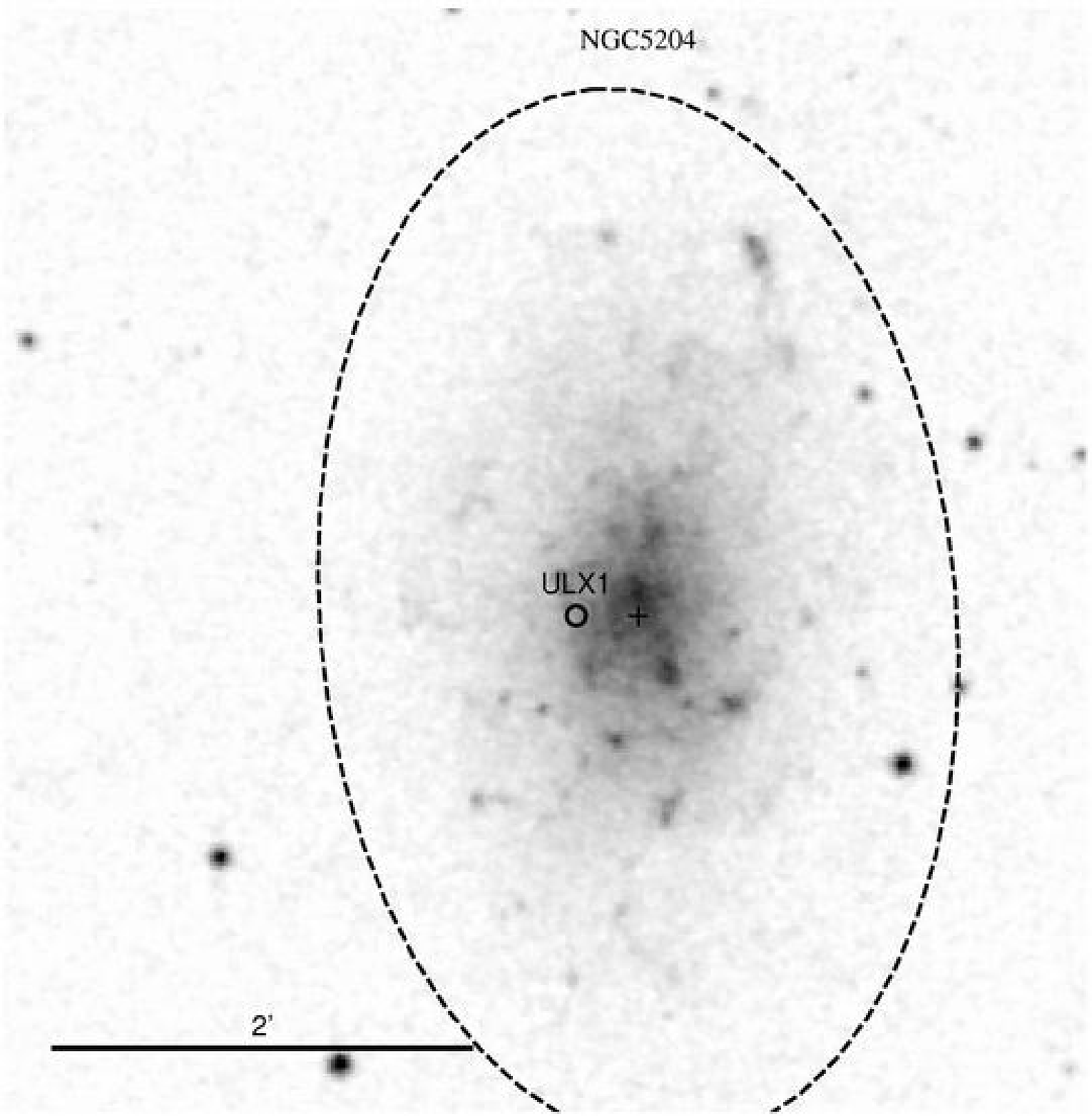}

\caption{The finding chart for the ULXs in NGC5204.}

\end{figure}
\begin{figure}
\plotone{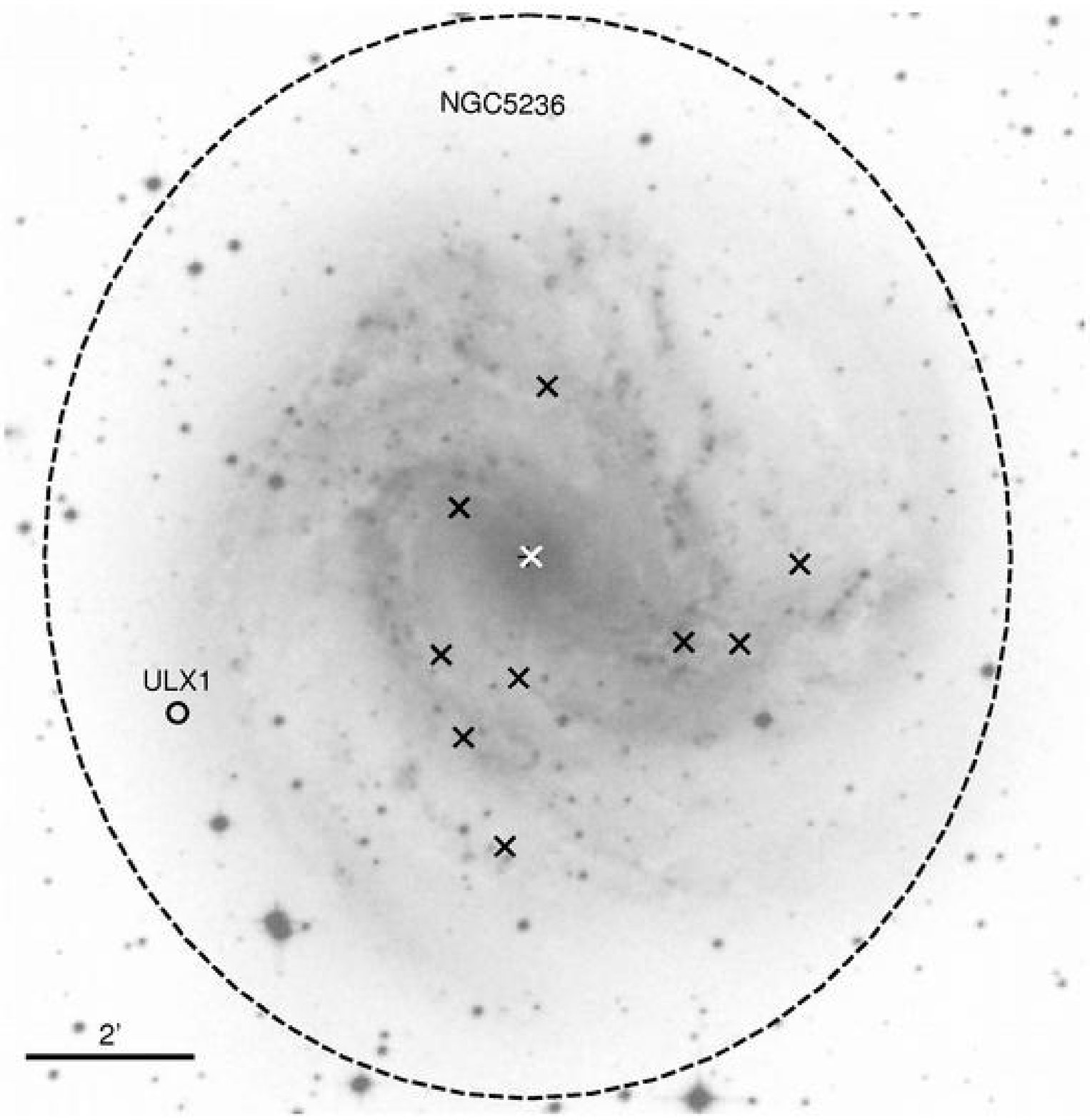}

\caption{The finding chart for the ULXs in NGC5236. }
\end{figure}
\begin{figure}
\plotone{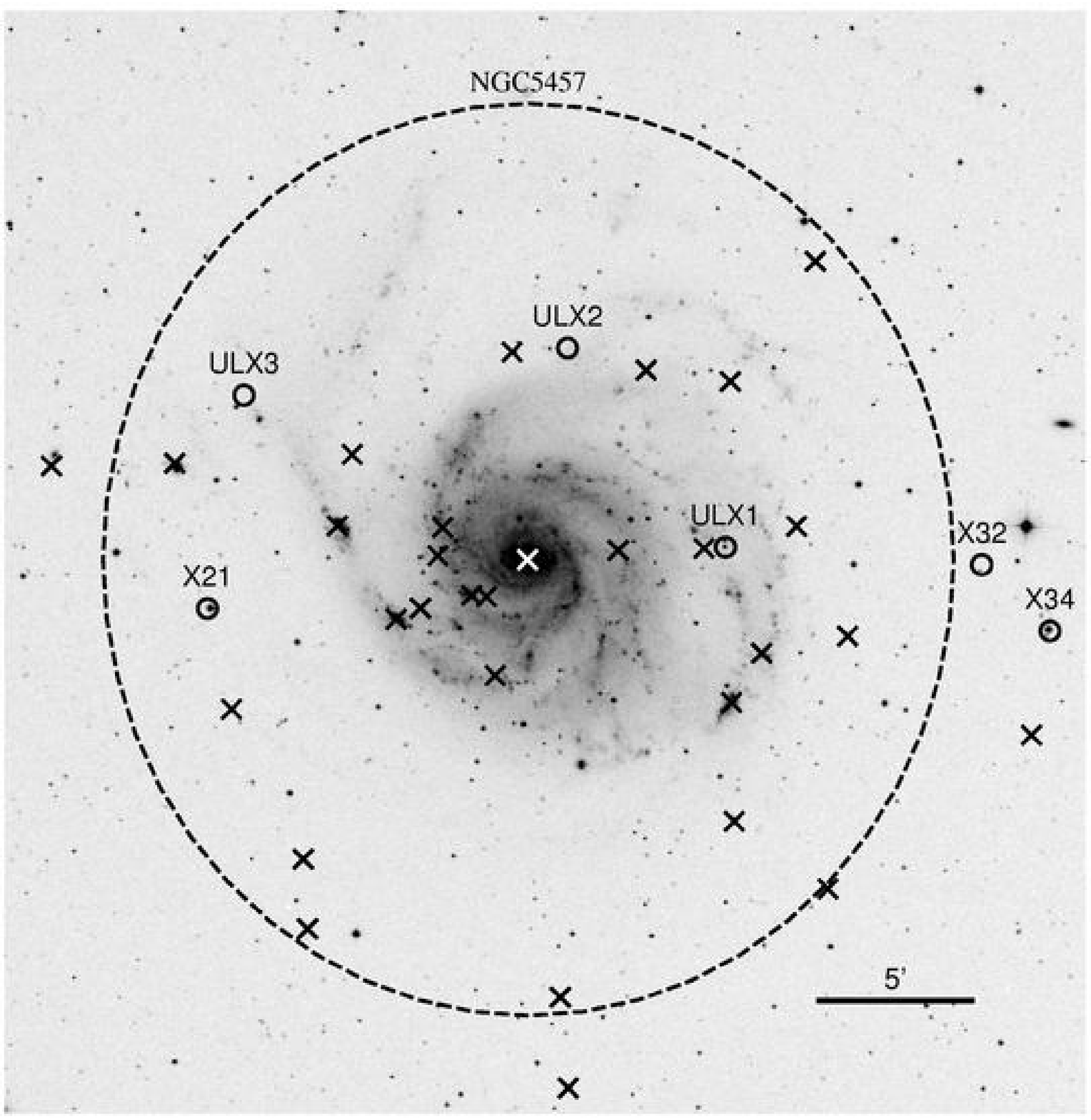}

\caption{The finding chart for the ULXs in NGC5457.}

\end{figure}
\begin{figure}
\plotone{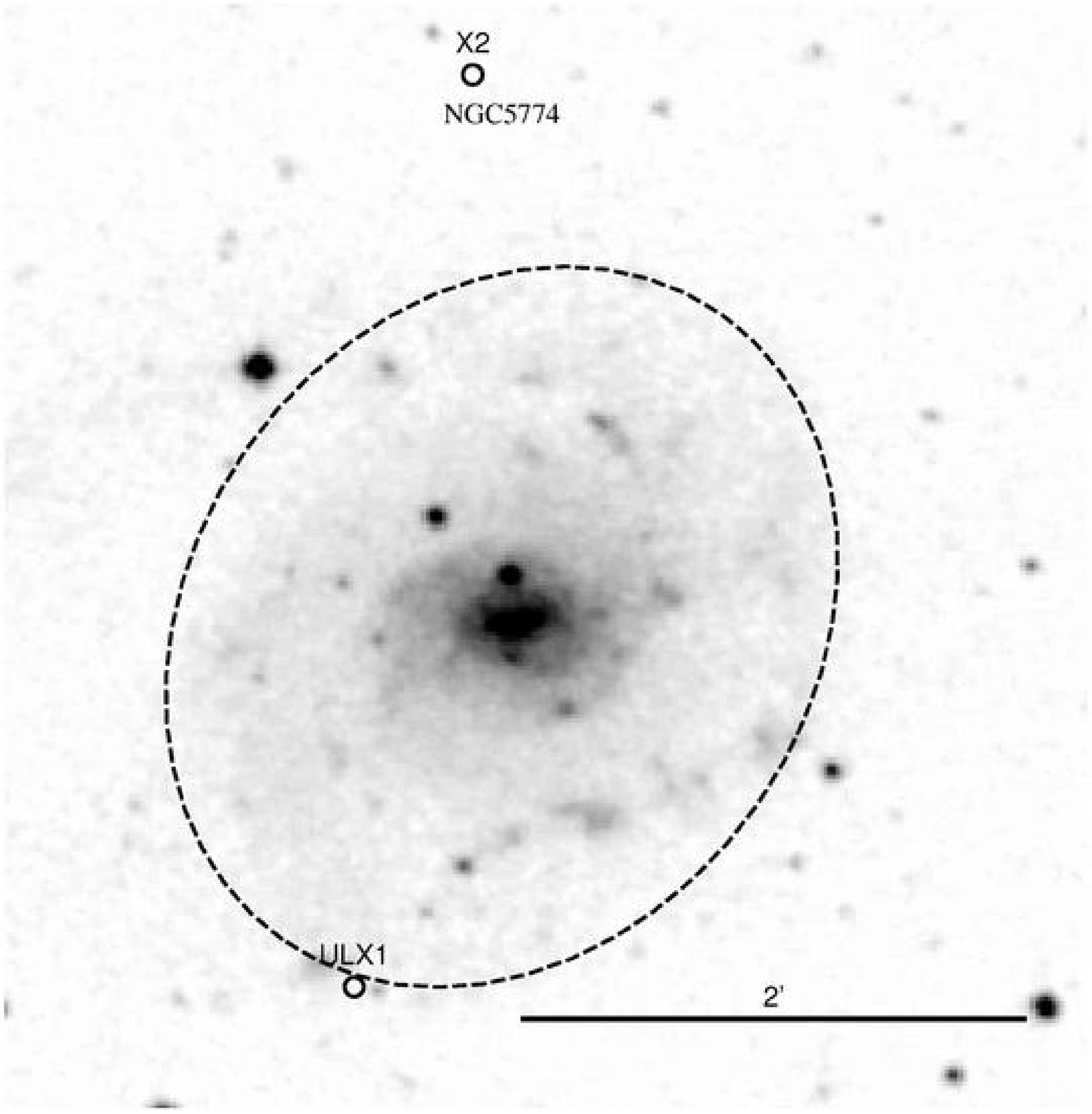}

\caption{The finding chart for the ULXs in NGC5774. }
\end{figure}
\begin{figure}
\plotone{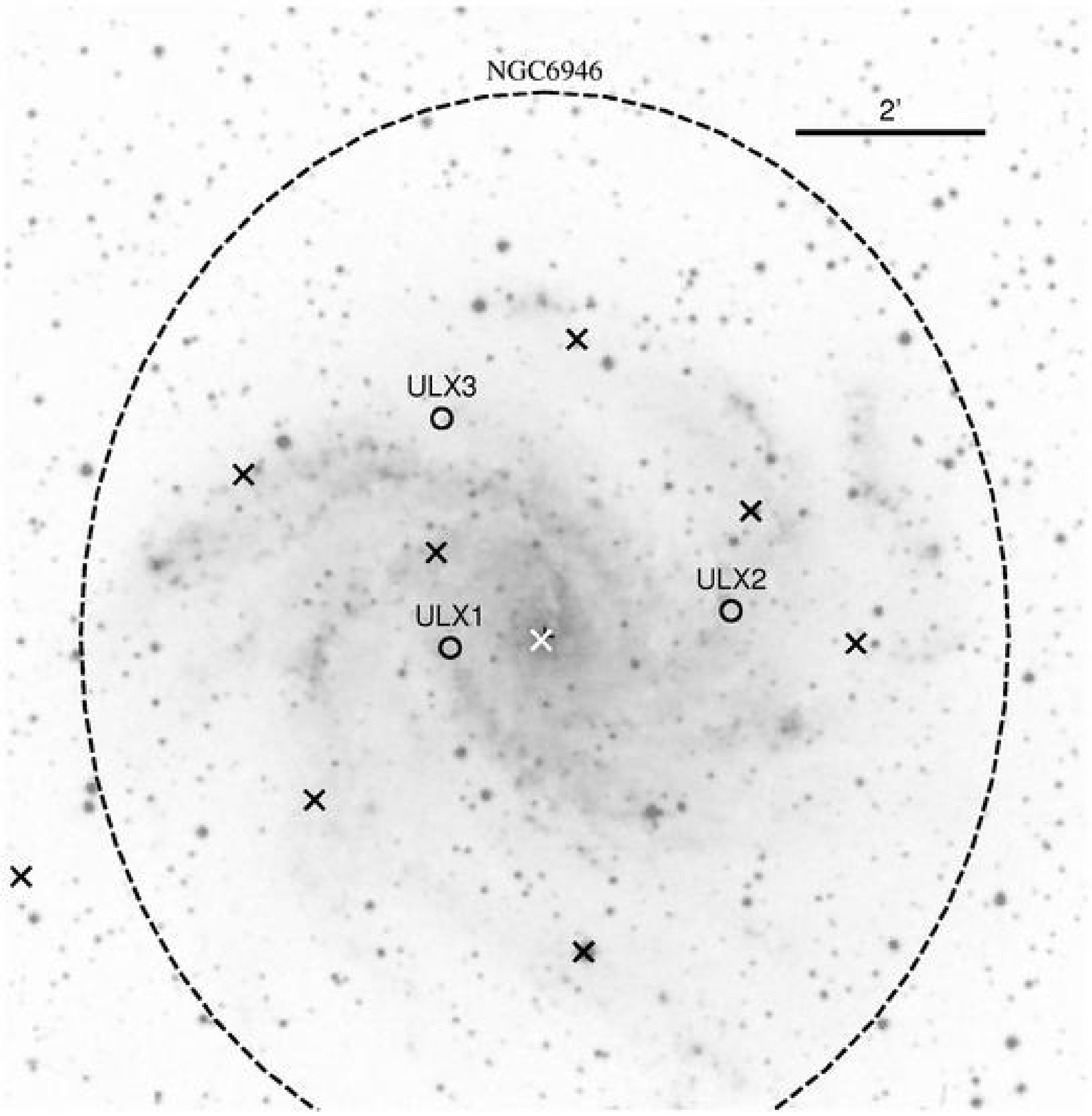}

\caption{The finding chart for the ULXs in NGC6946.}

\end{figure}
\clearpage

\begin{figure}
\plotone{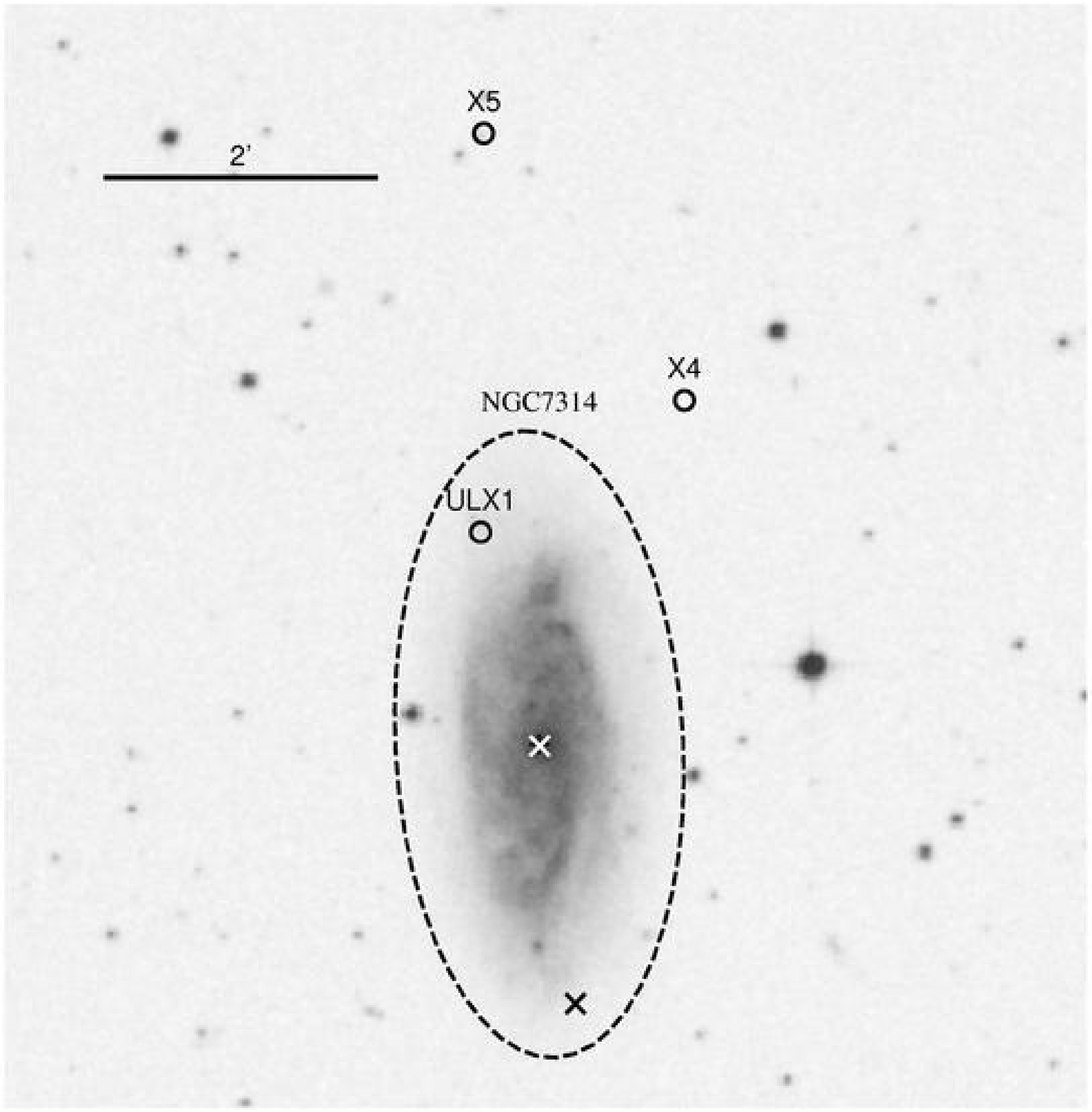}

\caption{The finding chart for the ULXs in NGC7314. }
\end{figure}
\begin{figure}
\plotone{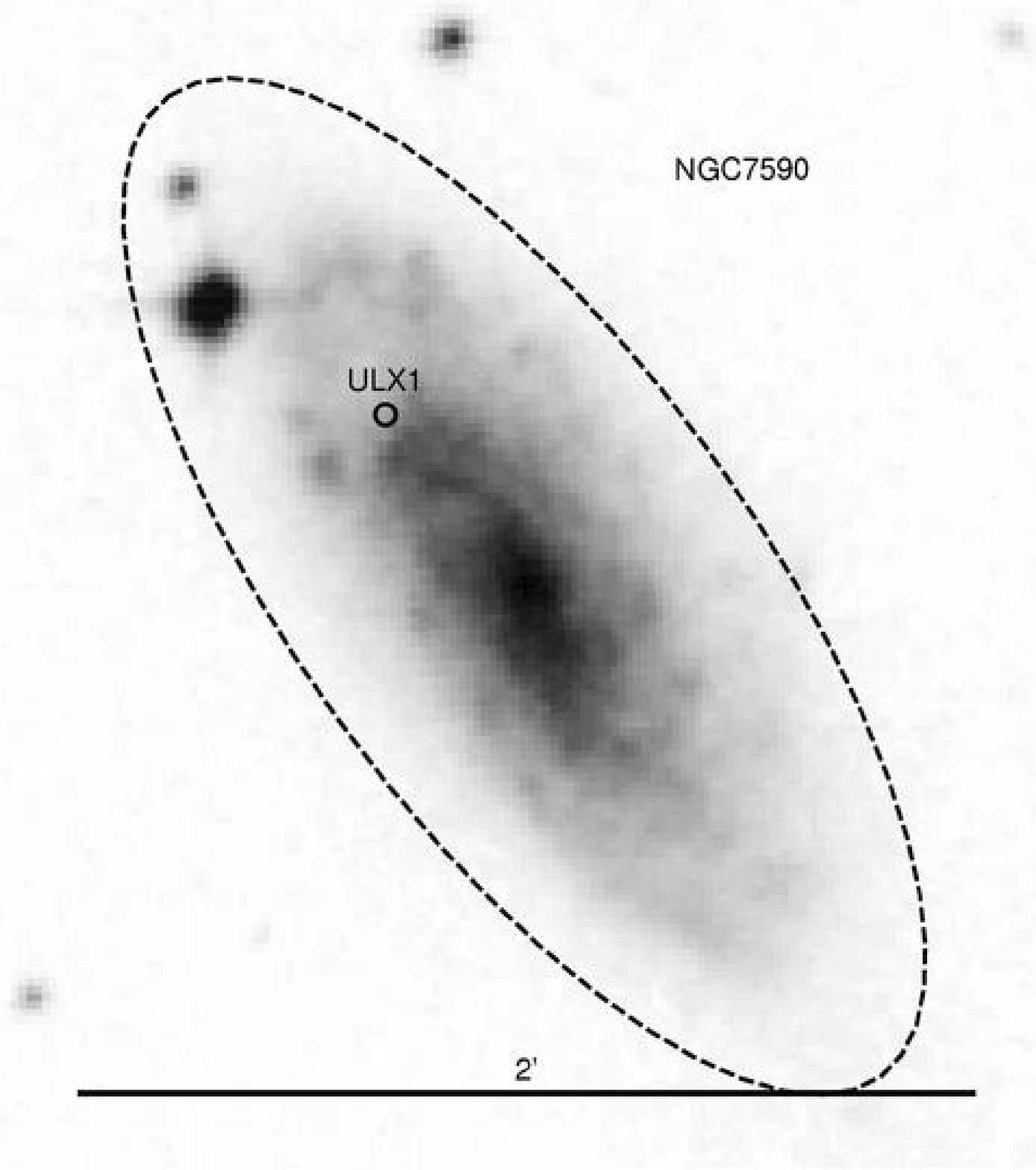}

\caption{The finding chart for the ULXs in NGC7590.}

\end{figure}
\begin{figure}
\plotone{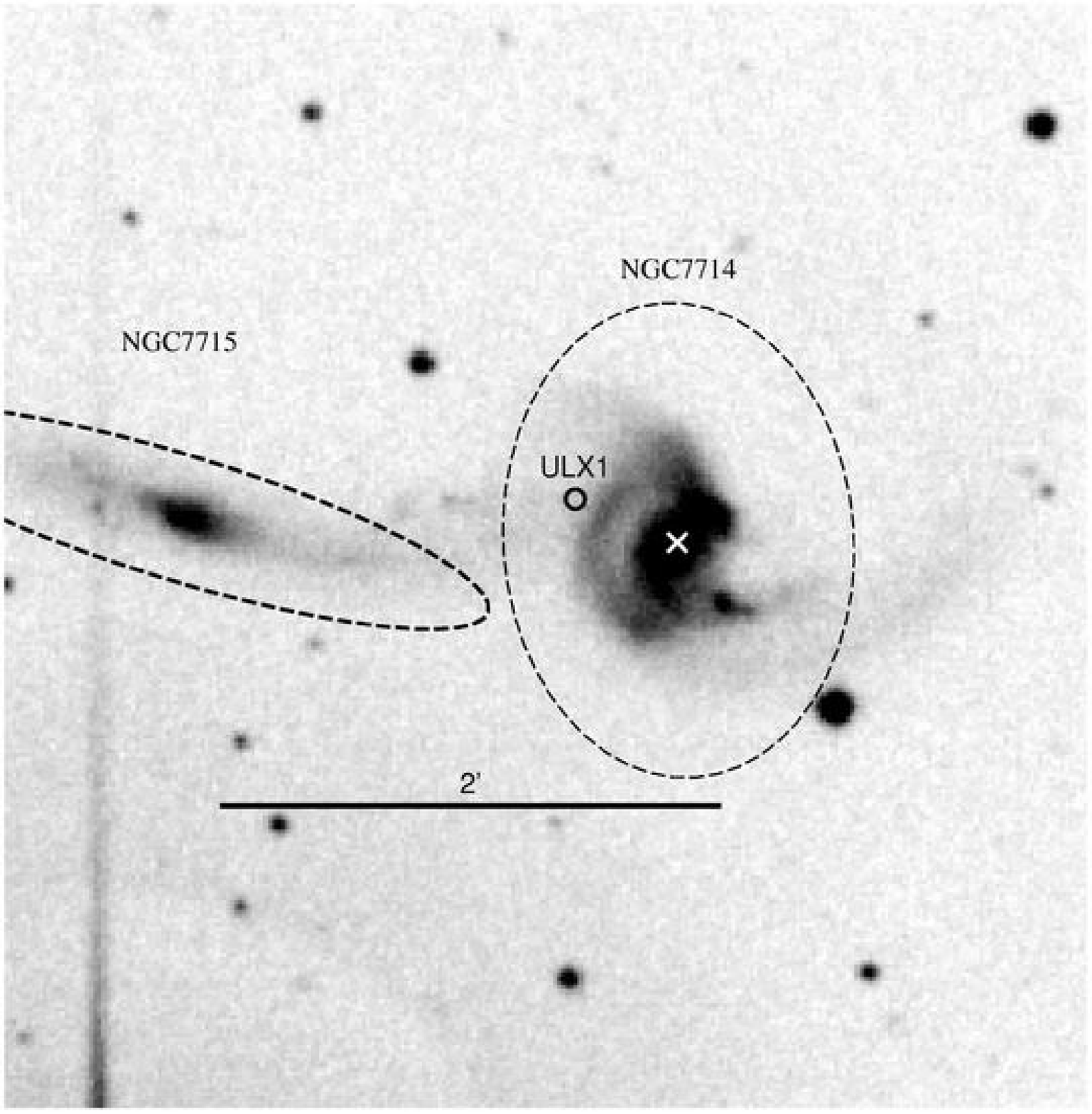}

\caption{The finding chart for the ULXs in NGC7714. }
\end{figure}
\begin{figure}
\plotone{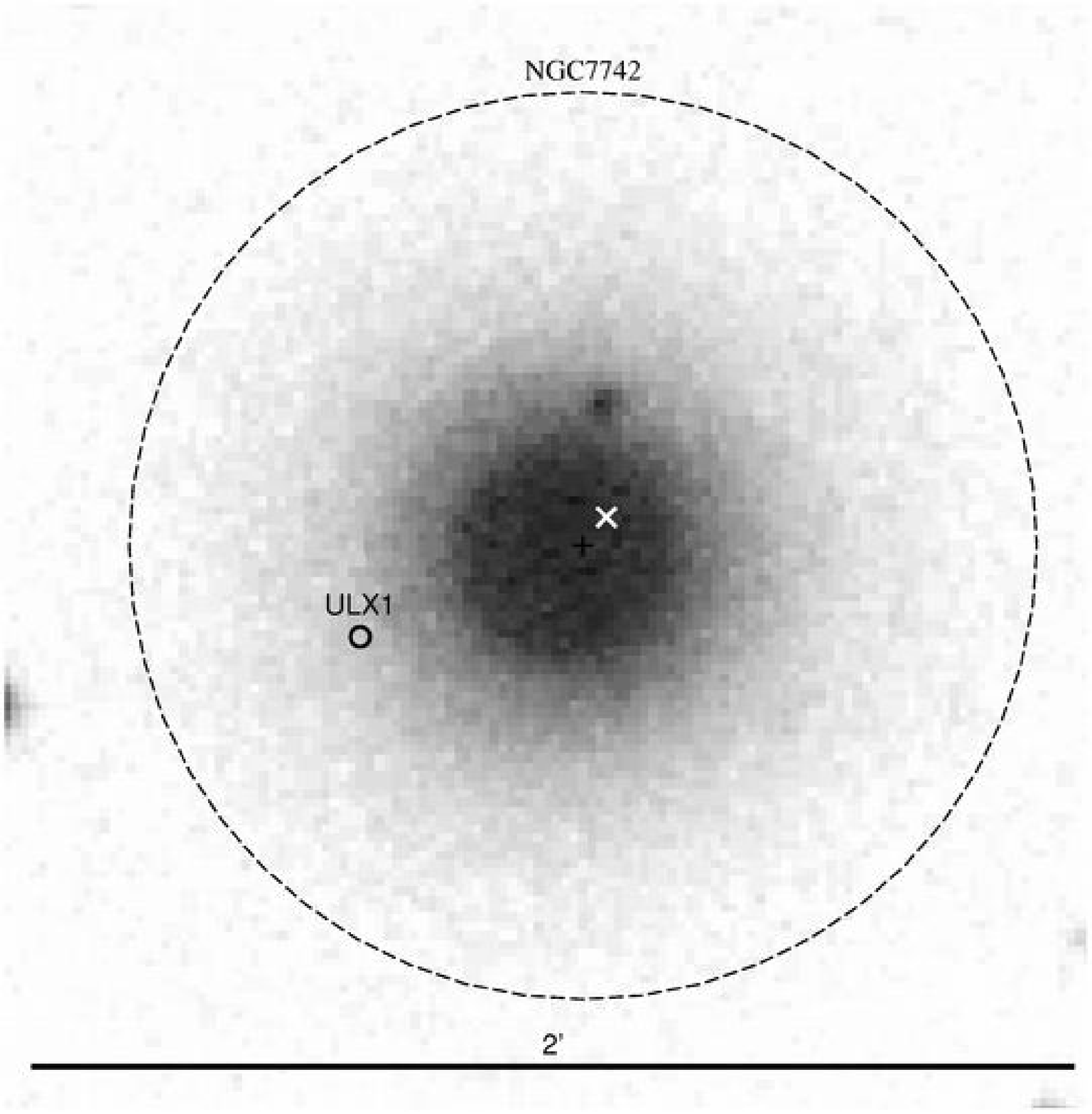}

\caption{The finding chart for the ULXs in NGC7742.}

\end{figure}


\begin{thebibliography}{}

\bibitem{} Angelini, L., Loewenstein, M., and Mushotzky, F., 2001, \apj, 557, L35


\bibitem{} Bauer, F.E., Brandt, W.N., Sambruna, R.M., Chartas, G., Garmire,
G.P., Kaspi, S., and Netzer, H., 2001, \aj, 122, 182



\bibitem{} Becker, R., Helfand, D., White R., Gregg M., and Laurent-Muehleisen S., 1997, \apj, 475, 479, version2003 (the FIRST survey catalog)



\bibitem{} Begelman, M.C., 2002, \apj, 568, L97



\bibitem{} Colbert, E. J. M. and Mushotzky, R. F. 1999, \apj, 519, 89



\bibitem[CP2002]{} Colbert, E. and Ptak, A. 2002, ApJS, 143, 25 (CP2002)



\bibitem{} Cutri, R., et al. 2003 (2MASS)



\bibitem{} Dickey, J, and Lockman, F., 1990, ARA\&A, 28, 215

\bibitem{} Dunne, B., Points, S., and Chu, Y., 2001, AJ, 119, 1172

\bibitem{} Freedman, W., Madore, B., Gibson, B., Ferrarese, L., Kelson, D. et al., 2001, \apj, 553, 47 (KP)

\bibitem{} Fabbiano, G, Gioia, I., and Trinchieri, G., 1989, ARA\&A, 27, 87


\bibitem{} Hasinger, G., Burg, R., Giacconi, R., Hartner, G., Schmidt, M., Trumper, J. and Zamorani, G. . 1993, \aap, 275,1


\bibitem{} Ho, L., Filippenko, A., and Sargent, W., 1995, ApJS, 98, 477

\bibitem{} Hog, E., Fabricius, C., Makarov, V.V., Urban, S., Corbin T., et al. 2000, A\&A 355, L27 (Tycho-2)



\bibitem{} Irwin, J., Bregman, J., and Athey, A., 2004, ApJL, 60, 143



\bibitem{} King, A. R., Davies, M. B., Ward, M. J., Fabbiano, G. and Elvis, M.  2001, \apj, 552, L109


\bibitem{} Kobuta, A., Mizuno, T., Makishima, K., et al. 2001, \apj, 547, L119


\bibitem{} La Perola, V., Peres, G., et al. 2001, \apj, 556, 47

\bibitem{} Liu, J., Bregman, J., and Seitzer, P., 2002, ApJL, 580, 31

\bibitem{} Liu, J., Bregman, J., and Irwin, J., 2002, ApJL, 581, 93

\bibitem{} Liu, J., Bregman, J., and Seitzer, P., 2004, ApJ, 602, 249

\bibitem{} Monet D.G., Levine S.E., Casian B., et al. 2003, AJ, 125, 984 (USNO-B1.0)



\bibitem{} Morse J., 1994, PASP, 106, 675



\bibitem{} Pakull, M, and Mirioni, L., 2002, astro-ph/0202488 (PM2002)


\bibitem{} Portegies Zwart, S., 2004, AAS203, 102.03, and private discussions

\bibitem{} Ptak, A., and Colbert, E., 2004, astro-ph/0401525

\bibitem[RW2000]{} Roberts, T. P. and Warwick, R. S. 2000, MNRAS, 315, 98 (RW2000)


\bibitem{} Roberts, T., and Colbert, E., 2003, MNRAS, 341, L49


\bibitem{} ROSAT Scientific Team, ROSAT NEWS No. 71, The ROSAT Consortium (2000)

\bibitem{} Stocke, J., Morris, S., et al. 1991, ApJS, 76, 813

\bibitem{} Swartz, D., Ghosh, K., Tennant, A., and Wu, K., 2004, astro-ph/0405498

\bibitem{} Tongue, T., and Westpfahl, D., 1995, AJ, 109, 2462

\bibitem{} Tonry J.L., Dressler A., Blakeslee J.P., Ajhar E.A., Fletcher A.B., et al., 2001, \apj, 546, 681 (SBF)



\bibitem{} TULLY R.B, 1988, {\it Nearby Galaxies Catalogue}, Cambridge University Press. (T88)



\bibitem{} Tully R.B., Shaya E.J., Pierce M.J., 1992, ApJS, 80, 479 (T92)


\bibitem{} Zampieri, L, Mucciarelli, P., et al., 2004, \apj, 603, 523



\end{thebibliography}
\end{document}